\documentclass[12pt,a4paper]{article}
\usepackage[utf8]{inputenc}
\usepackage{amsmath}
\usepackage{amsfonts}
\usepackage{amssymb}
\usepackage{geometry}
\usepackage{slashed}

\title{Sigma model approach to string theory}
\author{A.A. Tseytlin \\ 
\textit{\small P.N. Lebedev Physical Institute, Leninsky pr. 53, Moscow 117924, USSR}}
\date{\small July 1988}

\begin{document}

\maketitle

\begin{abstract}
A review of the $\sigma$-model approach to derivation of effective string equations of motion for the massless fields is presented. We limit our consideration to the case of the tree approximation in the closed bosonic string theory.
\end{abstract}

\

\

{


     1. Introduction 
     
\
     
    2. Sigma model and second quantized string theory   
    
    \

   3. Sigma model and first quantized string theory

    \

   4. Subtraction of M\"obius infinities and sigma model representation
   
    \ \ \ \ for string theory effective action 
   
   \

     5. Weyl anomaly in the sigma model

     \

     6. Sigma model Weyl invariance conditions  from  stationarity
     
     \ \ \ \  of ``central charge'' action

     \

     7. Perturbation theory results for Weyl anomaly coefficients

     \

     8. Concluding remarks 
}

\

\

\

\

{\it Published  in Int. J. Mod. Phys. A 4 (1989) 1257-1318  

and in  ``Superstrings '88'', Proceedings of the  ICTP  Trieste Spring School, 

 11-19 April 1988, 
World Scientific, 1989,  p.99-166}

\newpage

\section{Introduction}
\renewcommand{\theequation}{1.\arabic{equation}}
\setcounter{equation}{0}

\def \D  {{\cal D}}

A shortcoming of our present understanding of a (closed) string theory is the absence of a non-perturbative dynamical principle. What we do know is that conformal invariant 2d theories correspond to perturbative (tree) string vacua (in superstring theory they are likely to remain vacua to all orders in loop expansion if the background is supersymmetric). A proliferation of perturbative vacua is a disaster. It would be better to have complete equations of motion (which may provide additional restrictions on a vacuum) rather than perturbative criteria for stability of a string in a background.

What we are lacking is a non-perturbative definition of what a ``critical'' string theory is. We know how to develop a perturbation theory near a tree-level vacuum corresponding to a Weyl invariant 2d theory: the amplitudes are given (order by order in loop expansion) by the Polyakov path integrals which we divide by the full volume of the classical symmetries, i.e. the diffeomorphism and Weyl groups. The open questions are why the Weyl invariance condition is equivalent to the classical equations of motion of a ``critical'' theory and whether there exists some generalization of this condition, which is equivalent to the ``critical'' string theory equations of motion with loop contributions included.

A consideration of a generalized ``string'' $\sigma$-model, i.e. of the Polyakov path integral on arbitrary (non-Weyl invariant) backgrounds, may help one to answer these questions and may provide some clues to the problem of construction of covariant second quantized closed string theory. The $\sigma$-model is defined for an arbitrary number of space-time dimensions $D$ and for arbitrary couplings (background fields), and thus incorporates the features which should be characteristic to a background-independent closed string field theory. The usual restrictions of $D=26$ and $G_{\mu\nu}=\delta_{\mu\nu}$ correspond to the expansion near a particular tree-level bosonic string vacuum and hence should not appear in a fundamental formulation of string theory.\footnote{The bosonic string loop amplitudes are not well defined even on a flat $D = 26$ background. Modular infinities indicate that this background is not a true vacuum. Though the flat $D = 10$ space-time 
 is  a stable perturbative  vacuum of superstring theory, it is not a phenomenologically interesting one. }

The ``string'' $\sigma$-model is different from the ordinary $\sigma$-model defined on a fixed 2d space: at genus zero we are somehow to ``divide'' the $\sigma$-model objects by the M\"obius group volume while at higher genera we are to ``average'' over the moduli (with the measure which includes the contribution of the diffeomorphism group ghosts).

Though we do not know how to formulate the basic dynamical equation expressing a generalized Weyl invariance condition, it is likely that it should be equivalent to the vanishing of the full (loop-corrected) massless tadpole amplitudes on a background, i.e. to a generalized vacuum stability condition.

The $\sigma$-model approach helps one to clarify the general structure of the first-quantized string theory. In particular, the form of the vertex operators is fixed by the $\sigma$-model couplings (and by linearized Weyl invariance condition) and the $\sigma$-model partition function plays the role of the generating functional for string amplitudes (and hence the $\sigma$-model fixes the correct prescription for computing the amplitudes which preserves the necessary symmetries), etc. A reinterpretation of a first-quantized string theory as a 2d quantum field theory implies also that the dependence on a 2d UV cutoff can be consistently absorbed into a renormalization of the $\sigma$-model couplings and hence that 2d infinities are harmless in string theory.

It thus appears that the $\sigma$-model is more fundamental than the first-quantized string theory (vertex operators, on-shell amplitudes, etc.) corresponding to a particular conformal invariant background. Below we shall review some aspects of the $\sigma$-model approach to string theory. We shall first explain how the $\sigma$-model naturally appears in a second-quantized string theory (Sec. 2). Then (in Sec. 3) we discuss the general ideas behind the $\sigma$-model formulation of the first-quantized string theory. The case of the trivial topology corresponding to the tree approximation is analyzed in Sec. 4 where we formulate the $\sigma$-model analog of the subtraction procedure of M\"obius infinities in string tree amplitudes. This makes it possible to give a $\sigma$-model representation for the string theory effective action. The general properties of the $\sigma$-model (discussed in detail in Secs. 5-7) then imply that the effective action can be represented as an integral of the ``central charge'' coefficient and that the corresponding effective equations of motion are equivalent to the Weyl invariance conditions of the $\sigma$-model.

In Sec. 5 we obtain the operator expression for the Weyl anomaly in the $\sigma$-model. In Sec. 6 we argue that this Weyl anomaly relation can be derived as a stationarity condition for a ``central charge'' action functional. In Sec. 7 we summarize the results of some perturbative computations of the Weyl anomaly coefficients (``$\beta$-functions'') which provide a check of the statements made in Secs. 4 and 6. Section 8 contains some concluding remarks.

\section{Sigma model and second quantized string theory}
\renewcommand{\theequation}{2.\arabic{equation}}
\setcounter{equation}{0}

Below we shall discuss at a qualitative level how the $\sigma$-model approach may be related to a second quantized string field theory (see e.g. Refs. 1-8). We shall follow Refs. 9 and 10.

Let us first recall some facts about the path integral representation of an ordinary field theory. Consider the scalar $\varphi^3$ theory
\begin{equation}
    S = \frac{1}{2g^2} \int d^D x \Big(  \varphi \Delta \varphi + \frac{1}{3} \varphi^3 \Big), \qquad\qquad  \Delta = -\Box + m^2.
\end{equation}
Introducing a background field $\bar{\varphi}$ we define the effective action as the sum of all irreducible Feynman diagrams on the background. In the $\varphi^3$-theory, $\bar{\varphi}$ enters the diagrams only through the propagator $N=(\Delta + \bar{\varphi})^{-1}$, which has the following path integral representation
\begin{align}
    N(x_1, x_2) &= \int_{\partial C = (x_1, x_2)} \mathcal{D}C \exp(-I), \qquad \ \ \  I = I_0 + I_{int}, \\
    I_0 &= \int_0^1 dt (e^{-1} \dot{x}^\mu \dot{x}^\mu + e m^2), \\
    I_{int} &= \int_0^1 dt \, e\,  \bar{\varphi}(x(t)), \qquad 
     \qquad \mathcal{D}C = [de(t) dx^\mu(t)].
\end{align}
Here $e$ is a square root of the one-dimensional metric and the measure contains a gauge condition corresponding to the group of reparametrizations. In the ``proper-time'' gauge: $e=T=\text{const}$, $N = \int_0^\infty dT \int [dx] e^{-I}$. The analogous path integral representation is true for $\log \det(\Delta + \bar{\varphi})$ and hence for the full effective action
\begin{equation}
    \Gamma[\bar{\varphi}] = S + \frac{1}{2} \log \det(\Delta + \bar{\varphi}) + \sum_{\gamma_n} g^n d_n^{-1} \int d^D x_1 \dots d^D x_n N(x_1, x_2) \dots N(x_{n-1}, x_n).
\end{equation}
Here the sum goes over all irreducible graphs $\gamma_n$ with the number of vertices $n=2L-2$ ($L$ is the number of loops). $d_n$ is the dimension of a symmetry group of $\gamma_n$. Instead of summing over paths for each individual propagator and then integrating over the boundary points we can directly integrate over classes of graphs of a given topology. Then the factors of $d_n$ cancel out and different topologies appear to be summed over with an ``equal weight''
\begin{equation}
    \Gamma = S + \Gamma_q, \qquad\qquad  \Gamma_q = \sum_{n=0, 2...} g^n \int_{\gamma_n} \mathcal{D}C \exp(-I).
\end{equation}

Let us now turn to a string field theory. Since our discussion will be qualitative, we shall consider the closed bosonic string case; the arguments below can be made more rigorous in the case of the open string field theory of Witten [6]. A necessary requirement that a string field theory should satisfy is that the corresponding perturbation theory near a trivial vacuum should reproduce the results of the Polyakov path integral approach.

We shall assume the string field $\Phi[C] = \Phi[x^\mu(t), e(t)]$ is a functional of a parametrized loop and of a one-dimensional metric $(0 \le t \le 1)$ (instead of $e(t)$ we may use either the bosonic or fermionic ghost variables, see e.g. Refs. 3-8). We shall define the action for $\Phi$ (the propagator and the vertex) in terms of the Polyakov path integrals. The free propagator is thus given by a sum over cylindrical 2-surfaces
\begin{align}
    N_0[C_1, C_2] &= \int_{\partial M = (C_1, C_2)} \mathcal{D}M \exp(-I_0), \\
    I_0 &= \frac{1}{4\pi\alpha'} \int d^2 \sigma \sqrt{g}\, \partial^a x^\mu \partial_a x^\mu, \qquad 
    \mathcal{D}M = [dg_{ab}(\sigma) dx^\mu(\sigma)]. \nonumber
\end{align}
The 2-metric on $M$ is assumed to have boundary values consistent with 1-metrics on $C_1$ and $C_2$. By the definition of a ``critical'' string theory we always do not integrate over the conformal factor of the 2-metric (a choice of $g_{ab}$ corresponds to a choice of a Weyl gauge). One should of course properly account for the ghost contributions to the measure $M$ (and probably relate $e(t)$ to ghosts) in order to satisfy the composition rule. We shall disregard all such subtleties in the present discussion.

The inverse of $N_0$ defines the kinetic term in the string field theory action
\begin{equation}
    S_0 = \frac{1}{2g^2} \int \mathcal{D}C_1 \mathcal{D}C_2\ \Phi[C_1] \ \Delta[C_1, C_2] \ \Phi[C_2], \qquad\ \  \Delta = N_0^{-1}.
\end{equation}
As is well known, all compact orientable 2-manifolds can be glued out of 3-tube vertices (``pants''). Hence the fundamental closed string interaction should be a $\Phi^3$ one,
\begin{equation}
    S_{int} = \frac{1}{6g^2} \int \mathcal{D}C_1 \mathcal{D}C_2 \mathcal{D}C_3\  K[C_1, C_2, C_3]\  \Phi[C_1] \Phi[C_2] \Phi[C_3],
\end{equation}
\begin{equation}
    K[C_1, C_2, C_3] = \lim_{\epsilon \to 0} \int_{\partial M_\epsilon = (C_1, C_2, C_3)} \mathcal{D}M \exp(-I_0).
\end{equation}
The integration in (2.10) goes over ``infinitesimally short'' 3-tube surfaces (for analogous representation of the interaction vertex in the light-cone gauge, see Ref. 1; see also Ref. 6).

Using (2.7) and (2.10) we can represent all string field theory graphs in terms of sums over surfaces. Particular definitions of $N_0$ and $K$ in (2.7) and (2.10), which differ by a conformal transformation, should correspond to particular triangulations of moduli space in the Polyakov approach, see Ref. 11. Different choices of a string field theory action produce different off-shell extensions of the first-quantized theory, which probably correspond to different choices of a Weyl gauge.

Introducing a background functional $\bar{\Phi}[C]$ and expanding the action $S=S_0 + S_{int}$ near $\bar{\Phi}$ we conclude that all dependence on the background is contained in the propagator
\begin{equation}
    N = (\Delta + K \cdot \bar{\Phi})^{-1}, \qquad K \cdot \bar{\Phi} = \int \mathcal{D}C_3\  K[C_1, C_2, C_3]\  \bar{\Phi}[C_3].
\end{equation}
In analogy with the particle-theory case (cf. (2.2)) $N$ should admit a sum over surfaces representation
\begin{equation}
    N[C_1, C_2] = \int_{\partial M = (C_1, C_2)} \mathcal{D}M \exp(-I), \qquad I = I_0 + I_{int}[\bar{\Phi}].
\end{equation}
It is plausible that $I_{int}$ can be represented as an arbitrary local functional of $x^\mu$ and $g_{ab}$
\begin{align}
    I_{int} = \int d^2 \sigma \sqrt{g}\, \Big\{ & \psi(x) + \Big[\partial_a x^\mu \partial^a x^\nu H_{\mu\nu}(x) + \frac{i}{\sqrt{g}} \epsilon^{ab} \partial_a x^\mu \partial_b x^\nu B_{\mu\nu}(x) + R^{(2)} \phi(x)\Big] \nonumber \\
    & + \Big[\partial_a x^\mu \partial^a x^\nu \partial_b x^\lambda \partial^b x^\rho \mathcal{B}_{\mu\nu \lambda \rho}(x) + \mathcal{D}^2 x^\mu \partial_a x^\nu \partial^a x^\lambda \mathcal{B}_{\mu\nu\lambda}(x) \nonumber \\
    & + \mathcal{D}_a \partial_b x^\mu \partial^a x^\nu \partial^b x^\lambda \bar{\mathcal{B}}_{[\mu\nu]\lambda} + \mathcal{D}^2 x^\mu \mathcal{D}^2 x^\nu \mathcal{B}_{\mu\nu}(x) + R^{(2)} \partial_a x^\mu \partial^a x^\nu C_{\mu\nu}(x) \nonumber \\
    & + R^{(2)} \mathcal{D}^2 x^\mu C_\mu(x) + (R^{(2)})^2 C(x) + \text{terms with } \epsilon^{ab}\Big] + \dots \Big\}. \label{eq:2.13}
\end{align}
Here $R^{(2)}$ is the curvature of $g_{ab}$. We have included in (2.13) all possible independent local operators built out of $x^\mu$ and $g_{ab}$ and grouped them according to the dimensions of the corresponding fields (we assume that $[x^\mu]=\ell^0$, $[g_{ab}]=\ell^0$, $[\sigma]=\ell$).

 The fields involved in (2.13) are in correspondence with the fields which parametrize the background functional $\bar \Phi$. They represent the physical degrees of freedom as well as the auxiliary modes which are necessary in a covariant formulation [3]. This can be seen by choosing a special parametrization of $g_{ab}$ on the cylinder ($ds^2 = e^2(t,\tau) dt^2 + d\tau^2$) and counting the degrees of freedom. Note that the fields which multiply $R^{(2)}$ are necessarily auxiliary since they are absent in the lightcone gauge (in which $g_{ab}=\eta_{ab}$).

Defining the quantum part of the string field theory effective action $\Gamma_q$ as a sum of all connected
[9]  vacuum graphs in the $\Phi^3$-theory with the propagator (2.12), we should get in analogy with (2.6),
\begin{equation}
    \Gamma_q[\bar{\Phi}] = \sum_{\chi} g^{-\chi} \int_{\partial M_\chi = 0} \mathcal{D}M \exp(-I), \qquad I = I_0 + I_{int}.
\end{equation}
Here $\chi = 2 - 2n = 0, -2, \dots$ is the Euler number of the corresponding two-surface ($n$ is the number of handles; in contrast to the particle-theory case, here, all graphs with a given number of ``loops'' are topologically equivalent). Thus the loop contributions in string field theory can be expressed in terms of generalized partition functions of the ``$\sigma$-model'' (2.13) on higher genus surfaces.

Let us now demonstrate (again at a heuristic level) that the relation between the string field theory and the $\sigma$-model should be true even at the classical level (see also Ref. 10). Above we were considering the string field theory action in the form $S_0 + S_{int}$ corresponding to the expansion near a flat space vacuum. Let us suppose (following Refs. 7 and 8) that an ``unexpanded'' background-independent form of a string field theory action is given simply by a purely ``cubic'' functional
\begin{equation}
    S[\hat{\Phi}] = S_{int}[\hat{\Phi}] = \frac{1}{g^2} \int K \, \hat{\Phi} \hat{\Phi} \hat{\Phi},
\end{equation}
where $S_{int}$ was defined in (2.9) and (2.10). Let us stress again that in (2.10) we do not integrate over the conformal factor of the two-metric but simply fix a particular metric on $M_\epsilon$ as a Weyl gauge. Different choices of a Weyl gauge imply different off-shell continuations of string field theory (the corresponding actions should be related by field redefinitions, see  also Ref.  12). 

 Next, we shall assume that $\hat{\Phi}$ can be parametrized in terms of local fields in the following non-standard way
\begin{equation}
    \hat{\Phi}[C] = \int_{\partial M=C} [dx] \  e^{-I}, \qquad I = \frac{1}{4\pi\alpha'} \int d^2 \sigma \sqrt{g}\, \Big[\partial^a x^\mu \partial_a x^\nu G_{\mu\nu} + \dots\Big].
\end{equation}
Here $I$ is the most general $\sigma$-model action which has already appeared in (2.12) and (2.13) ($G_{\mu\nu}=\delta_{\mu\nu} + 4\pi\alpha' H_{\mu\nu}$). The functional integration in (2.16) goes over disc-like two-surfaces with the boundary $C$. It should be noted that (2.16) represents an arbitrary functional of $C$: the sets of fields $A'=\{G_{\mu\nu}, B_{\mu\nu}, \phi\}$, $B'=\{\psi, \mathcal{B}_{\mu\nu\lambda\rho}, \dots\}$ can be put in one-to-one correspondence with the set of ``massless'' and ``massive'' components $(A, B)$ of $\hat{\Phi}$ in the standard parametrization based on the normal mode expansions of $x^\mu(t)$ and $e(t)$.

To see this, one may choose a conformal gauge $g_{ab}= e^{2\rho} \delta_{ab}$. Then (2.13) 
 reduces to a sum of products of the fields times the vertex operators in a $D+1$ dimensional theory on a flat world sheet. The exact form of the transformation from $(A',B')$ to $(A,B)$ is, of course, nonlinear and nonlocal.
The representation (2.16)  is natural in view of the fact that the string field corresponding to the trivial vacuum $(G_{\mu \nu} = \delta_{\mu \nu}, \ B_{\mu\nu}, ...=0)$ admits a path integral representation (see Refs. 14 and 15). Also, it is likely that a classical vacuum string field corresponds to a ground state of a conformally invariant 2d theory [16]  and hence  given by (2.16)
 with  $I$  taken at a Weyl invariant point in the coupling constant space.

Substituting (2.16) into the action (2.15),(2.10) we conclude that the 3-vertex $K$ ``glues'' three disc-like surfaces together to form a closed sphere-like surface. Assuming that it is possible to ignore the difference between $I_0$ and $I$ on an infinitesimal 3-tube surface we get
\begin{align}
    S &= \frac{1}{g^2} \int [dx] \ e^{-I}, \\
    I &= \frac{1}{4\pi\alpha'}
     \int d^2 \sigma \sqrt{g}\, \Big\{ \Lambda^2 \psi(x) + \Big[{\partial}_a x^\mu \partial^a x^\nu G_{\mu\nu}(x) + i \frac{\epsilon^{ab}}{\sqrt{g}} {\partial}_a x^\mu {\partial}_b x^\nu B_{\mu\nu}(x) + \alpha' R^{(2)} \phi(x)\Big] \nonumber 
    \\
    & \qquad\qquad  + \Lambda^{-2} 
    \Big[\partial_a x^\mu \partial^a x^\nu \partial_b x^\lambda \partial^b x^\rho \mathcal{B}_{\mu\nu\lambda\rho} + \dots\Big]
     + \dots \Big\}.
\end{align}
Here $I$ is the same as in (2.12), (2.13) and (2.16) (we have only rescaled the fields to make them dimensionless). $\Lambda$ is a covariant 2d UV cutoff ($g_{ab} \Delta \sigma^a \Delta \sigma^b \ge \Lambda^{-2}$). The fields in (2.18) are the bare ($\Lambda$-dependent) couplings of the $\sigma$-model. The integration in (2.17) goes over 2-surfaces with the topology of a sphere.

The same reasoning can be repeated (in a more rigorous form) for the open string theory case. Then $C$
 is an open segment and $\hat \Phi$  is given by a path integral over disk-like surfaces (with C belonging to the boundary of the disk)  of $e^{-I}$, where  $I$  contains  also  the boundary terms $\int ds [ A(x) + i \dot x^\mu A_\mu (x) + ... ]$.
 Note that (2.16)  implies that the open string field should be parameterized in terms of local fields corresponding to
  both open and closed string spectra, cf.  [8]. The integration in the open string analog of  (2.17)  goes over disk-like surfaces
   with the Neumann boundary conditions.

To understand the meaning of (2.17) let us consider again the $\Phi^3$-action expanded near a trivial vacuum
\begin{align}
    S &= \frac{1}{2g^2} \int \Big(\Phi \Delta \Phi + \frac{1}{3} \Phi^3\Big) \nonumber \\
    &= \frac{1}{2g^2} \int d^D x \Big[A(-\Box)A + B(-\Box+m^2)B + A^3 + A^2 B + AB^2 + B^3\Big].
\end{align}
Here we have assumed that the string field $\Phi$ is represented using the standard normal mode expansion in terms of a finite number of massless fields $A$ and an infinite number of massive fields $B$. For simplicity, we consider $A$ and $B$ as two scalar fields and ignore a complicated derivative structure of the 3-vertices. 

To find the effective action which describes low-energy phenomena we are to integrate over $B$ in $\int [dA][dB] \exp(-S)$. At the classical level this amounts to the elimination of $B$ from (2.19) with the help of the classical equations of motion. Let
\begin{equation}
    B_{cl}(A) = -(-\Box+m^2)^{-1} A^2 + \dots
\end{equation}
be a solution of $\delta S/\delta B=0$.
 If we make a change of the massive variables, $B = \tilde{B} + B_{cl}(A)$, the action (2.19) takes the form
\begin{align}
    S &= S_0[A] + S_1[\tilde{B}, A], \\
    S_0[A] &= \frac{1}{2g^2} \int d^D x [A(-\Box)A + A^3 + A^2 (-\Box+m^2)^{-1} A^2 + \dots], \\
    S_1[\tilde{B}, A] &= \frac{1}{2g^2} \int d^D x \Big[\tilde{B}[-\Box+m^2+A+B_{cl}(A)]\tilde{B} + \tilde{B}^3\Big].
\end{align}
The transformation $B \to \tilde{B}$ has eliminated the $A^2 B$ coupling in $S$ 
(so that  in contrast to  B the field $\tilde B$   cannot decay in $A$)  and thus decoupled the massless sector from the massive one. Hence $S_0$ is the low-energy (tree) effective action for the massless fields.

By definition, the field theory (2.19) should reproduce the string as matrix. Hence $S_0$  should reproduce the massless sector  of  the (tree)  string $S$-matrix.
This condition is usually taken as a definition of the effective action [17,18]. Note that (2.21) 
 does not correctly reproduce the string as matrix in the $\tilde B$ sector. The reason is that the transformation 
 $B \to \tilde B$  being non-local spoils the condition of the equivalence theorem (see, e.g., Ref. [19]). It  is 
  also important to stress that the standard choice of the massive fields $B$  is not a proper 
  one from the low-energy point of view:
  $B_{cl(A)}$  is certainly non-vanishing for any non-trivial vacuum configuration of the massless field (corresponding to a compactification, etc.). It is  $\tilde B= 0$ that is consistent with  $A\not=0$.

Returning to (2.16),(2.17) it is natural to anticipate that the fields $\{G_{\mu\nu}, \phi, B_{\mu\nu}\}$ should correspond to the $A$'s while the massive fields $\{\psi, \mathcal{B}, \dots \}$ should correspond to the $\tilde{B}$'s rather than the $B$'s\footnote{The argument in  favor of this claim  is  that   while  (2.19)  expressed in terms of $A$  and $B$   is not covariant   under the standard   general coordinate transformations  (with $B$ transforming simply as tensors)   the $\sigma$-model is.}
(``to correspond'' here  means ``to be related by a   nonlinear  though homogeneous transformation''). 
Hence setting in (2.17), (2.18) all the massive fields equal to zero we should get the expression for the {\it effective action} (EA). The validity of such a $\sigma$-model partition function representation for the EA was conjectured in Ref. 20.

It turns out that Eq. (2.17) does give the expression for the EA in the open (super)string theory [21,22].
 However, (2.17) should be modified in the closed (super)string case: as we shall see in Sec. 4, the corresponding EA is given by the derivative $\partial / \partial \log \Lambda$ of the bare partition function $\int [dx] \exp(-I)$ [23].
 The difference between  the open    and closed string cases  is due  the different  structure of the M\"obius   groups corresponding to the disc  and the sphere (see Sec. 4). 

Loop corrections to string theory effective action are given by the generalized $\sigma$-model partition functions on surfaces of corresponding genera,\footnote{Since the M\"obius   group  is trivial  for higher  genus  surfaces,  (2.14) should be true  for both open   and closed   string theories.}
 i.e. by (2.14) with massive fields set equal to zero. Thus defined EA should reproduce the massless sector of the full loop-corrected string S matrix [24].  A vacuum of a string field theory is a solution of the effective equations $\frac{\delta \Gamma}{\delta \Phi} = 0$. Since varying $I$ with respect to a massless field $\varphi^i$ (with $A=\{\varphi^i\}$) yields a (massless) vertex operator $V_i$, the quantum effective equations of motion should take the form
\begin{equation}
    \sum_{n=0}^\infty \langle V_i \rangle_n = 0,
\end{equation}
where the sum goes over genera and the $\langle \dots \rangle_n$ are computed for an arbitrary ``massless'' background. The vacuum values of the massless fields are determined by solving (2.24). Since the propagator and the 3-vertex of the string field theory (2.7)--(2.10) 
are defined by path integrals in which one integrates over the string (and ghost) coordinates and the moduli but not over the conformal factor of 2-metric, the resulting path integral (2.14) and hence $\langle \dots \rangle_n$ are also computed with a Weyl gauge fixed, i.e.
\begin{equation}
    \langle \dots \rangle_n = \int [d\tau]_n \int [dx] e^{-I}.
\end{equation}
Here $\{\tau\}$ are moduli parameters and $I$ depends on a fixed 2-metric $\hat{g}_{ab}(\tau)$.

Note that the prescription of how to split $g_{ab}$  into a conformal factor and the  moduli in a way universal for all $n$  is fixed by the  string field  theory action (i.e.  by the definitions of  $N_0$  and  $K$ in (2.8)--(2.10), which correspond to a particular triangulation of the moduli space [11]).\footnote{Different choices of  $N_0$  and  $K$ corresponding to different Weyl group gauges should lead to the same physical vacua.} The basic consistency conjecture is that (2.24)  is equivalent to a generalized Weyl invariance condition, i.e. that a background which satisfies (2.24) should be independent of a Weyl gauge chosen. 

 In order to study the solutions of (2.24) we need to understand how to compute $\langle V_i \rangle_0$ in the tree approximation. The problem is that the usual prescription of the division of the tree-level correlators by the volume of the M\"obius group cannot be directly applied in the case of an arbitrary background. The correct prescription (see Sec. 4) is to define $\langle V_i \rangle_0$ as the derivative $\partial / \partial \log \Lambda$ of the expectation value of the bare vertex operator.

To summarize, let  us suppose that there exists a background-independent formulation of string field theory such that the action (2.15) does  not  ``know'' about particular properties of space-time  (its dimension  $D$, topology, metric, etc.). A space-time interpretation itself then should be a property of a particular solution. This is nicely reflected in the sigma model representation. The sigma model action (2.18) is defined for arbitrary $D$ and arbitrary values of local fields. The values of $D$ and of the fields are fixed by the string field theory equations of motion which should be equivalent to the condition of Weyl invariance of the sigma model. Since the sigma model is naturally defined ``off-shell'', we may anticipate that there exists a smooth ``off-shell''  (off $D=26$, etc.) continuation of string amplitudes which is closely related to the sigma model. In particular, the $N =0,1,2$-point amplitudes should not vanish identically but should rather start with  $D-26$  (see Sec. 4).


\section{Sigma model and first quantized string theory}
\renewcommand{\theequation}{3.\arabic{equation}}
\setcounter{equation}{0}

Let us now see how the $\sigma$ model appears in the sum over surfaces approach. The first-quantized approach is based on the representation of string amplitudes as the Polyakov sums over surfaces. One usually considers the amplitudes in a flat background and imposes $D=26$ in order to cancel the ``local'' Weyl anomaly. To study admissible string vacua one may consider propagation of a string in a nontrivial background. If the background fields are massless the string propagation is described by the renormalizable $\sigma$ model
\begin{equation}
    I = \frac{1}{4\pi\alpha'} \int d^2\sigma \sqrt{g}\,  \Big[ \partial^a x^\mu \partial_a x^\nu G_{\mu\nu}(x) + \frac{i}{\sqrt{g}} \epsilon^{ab} \partial_a x^\mu \partial_b x^\nu B_{\mu\nu}(x) + \alpha' R^{(2)} \phi(x)  \Big].
\end{equation}
The necessary requirement that a tree-level vacuum should satisfy is the Weyl invariance of the corresponding string theory on the sphere, i.e. of the $\sigma$ model (3.1) [25]. Since the tadpoles vanish in a conformal invariant theory on a sphere [26,27] the conformal invariance is the condition of stability of a tree-level vacuum.

To study string loop corrections to the equations of motion (i.e. the conditions on ``nonperturbative'' vacua) we are to consider a generalized $\sigma$ model defined on surfaces of all genera. The corresponding partition function, i.e. the partition function of a string propagating in a nontrivial background is
\begin{align}
    Z = &\sum_{n=0}^\infty Z_n, \qquad \qquad Z_0 = \Omega^{-1} \int [dx] \ e^{-I} \ , \\
     & \qquad Z_n = \int [d\tau]_n \int [dx] \ e^{-I}, \qquad n=1, 2, \dots \nonumber
\end{align}
where $I$ is given by (3.1), ``$\Omega^{-1}$'' indicates some prescription corresponding to subtraction of M\"obius infinities in on-shell amplitudes, and $\{\tau\}$ are the moduli parameters ($[d\tau]_n$ includes the contribution of ghosts and a factor of inverse volume of the modular group). In addition to the reparametrization gauge $g_{ab} = e^{2\rho}\hat{g}_{ab}(\tau)$ we fix also the Weyl gauge $\rho=0$. To make the split of the metric into a conformal factor and the moduli in some way universal for all genera we may assume, e.g., that $\hat{g}$ satisfies $R^{(2)}(\hat{g}) = \text{const}$. We hope that the physical results for a nonperturbative vacuum will be independent of a  choice of a Weyl gauge.

It should be noted that the complications of ``$\Omega^{-1}$'' and ``$\int [d\tau]$'' in (3.2) which are absent in the ordinary $\sigma$ model are characteristic to string theory. In fact, while the ordinary $\sigma$ model objects are defined on a 2-space of a fixed topology and metric, the string theory objects are given by the path integrals over $x^\mu$ {\it and}
 the 2-metric. Even though the conformal factor should not be integrated over in a ``critical'' string theory, we are still to include the integrals over moduli and to divide the measure in  $\int [dg_{ab}] [dx]$ by the volume of the full diffeomorphism group. Thus the ``string'' $\sigma$ model (3.2) is quite a nonstandard one.

As we already noted, a condition that the vacuum fields in (3.1) are to satisfy should be the vanishing of the full tadpole amplitudes in a background (see (2.24)). This equation should be equivalent to a generalized Weyl invariance condition for (3.2), i.e. to the vanishing of the generalized ``$\beta$ functions'' which include the contributions of modular divergences [25, 28-30] (see Ref. 24 for details).

In addition to the study of string propagation in background fields there is another (related) line of reasoning which leads one to consider the generalized partition function (3.2). The massless string scattering amplitudes in a flat background are given by
\begin{equation}
    A_N = \sum_{n=0}^\infty \langle V_1 \dots V_N \rangle_n,
\end{equation}
\begin{equation}
    \langle \dots \rangle_0 = \Omega^{-1} \int [dx] \ e^{-I_0} \dots, \qquad \langle \dots \rangle_n = \int [d\tau]_n \int [dx] \ e^{-I_0} \dots,
\end{equation}
where $V_i$ are the ``massless'' vertex operators (we suppress the dependence on the string coupling constant). Let us introduce the generating functional for scattering amplitudes by multiplying $V_i$ by the corresponding fields $\varphi^i$ and exponentiating
\begin{equation}
    Z[\varphi] = \sum_{n=0}^\infty \langle e^{-\varphi^i V_i} \rangle_n.
\end{equation}
This object is defined for arbitrary values of $\varphi^i$. Expanding (3.5) in powers of $\varphi^i$ and setting $\varphi^i$ equal to its ``in'' value ($\Box \varphi^i_{in} = 0$) we obtain the on-shell amplitudes (3.3). The basic observation is that $I_0 + \varphi^i V_i$ coincides (modulo field re-definitions) with the $\sigma$ model action (3.1) [20] and hence (3.5) is the same as the partition function (3.2).

The point which we would like to stress is that the generating functional (3.2) is a more general object than the vertex operators and their correlators. In fact, the structure of the vertex operators is dictated by the $\sigma$ model. We start with the most general renormalizable $\sigma$-model (3.1). The fields in (3.1) can of course be re-defined in a regular way. The choice of $\tilde{G}_{\mu\nu} = \delta_{\mu\nu} + h_{\mu\nu}$ for which the graviton and the dilaton turn out to be decoupled in the propagator, is $\tilde{G}_{\mu\nu} = G_{\mu\nu} \exp \big(-\frac{4\phi}{D-2}  \big)$ [20]. Then expanding (3.1) in powers of $\phi$ we get
\begin{align}
    I &= \frac{1}{4\pi\alpha'} \int d^2\sigma\sqrt g  \Big[ \partial^a x^\mu \partial_a x^\mu + \partial^a x^\mu \partial_a x^\nu h_{\mu\nu}\nonumber\\   &\qquad \qquad\qquad \qquad + \frac{4}{D-2} \Big( \partial^a x^\mu \partial_a x^\mu + \frac{\alpha'}{4} (D-2) R^{(2)} \Big) \phi + \dots \Big] \nonumber \\
    &= I_0 + V_h \cdot h + V_\phi \cdot \phi + \dots.
\end{align}
$V_\phi$ is indeed the correct vertex operator for the dilaton (see, e.g., Refs. 24 and 31). Being substituted into $Z$ (3.2) higher order terms in $\phi$ in (3.6) produce additional ``contact'' contributions to the amplitudes (defined as coefficients in the expansion of $Z$ in powers of $h$ and $\phi$). These contributions may be ignored if the amplitudes are computed for nonzero momenta using the analytic continuation [32]. However, in general, they must be included in order to maintain the correspondence between the $\sigma$ model and the S matrix approaches.\footnote{For example,  it is necessary to account  for the contact terms in order to prove  that the  string coupling  constant   can be absorbed into  a constant part of the dilaton [33,20].}
 In superstring theory, the account of the contact contributions dictated by the $\sigma$ model is essential for gauge invariance and supersymmetry of the resulting amplitudes [32, 22].

A general point of view (which is supported by heuristic string field theory considerations of Sec. 2) can thus be summarized as follows. Start with the most general 2d theory in a curved background. The couplings of the $\sigma$ model correspond to the (off-shell) modes of the string. Choosing a proper basis of the fields we find the corresponding vertex operators directly from the $\sigma$ model action $I$. All nonlinear terms in $I$ which appear after the field re-definitions are to be accounted for in computation of the amplitudes in order to preserve the symmetries of the theory.

Given that the $\sigma$ model partition function (3.2) can be interpreted as the generating functional for massless string scattering amplitudes, it is natural to try to relate it to the effective action (EA) for the massless string fields. By definition, the EA is a field theory action which has a {\it tree} S matrix reproducing the massless sector of the full (loop-corrected) string S matrix [17, 18, 34, 24]. To obtain the vertices in the EA from string amplitudes we are to subtract the massless tree exchanges from the latter. Suppose that $Z$ (3.2) is computed with a UV cutoff (which simultaneously regularizes also the integrals over the moduli). Since both ``local'' and ``modular'' infinities can be interpreted as being due to massless tree exchanges, their renormalization should correspond to the subtraction of massless poles in the amplitudes (see Refs. 25, 35 and 34).

Let $Z$ be the bare partition function (3.2). Then the generating functional for string-theory massless scattering amplitudes is $\mathcal{S}_{s.t.}[\varphi_{in}] = Z[\varphi_{in}]$, where $\varphi^i_{in}$ are the on-shell values of the massless fields $\varphi^i$. If the EA is
\begin{equation}
    S[\varphi] = \frac{1}{2} \int \varphi \Delta \varphi + S_{int}[\varphi], \qquad \Delta = -\Box,
\end{equation}
then the generating functional for the field theory tree amplitudes is given by (see, e.g., Ref. 36)
\begin{equation}
    \mathcal{S}_{f.t.}[\varphi_{in}] = S_{int}[\varphi_{cl}] + \frac{1}{2}  \Big( \frac{\delta S_{int}}{\delta \varphi} \Delta^{-1} \frac{\delta S_{int}}{\delta \varphi}  \Big)_{\varphi_{cl}},
\end{equation}
where $\varphi_{cl}(\varphi_{in})$ is the classical solution, $\varphi_{cl} = \varphi_{in} - \Delta^{-1}  \Big( \frac{\delta S_{int}}{\delta \varphi}  \Big)_{\varphi_{cl}}$. By the definition of the EA we should have: $\mathcal{S}_{s.t.} = \mathcal{S}_{f.t.}$. If $Z$ is regularized by a UV cutoff $\varepsilon$ and is computed using the $\sigma$ model loop expansion (which corresponds to the expansion of the amplitudes in powers of momenta) then
\begin{equation}
    Z[\varphi_{in}, \varepsilon] = Z_R[\varphi_{in}] + O(\log \varepsilon) = Z_R[\varphi_R],
\end{equation}
where the subscript ``$R$'' indicates a renormalized quantity. In order to compare $Z$ with $\mathcal{S}_{f.t.}$ we are to regularize (3.8) by introducing the cutoff
\begin{equation}
    \Delta^{-1} \to \int_\varepsilon^1 \frac{dt}{t} t^\Delta = -\log \varepsilon - \frac{1}{2} \Delta (\log \varepsilon)^2 + \dots.
\end{equation}
Then the second term in (3.8) becomes $O(\log \varepsilon)$ and we get
\begin{align}
    \mathcal{S}_{f.t.}[\varphi_{in}, \varepsilon] &= S_{int}[\varphi_{in}] + O(\log \varepsilon) \nonumber \\
    &= \mathcal{S}_{s.t.}[\varphi_{in}, \varepsilon] = Z_R[\varphi_{in}] + O(\log \varepsilon).
\end{align}
Hence finally $S_{int}[\varphi_{in}] = Z_R[\varphi_{in}]$. If there is a gauge symmetry relating the kinetic and the interaction terms in $S$ and if $Z$ is defined to be gauge-invariant off-shell then (modulo field redefinitions which represent ambiguity in off-shell continuation)
\begin{equation}
    S[\varphi] = Z_R[\varphi].
\end{equation}
The above argument thus suggests that the effective action can be represented as a renormalized partition function of the $\sigma$ model. The important clarification, however, is in order. While there are no principal problems with higher loop contributions to $Z$, we are still to define what is meant by ``$\Omega^{-1}$'' in the tree-level partition function $Z_0$ in (3.2). It is the ``stringy'' partition function $Z_0$ and not the ordinary $\sigma$ model partition function on the sphere $Z_0 = \int [dx] \ e^{-I}$ that should be equal to the EA according to (3.12). In the next section we shall present a prescription of how to define $Z_0$ and the corresponding tree amplitudes as objects pertaining to the $\sigma$ model on the sphere. As a result, we shall find an explicit representation for the closed string tree-level effective action in terms of the integral of a ``central charge'' coefficient.


\section{Subtraction of M\"obius infinities and sigma model representation for string theory effective action}
\renewcommand{\theequation}{4.\arabic{equation}}
\setcounter{equation}{0}

As is well-known, the tree-level closed string amplitudes given by correlators of on-shell vertex operators on the sphere (and defined by analytic continuation) contain the M\"obius infinities. The latter are usually subtracted by fixing the M\"obius gauge [37] (e.g. fixing 3 out of $N$ points of integration). A division by a M\"obius group volume $\Omega$ is necessary in the Polyakov path integral approach [38-40]. This M\"obius gauge fixing prescription looks strange from the $\sigma$ model point of view: in the $\sigma$ model approach one considers UV regularized correlators of composite (vertex) operators which are not M\"obius invariant (cf. Ref. 41). Our aim will be to find a $\sigma$ model analog of the ``$1/\Omega$'' prescription, i.e. to represent the string amplitudes as objects in the $\sigma$ model.

 The $\sigma$ model counterparts of the string amplitudes will not only satisfy the ``correspondence principle'', i.e. will reproduce the standard expressions in the case of a flat $D=26$ background and on-shell momenta, but will also provide a reasonable extension of the amplitudes off the conformal point (e.g. ``off''   the flat $D=26$ background). In particular, we shall find that the $N=0, 1, 2$ amplitudes computed according to the $\sigma$ model prescription do not vanish in general but start instead with $D=26$ terms. This is consistent with an expectation that the $\sigma$ model approach provides an ``off-shell'' extension of the string amplitudes.

It should be stated clearly that we consider the $\sigma$ model only perturbatively in loop ($\alpha'$) expansion.\footnote{We thus disregard non-perturbative sigma model infinities which correspond to massive poles in the amplitudes [42]. The renormalizable  sigma model (3.1) is renormalizable  only within the loop expansion, i.e. within the expansion and powers of derivatives of the fields. }
 This corresponds to an expansion of massless string amplitudes in powers of momenta. To make this expansion regular one may introduce a UV regularization in the 2d propagator thus trading the massless poles in the amplitudes for logarithmic UV infinities $ \Big( \frac{1}{\alpha' k^2} \to -\log \varepsilon - \frac{1}{2} \alpha' k^2 (\log \varepsilon)^2 + \dots, \text{cf. (3.10)}  \Big)$. It is such regularized and expanded in powers of momenta expression for a massless string amplitude that we are going to represent in terms of a $\sigma$ model correlator. Hence the ``off-shell'' extension provided by the $\sigma$ model is applicable only in the vicinity of the massless ``mass shell'', i.e. in the low-energy region. This is, however, sufficient for the purpose of understanding why the $\sigma$ model and the string S matrix approaches lead to equivalent effective equations of motion.

Let us first briefly consider the case of the \textit{open string theory}. Introducing a UV regularization one finds that the M\"obius infinities in the correlators of vertex operators on the disc $\langle V_1 \dots V_N \rangle$ are regularized and can be identified with a part of power (linear) UV infinities. In particular, the M\"obius group volume for the disc is $\Omega \sim \langle V_1 V_2 V_3 \rangle = \frac{a_1}{\varepsilon} + a_2$, $\varepsilon \to 0$ (see Refs. 21 and 31). The presence of power infinities in low-energy expansion for regularized massless (vector) amplitudes is related to the presence of the tachyon in the open bosonic string theory. Renormalization of the tachyon field corresponds to a subtraction of power infinities (and, in particular, of the M\"obius infinities). Then the $\sigma$ model prescription for the open string massless amplitudes is
\begin{equation}
    A_N = \Omega_R^{-1} \langle V_1 \dots V_N \rangle_R,
\end{equation}
where ``$R$'' indicates a subtraction of power infinities. Since $\Omega_R = \text{finite}$, the regularized amplitudes are proportional to the regularized correlators (with ``tachyonic'' power infinities subtracted out). Hence the open string theory tree effective action is given simply by the completely renormalized $\sigma$ model partition function
\begin{equation}
    S \sim Z_R
\end{equation}
(see Refs. 21 and 22 for details).

The situation in the open superstring theory is even simpler: the 2d supersymmetry implies the automatic cancellation of 2d power UV infinities and hence of M\"obius infinities [22]. In particular, the (supersymmetrically regularized) volume of the super M\"obius group of the disc is finite and again $S \sim Z_R$ (here ``$R$'' indicates a renormalization of remaining logarithmic infinities).

The case of the \textit{closed string theory} is more subtle. The regularized expression for the volume of the M\"obius group of the sphere contains a logarithmic infinity, $\Omega = \frac{c_0}{\varepsilon^2} + \frac{c_1}{\varepsilon^2} \log \varepsilon + c_2 \log \varepsilon + c_3$ (see, e.g., Refs. 43, 31). One could expect that like the open string amplitudes, the closed string amplitudes could be obtained by subtracting power infinities from the regularized expressions for $\langle V_1 \dots V_N \rangle$ and $\Omega$ and by computing the ratio $\langle V_1 \dots V_N \rangle_R / \Omega_R, \ \Omega_R = c_2 \log \varepsilon'$, \ $\varepsilon' \sim \varepsilon$ (cf. Refs. 43, 44). However, this ``$1/\log \varepsilon$'' prescription is not a correct one (in particular, it is not renormalization group invariant since the result of its application changes under a rescaling of $\varepsilon$). As we shall see below, the correct prescription to obtain amplitudes in the closed string theory is to take the \textit{derivative} $\partial / \partial \log \varepsilon$ of the regularized correlators.

Let us start with a demonstration of the presence of logarithmic M\"obius infinities in the regularized correlators of the closed string vertex operators on the sphere. For simplicity we shall consider the correlators on the complex plane assuming an IR cutoff at large distances ($|z| \le r,\  r \to \infty$). The UV infinities can be regularized by introducing, e.g., a short distance cutoff in the 2d propagator $\sim \log (|z-z'|^2 + a^2)$, $a \to 0$. Since in contrast to the standard M\"obius gauge fixed expression for the amplitude the integrals in $\langle V_1 \dots V_N \rangle$ go over all $N$ Koba-Nielsen variables $z_i$, $\langle V_1 \dots V_N \rangle$ may have additional ``momentum-independent'' singularities corresponding to the two limits:

(1) all $N$ points $z_i$ are close to each other

(2) some $N-1$ points are close to each other.

$\langle V_1 \dots V_N \rangle$ is quadratically divergent in the first limit, factorizing into the product of the $N+1$ point amplitude ($N$ original particles plus a zero-momentum tachyon), the tachyon propagator at zero momentum (which is quadratically divergent) and the one-tachyon correlator $\langle V_T(0) \rangle$, $V_T(0) \sim \int d^2 z \sim r^2$.

 The logarithmic singularity appearing in the second limit is due to the ``external leg'' correction: $\langle V_1 \dots V_N \rangle \sim \sum_i A_N \log \varepsilon \langle V_i V_i \rangle$, where $A_N$ is the $N$ point amplitude (i.e. $\langle V_1 \dots V_N \rangle$ with 3 points fixed), $\langle V_i V_i \rangle$ is the ``mass correction'' and $\log \varepsilon$ stands for the on-shell value of the propagator of the $i$th particle. 
 
 Though we are interested in the massless amplitudes, it is easier to prove the above statements by considering the example of the $N$ tachyon correlator ($\alpha' k^2_i = 4, \ z_{ij} \equiv z_i - z_j$)
\begin{equation}
    \langle V_1 \dots V_N \rangle \sim K_N = \int d^2 z_1 \dots d^2 z_N \prod_{1 \le i < j \le N} |z_{ij}|^{\alpha' k_i k_j}
\end{equation}
(we do not explicitly indicate the cutoffs). Consider, e.g., the second limit: $z_2 \approx z_i$, $i=3, \dots, N$. Changing the variables $z_3 = z_2 + u, \ z_n = z_2 + u w_n, \ n=4, \dots, N$ we obtain
\begin{align}
    K_N &= \int d^2 z_1 d^2 z_2 d^2 u d^2 w_4 \dots d^2 w_N |u|^{2(N-3) + \sum_{2 \le n < m \le N} \alpha' k_n k_m} |z_{12}|^{\alpha' k_1 k_2} \nonumber \\
    &\qquad \times |z_{12} - u|^{\alpha' k_2 k_3} \prod_{n=4}^N |z_{12} - u w_n|^{\alpha' k_1 k_n} |w_n|^{\alpha' k_2 k_n} |1-w_n|^{\alpha' k_3 k_n} \nonumber \\
    &\qquad \times \prod_{4 \le n < m \le N} |w_{nm}|^{\alpha' k_n k_m} \stackrel{u \to 0}{\sim} \int \frac{d^2 z_1 d^2 z_2}{|z_{12}|^4} \int_{|u| \ge a} \frac{d^2 u}{|u|^2} A_N\,  \sim \, \log \varepsilon\,  A_N,
\end{align}
\begin{equation}
    A_N = \int d^2 w_4 \dots d^2 w_N \prod_{n=4}^N |w_n|^{\alpha' k_2 k_n} |1-w_n|^{\alpha' k_3 k_n} \prod_{4 \le n < m \le N} |w_{nm}|^{\alpha' k_n k_m}.
\end{equation}
Here $\varepsilon$ is the ratio of the UV and IR cutoffs and $A_N$ is the standard Shapiro-Virasoro amplitude (computed in the M\"obius gauge $w_1 = \infty, w_2 = 0, w_3 = 1$). We have used that $\alpha' \sum_{2 \le n < m \le N} k_n k_m = -2(N-1)$ and that the 2-tachyon correlator is
\begin{equation}
    K_2 = \int_{|z| \le r, |z_{12}| \ge a} \frac{d^2 z_1 d^2 z_2}{|z_{12}|^4} = \frac{b_1}{\varepsilon^2} + b_2, \ \ \ \qquad \varepsilon = \frac{a}{r} \to 0.
\end{equation}
Equations (4.4)-(4.6) imply, in particular, that the 3-point correlator, which should be proportional to the regularized M\"obius group volume, is given by
\begin{equation}
    K_3 = \frac{c_0}{\varepsilon^2} + \frac{c_1}{\varepsilon^2} \log \varepsilon + c_2 \log \varepsilon + c_3.
\end{equation}
Thus we see that the coefficient of the ``momentum-independent'' logarithmic singularity in the $N$ point correlator is proportional to the $N$ point amplitude. However, the amplitudes cannot be in general (for $N>3$) represented as $\langle V_1 \dots V_N \rangle_R / \Omega_R \sim \langle V_1 \dots V_N \rangle_R / \log \varepsilon$, where ``$R$'' indicates a subtraction of power infinities. The reason is that the regularized expression for $\langle V_1 \dots V_N \rangle$ contains not only the ``M\"obius'' but also the ``physical'' logarithmic infinities corresponding to the massless poles in transferred momenta. 

Consider the correlators of the massless vertex operators and suppose that they are computed using the expansion in powers of the momenta. Then neglecting the ``tachyonic'' power infinities we get
\begin{equation}
    \langle V_1 \dots V_N \rangle_R = B_N^{(0)} + B_N^{(1)} \log \varepsilon + ... + B_N^{(n)} \log^n \varepsilon + \dots.
\end{equation}
$B_N$ are power series  in momenta and, in view of the above argument, $B_N^{(0)} \ne 0$ only for $N \le 3$. One of the $n$ of 
$ \log \varepsilon$ factors in the $\log^n \varepsilon$ term in (4.8) has a meaning of an ``overall'' M\"obius infinity and hence is to be cancelled out in order to obtain the string amplitude. However, all the $n$ of  $ \log \varepsilon$ factors in $\log^n \varepsilon$ appear in (4.8) on an equal footing. Hence the symmetry factor of $n$ is to be included into the result. This is the same as taking the derivative $\partial / \partial \log \varepsilon$ of (4.8). Thus the $\sigma$ model expression for the closed string massless amplitude is [23]
\begin{equation}
    A_N = \frac{\partial}{\partial \log \varepsilon} \langle V_1 \dots V_N \rangle_R = B_N^{(1)} + 2 B_N^{(2)} \log \varepsilon + \dots.
\end{equation}
Equation (4.9) gives a regularized expression for the amplitude. The $\log^n \varepsilon$ terms in (4.9) should group together to form the massless poles of the amplitude according to the standard rule: $1/\alpha' k^2 = \int_\varepsilon^1 dt \, t^{\alpha' k^2 - 1} = -\log \varepsilon - \frac{1}{2} \alpha' k^2 \log^2 \varepsilon + \dots$.


There is an additional observation supporting the claim that ``$\partial / \partial \log \varepsilon$'' is the correct prescription. The $\sigma$ model corresponding to the massless correlators is a (perturbatively) renormalizable theory. Hence its basic objects should be renormalization group invariant. The naive $1/\log \varepsilon$ prescription is not RG invariant (the result of its application depends on a choice of the cutoff $\varepsilon$ if $N \ge 4$, i.e. if (4.9) contains $\log^n \varepsilon$ terms with $n > 1$). The $\partial / \partial \log \varepsilon$ prescription is certainly RG invariant and hence the ``amplitudes'' (4.9) are well-defined from the $\sigma$ model point of view.

It is remarkable that this $\partial / \partial \log \varepsilon$ prescription gives sensible results when applied also to the case of the $N=0, 1, 2$ point amplitudes in the flat background. The flat space string vacuum partition function on the sphere (computed in a Weyl gauge and not divided by $\Omega$) is given by [38, 39] (see also Ref. 20)
\begin{equation}
    Z_0 = c \int d^D y\  \exp \Big[-\frac{1}{3} (D-26) (\log \varepsilon + b)\Big].
\end{equation}
Here $y^\mu$ is the zero mode of the string coordinate. The value of $b = \frac{1}{32\pi} \int R^{(2)} \Delta^{-1} R^{(2)}$ is ambiguous, being dependent on a choice of the Weyl gauge (i.e. on a choice of the 2-metric on the sphere). In (4.10) we have dropped a  power infinity (assuming, e.g., that it is cancelled by the bare tachyon). To obtain the correct result for the coefficient of the logarithmic infinity in (4.10) one should carefully subtract the contribution of the six zero modes of the ghost operator (which correspond to the conformal Killing vectors). Applying $\partial / \partial \log \varepsilon$ to (4.10) we find the following expression for the $N=0$ amplitude
\begin{equation}
    A_0 = \frac{\partial}{\partial \log \varepsilon} Z_0 = -\frac{1}{3} (D-26) Z_0.
\end{equation}
Similar results are found for $A_1$ and $A_2$: $A_N = (D-26) a_N$, $N=0, 1, 2$. 

Note that the normalized correlators $Z_0^{-1} \langle V_1 \rangle$ and $Z_0^{-1} \langle V_1 V_2 \rangle$ do not contain the $\log \varepsilon$ terms so that $\partial / \partial \log \varepsilon$ acts only on $Z_0$. In particular, the graviton and dilaton tadpoles on the sphere are thus proportional to $D-26$. The $N > 2$ point amplitudes contain additional contributions coming from $\log \varepsilon$ terms in $Z_0^{-1} \langle V_1 \dots V_N \rangle$. 
For example, the 3-graviton amplitude reads: $A_3 = (D-26)p_0 + p_2 + p_4 + p_6$, where $p_n$ is of $n$th power in momenta.

These results are in perfect agreement with the effective action which appears in the $\sigma$ model approach: $S = \int d^D y \sqrt{G} e^{-2\phi} (D-26 - \frac{3}{2} \alpha' R + \dots)$ (see Secs. 6 and 7). We would like to emphasize that the standard proof of the vanishing of the tree-level closed string amplitudes $A_N$ for $N=0, 1, 2$ based on the ``$1/\Omega$'' prescription (see e.g. Ref. 40) is applicable only in the flat $D=26$ space and hence cannot be extended to arbitrary $D$ and nontrivial (nonconformal) backgrounds. Such an extension, provided by our prescription (4.9), is, however, of crucial importance for the study of string loop corrections to the equations of motion: to find a nonperturbative string vacuum one is to go out of the tree-level conformal point [25, 28, 30, 24] (see also below).

It is now straightforward to obtain the expression for the massless tree-level effective action corresponding to the prescription (4.9). The massless correlators (4.8) appear as the coefficients in the expansion of the regularized $\sigma$ model partition function $Z(\varphi_0(\varepsilon), \varepsilon)$ in powers of the fields ($\varphi_0^i$ are the bare fields). Hence the regularized amplitudes (4.9) are generated by $\partial Z / \partial \log \varepsilon$. As we have already discussed above the effective action is built out of the amplitudes with massless exchanges subtracted out. This subtraction is equivalent to dropping all $\log^n \varepsilon$ terms in $A_N$ (4.9), i.e. to renormalizing the logarithmic infinities in the $\sigma$ model. Thus the closed string EA is given by
\begin{equation}
    S =\Big. \frac{\partial}{\partial \log \varepsilon} Z\Big|_{\log \varepsilon = 0}
\end{equation}
(as always, we assume that power infinities are omitted in $Z$). 

Let $\varphi^i(\mu)$ be the renormalized couplings and $\beta^i = d\varphi^i / d\log\mu$ the corresponding $\beta$ functions. The renormalizability of the $\sigma$ model implies that ($t = \log \mu$)
\begin{equation}
    Z(\varphi_0(\varepsilon), \varepsilon) = Z_R(\varphi(\mu), \mu), \qquad \frac{dZ_R}{dt} =  \Big( \frac{\partial}{\partial t} + \beta^i \frac{\partial}{\partial \varphi^i}  \Big) Z_R = 0.
\end{equation}
It is easy to prove that $\frac{\partial Z}{\partial \log \varepsilon} = \frac{\partial}{\partial t} Z_R$. In fact, $\frac{\partial \beta}{\partial t} = 0$ and Eq. (4.13) imply that $\frac{d}{dt} \frac{\partial}{\partial t} Z_R = 0$, i.e. that $\partial Z_R / \partial t$ is not renormalized (is RG invariant). Hence (4.12) can be re-written as
\begin{equation}
    S =\Big. \frac{\partial}{\partial t} Z_R\Big|_{t=0} = -\Big. \beta^i \frac{\partial}{\partial \varphi^i} Z_R\Big|_{t=0}.
\end{equation}
The choice of $t=0$ in (4.14) is not essential. In fact, defining
\begin{equation}
    S = \frac{\partial}{\partial t} Z_R = - \beta^i \frac{\partial}{\partial \varphi^i} Z_R
\end{equation}
for arbitrary $t$ we conclude that $\frac{dS}{dt} = 0$, i.e. that $S$ is RG invariant. Hence the actions $S(\varphi(\mu), \mu)$ corresponding to different choices of $\mu$ are related by field re-definitions: $S(\varphi, \mu) = S'(\varphi', \mu')$,\   $\varphi'(\mu') = \varphi(\mu) + \beta_1 \log \mu/\mu' + \dots$, and hence are equivalent [13, 34]. The $t=0$ choice (4.14) is singled out by the fact that, as we shall prove in Sec. 6, it enables one to re-write (4.14) in the following explicit form [23, 45] (see (6.35)--(6.40))
\begin{equation}
    S = - c_0  \Big( \beta^G \cdot \frac{\partial}{\partial G} + \beta^\phi \cdot \frac{\partial}{\partial \phi}  \Big) \int d^D y\,  \sqrt{G}\,  e^{-2\phi},
\end{equation}
or
\begin{equation}
    S = 2c_0 \int d^D y\ \sqrt{G} \ e^{-2\phi}\ \tilde{\beta}^\phi, \qquad\qquad  \tilde{\beta}^\phi = \bar{\beta}^\phi - \frac{1}{4} \bar{\beta}^G_{\mu\nu} G^{\mu\nu},
\end{equation}
where $\tilde{\beta}^\phi$ is the ``central charge'' coefficient of the $\sigma$ model. 

The crucial observation that leads to (4.16) is that re-defining the dilaton one can represent the $\sigma$ model partition function $Z_R|_{t=0}$ simply as $c_0 \int d^D y\  \sqrt{G} \ e^{-2\phi}$. What we have thus proved is that there exists a choice of the fields (or a choice of a renormalization scheme in the $\sigma$ model) for which the EA can be represented as the integral (4.17) of the ``central charge'' coefficient. Equivalent EA's (reproducing the same massless string amplitudes) are related to (4.17) by field redefinitions.\footnote{It should be 
understood that the beta functions in (4.17) correspond to the regularisation scheme which is related, e.g., to ``dimensional regularisation with minimal subtraction'' scheme by some coupling redefinition (which is difficult to specify explicitly, see Sec. 6).
Still,  we would like to stress that the representation (4.17) is quite informative, since ``ambiguous'' structures in the beta functions and the effective action can be easily classified and separated out in perturbation theory (see, e..g.,  Ref.  52). }

The remarkable ``coincidence'' is that the action (4.17) is distinguished also by the following property: its equations of motion are equivalent to the Weyl invariance conditions for the $\sigma$ model (see Sec. 6). We thus get a proof of the conjecture that the tree-level effective equations of motion for the massless string modes are \textit{equivalent to the conditions of Weyl invariance of the corresponding $\sigma$ model}.\footnote{For other attempts  to prove this conjecture  see e.g. [25,35,46-48,42].} There have been little doubts that this ``equivalence conjecture'' should be true since its validity is necessary for the consistency of the ``critical'' string theory, and since it has passed several nontrivial perturbative checks (see, e.g., [27, 34, 49-52] and references therein).

As we have already emphasized, the $\partial / \partial \log \varepsilon$ prescription (4.9) can be applied to the computation of string tree amplitudes in a nontrivial (nonconformal) background. Consider, for example, the massless tadpoles: $A_{1i} = \frac{\partial}{\partial \log \varepsilon} \langle V_{i0} \rangle$. Here $V_{i0}$ is the (bare) vertex operator corresponding to the field $\varphi_0^i$, $V_{i0} = \frac{\partial I}{\partial \varphi_0^i}$, and $\langle \dots \rangle = \int [dx] e^{-I} \dots$, where $I$ is the $\sigma$ model action (3.1). Using that $\langle V_{i0} \rangle = \partial Z / \partial \varphi_0^i$ (see Sec. 5) we can relate $A_{1i}$ to the first derivative of the effective action (4.15)
\begin{equation}
    A_{1i} = \frac{\partial}{\partial \log \varepsilon} \frac{\partial Z}{\partial \varphi_0^i} = \frac{\partial}{\partial \varphi_0^i} \frac{\partial Z}{\partial \log \varepsilon} = \frac{\partial}{\partial \varphi_0^i} \frac{\partial Z_R}{\partial t} = \frac{\partial \varphi^j}{\partial \varphi_0^i} \frac{\partial S}{\partial \varphi^j}
\end{equation}
($\partial \varphi^j / \partial \varphi_0^i$ is the renormalization matrix for the composite operator $V_i$, cf. (5.9)). Hence the vacuum stability conditions $A_{1i} = 0$ are indeed equivalent to the effective equations of motion $\frac{\partial S}{\partial \varphi^i} = 0$!

To find the string loop corrected equations of motion we are to equate to zero the full (summed over genera) tadpole amplitude on a nontrivial background [25, 44]
\begin{equation}
    A_{1i}^{tot} = \sum_{n=0}^\infty \langle V_i \rangle_n = 0.
\end{equation}
Since the M\"obius group is trivial for genus $n > 0$ surfaces, the $n > 0$ tadpoles are given simply by the $\varphi$ derivatives of the corresponding partition functions. Hence the generalized vacuum stability conditions $A_{1i}^{tot} = 0$ should be equivalent to the equations of motion for the following loop corrected effective action
\begin{align}
    S &= \frac{\partial}{\partial t} Z_{0R} + \sum_{n=1}^\infty Z_{nR} \nonumber \\
    &= c_0 \int d^D y \sqrt{G} \Big[e^{-2\phi}(D-26 - \frac{3}{2}\alpha' R + \dots) + (\lambda_1 + \dots) + e^{2\phi}(\lambda_2 + \dots)\Big] \nonumber \\
    &\qquad\qquad \ \ \ \ \ \ \ \ \ \  + \text{nonlocal contributions}.
\end{align}
The sum of the partition functions should be renormalized with respect to both ``local'' and ``modular'' infinities [25, 28, 24]. Computing, e.g., the sum of the massless tadpoles on the sphere and on the torus (using our $\partial / \partial \log \varepsilon$ prescription in the case of the sphere) we do get the equation corresponding to (4.20).\footnote{The   naive ``$1/\log \varepsilon$''   prescription used in Ref. 44  fails to reproduce  the finite  and consistent  equations of motion.}

The results of this section have a straightforward generalization to the case of the closed superstring theory.
 Since the 2d supersymmetry implies the cancellation of power infinities,\footnote{The 2d supersymmetry is present in the sigma model defined on a complex plane. However,  the  theory on the plane contains IR divergences. To have a consistent theory we are to define it on a compact space, e.g.  on the standard sphere. In this case the NS ``boundary conditions'' break down the supersymmetry. Like in the finite temperature case the supersymmetry is still present at small distances, implying the cancellation of power infinities.
It is interesting to note that it is a breakdown of supersymmetry on a sphere (in the NS sector) that makes the corresponding partition function non-trivial. For a detailed discussion of the analogous facts in the open superstring case see Ref. 22.   }
  the superM\"obius volume corresponding to the sphere is simply proportional to $\log \varepsilon$ and it is not necessary to renormalized power infinities in Eq. (4.9).

\section{Weyl Anomaly in the Sigma Model}
\renewcommand{\theequation}{5.\arabic{equation}}
\setcounter{equation}{0}

In Secs. 5, 6 and 7 we shall ``forget'' about string theory and study some general properties of the perturbatively renormalizable bosonic $\sigma$ model (3.1) defined on a curved 2-space of  spherical topology. Our aim in this section will be to derive the operator expression for the Weyl anomaly in the theory (3.1) [53, 54] and to analyze some of its consequences.

\subsection{}

 Let $\{\varphi_0^i\}$ be a set of bare couplings present in a bare $\sigma$ model action (in this section we shall use subscript 0 to denote the bare quantities)
\begin{equation}
    I = \int d^d \sigma V_{i0} \cdot \varphi_0^i.
\end{equation}
Here $V_{i0}$ are the corresponding bare composite operators. It is useful to consider $\varphi^i$ as functions on $R^D$ with coordinates $\{y^\mu\}$. The dot in (5.1) stands for the scalar product: $f \cdot h = \int d^D y f(y) h(y)$. We shall sometimes omit the dot assuming that the integration over $y^\mu$ is implied by summation over $i$. For example, in the case of the bosonic $\sigma$ model with only the metric coupling
\begin{equation}
    \varphi^i = \{G_{\mu\nu}(y)\}, \qquad V_i = \{V_G\}, \qquad V_G^{\mu\nu} \sim \sqrt{g}\, \partial^a x^\mu \partial_a x^\nu \delta^{(D)}(y - x(\sigma)).
\end{equation}
The momentum transforms of $V_i$ are the massless vertex operators in the string theory, e.g.
\begin{equation}
    V_G(y) = \int \frac{d^D k}{(2\pi)^D} e^{-iky} \tilde{V}_G(k), \qquad \tilde{V}_G^{\mu\nu}(k) \sim \sqrt{g}\, \partial^a x^\mu \partial_a x^\nu e^{ikx}.
\end{equation}
We shall use dimensional regularization and choose the renormalized couplings to be dimensionless. Then the bare couplings $\varphi_0^i$ have the mass dimension $\varepsilon = d-2$ and $V_{i0}$ have dimension 2. In the minimal subtraction scheme
\begin{equation}
    \varphi_0^i = \mu^\varepsilon \Big[ \varphi^i + \sum_{n=1}^\infty \frac{1}{\varepsilon^n} T_n^i(\varphi) \Big].
\end{equation}
The RG $\beta$ functions are defined as usual
\begin{equation}
    \hat{\beta}^i \equiv \frac{d\varphi^i}{dt} = -\varepsilon \varphi^i + \beta^i, \qquad t = \log \mu,
\end{equation}
\begin{equation}
    \beta^i = -T_1^i + \varphi^j \frac{\partial}{\partial \varphi^j} T_1^i, \qquad \frac{d\varphi_0}{dt} = 0. \nonumber
\end{equation}
Here $\partial / \partial \varphi^i$ denotes the functional derivative
\begin{equation}
    f \cdot \frac{\partial F}{\partial \varphi} = \int d^D y f(y) \frac{\delta F[\varphi]}{\delta \varphi(y)} = (F[\varphi+f])_{\text{linear piece in } f}.
\nonumber \end{equation}
`` $\hat{ }$ '' on a finite quantity indicates that it is defined in $d$ dimensions, i.e. that the $\varepsilon$ dependent terms are retained in it.

 Let us define the partition function
\begin{equation}
    Z_0(\varphi_0, \varepsilon) = e^{-W_0(\varphi_0, \varepsilon)} = \int [dx] \ e^{-I}, \qquad \langle \dots \rangle = Z_0^{-1} \int [dx] \ e^{-I} \dots.
\end{equation}
For simplicity we assume here that the 2d space has the topology of a plane, i.e. that there is no normalizable zero mode of $x^\mu$ (the trivial measure $[dx]$ is covariant in dimensional regularization). The renormalizability of the model implies that $Z_0$ and $W_0$ must be finite, i.e. $W_0(\varphi_0, \varepsilon) = W(\varphi, \mu)$. Since the bare couplings are independent of $\mu$
\begin{equation}
    \frac{dW}{dt} = 0, \qquad \text{i.e.} \qquad \frac{\partial W}{\partial t} + \hat{\beta}^i \frac{\partial W}{\partial \varphi^i} = 0.
\end{equation}
Let us consider the renormalization of the composite operators $V_i$. We define the renormalized operators or ``normal products'' $[V_i]$ (see e.g. Ref. 55) so that
\begin{equation}
    \int d^d \sigma\  [V_i] = \frac{\partial I}{\partial \varphi^i}.
\end{equation}
Then $\langle \int d^d \sigma\ [V_i] \rangle$ is equal to $\frac{\partial W}{\partial \varphi^i}$ and hence is finite. In general
\begin{equation}
    [V_i] = V_{j0} Z^j{}_i, \qquad Z^j{}_i = \mu^\varepsilon  \Big( \delta^j_i + \sum_{n=1}^\infty \frac{1}{\varepsilon^n} X_{nj}^i(\varphi)  \Big),
\end{equation}
where $Z^j{}_i$ is a bilocal operator on the functions on $R^D$. Here we have assumed that $\{V_{i0}\}$ comprise the complete set of dimension 2 operators modulo the equation of motion operator $E = E^\mu(x) \delta I / \delta x^\mu$. Since in dimensional regularization $\langle E \rangle = 0$ we disregard $E$ in the operator relations. Contracting (5.9) with a set of arbitrary functions $F^i(y)$ we have
\begin{equation}
    [V_i] F^i = V_{i0} \tilde{F}^i, \qquad \tilde{F}^i = Z^i{}_j F^j.
\end{equation}
According to (5.1), (5.8) $\int d^d \sigma [V_i] F^i = \int d^d \sigma V_{i0} \frac{\partial \varphi_0^i}{\partial \varphi^j} F^j$ and hence
\begin{equation}
    [V_i] F^i = V_{i0} \frac{\partial \varphi_0^i}{\partial \varphi^j} F^j + \partial_a (\Omega_i^a F^i),
\end{equation}
where we have included a total derivative term which may appear in passing from the integrated to the local expression. Since $\{V_{i0}\}$ forms a complete set of dimension 2 operators, it should be possible to rewrite the total derivative term in (5.11) as
\begin{equation}
    \partial_a (\Omega_i^a F^i) = V_{i0} \Lambda_j^i F^j, \qquad \Lambda_j^i = \mu^\varepsilon \sum_{n=1}^\infty \frac{1}{\varepsilon^n} Q_{nj}^i(\varphi).
\end{equation}
Then comparing (5.9), (5.11), (5.12), we conclude that
\begin{equation}
    Z^i{}_j = \frac{\partial \varphi_0^i}{\partial \varphi^j} + \Lambda_j^i, \qquad X_{nj}^i = \frac{\partial T_n^i}{\partial \varphi^j} + Q_{nj}^i.
\end{equation}
Note that
\begin{equation}
    V_{i0} = [V_j] Z^{-1j}{}_i, \qquad Z^{-1j}{}_i = \mu^\varepsilon  \Big( \delta^j_i - \frac{1}{\varepsilon} X_{1i}^j + O \Big(\frac{1}{\varepsilon^2} \Big)  \Big).
\end{equation}
The anomalous dimension matrix for  composite operators is defined as  $(dV_{i0}/dt = 0)$
\begin{equation}
    \frac{d}{dt} [V_i] = [V_j] \hat{\Gamma}^j{}_i, \qquad \qquad \hat{\Gamma}^j{}_i = Z^{-1j}{}_k \frac{d}{dt} Z^k{}_i.
\end{equation}
Observing that $\frac{d}{dt} = \frac{\partial}{\partial t} + \hat{\beta}^i \frac{\partial}{\partial \varphi^i}$ we find
\begin{equation}
    \hat{\Gamma}^i{}_j = \varepsilon \delta^i_j - \varphi^k \frac{\partial}{\partial \varphi^k} X_{1j}^i = \frac{\partial \hat{\beta}^i}{\partial \varphi^j} - P^i{}_j,
\qquad \ \ \ 
    P^i{}_j = \varphi^k \frac{\partial}{\partial \varphi^k} Q_{1j}^i.
\end{equation}
An alternative derivation of the equation: $\hat{\Gamma}^i{}_j = \frac{\partial \hat{\beta}^i}{\partial \varphi^j} + (\text{total derivative term})$ is based on integrating (5.15) over 2-space, taking $\langle \dots \rangle$ and using (5.8). Then we find that $\frac{d}{dt} \frac{\partial W}{\partial \varphi^i} = \frac{\partial W}{\partial \varphi^j} \hat{\Gamma}^j{}_i$. The final step is to employ the relation
\begin{equation}
    \frac{d}{dt} \frac{\partial W}{\partial \varphi^i} = - \frac{\partial W}{\partial \varphi^j} \frac{\partial \hat{\beta}^j}{\partial \varphi^i}
\end{equation}
which follows from (5.7) after applying $\partial / \partial \varphi^i$.

Now let us derive the expression for the Weyl anomaly. The trace of the energy-momentum tensor is defined as usual
\begin{equation}
    T_a^a = \frac{2}{\sqrt{g}} g^{ab} \frac{\delta I}{\delta g^{ab}}, \qquad \qquad \langle T_a^a \rangle = \frac{2}{\sqrt{g}} g^{ab} \frac{\delta W}{\delta g^{ab}}.
\end{equation}
Note that $\langle T_a^a \rangle$ is the expectation value computed in a trivial vacuum ($x^\mu=0$). In a general case of an arbitrary background we are to replace $W$ in (5.18) by the 2d effective action of the $\sigma$ model (the sum of 1-PI graphs in a background, see, e.g., Refs. 56 and 57). $T_a^a$, being a finite operator (its expectation value is a derivative of a finite quantity over a finite one) of dimension $d$, must be expressible in terms of the finite operators $[V_i]$. We shall assume that
\begin{equation}
    2 g^{ab} \frac{\delta}{\delta g^{ab}} V_{i0} = -\varepsilon V_{i0} + \partial_a \omega_i^a,
\end{equation}
where the total derivative term accounts for possible non-invariance of the classical (scale-invariant in $d=2$) action under the Weyl transformations. Using the bare equations of motion we can represent this term as a linear combination of $V_{i0}$ (cf. (5.12))
\begin{equation}
    \partial_a \omega_i^a = V_{j0} \lambda_i^j(\varphi_0), \qquad \lambda_i^j(\varphi_0) = \lambda_i^j(\varphi) + O \Big(\frac{1}{\varepsilon} \Big).
\end{equation}
Then
\begin{equation}
    \sqrt{g}\, T_a^a = V_{i0} \psi^i, \qquad \psi^i = -\varepsilon \varphi_0^i + \lambda_j^i(\varphi_0) \varphi_0^j,
\end{equation}
\begin{equation}
    \psi^i = \mu^\varepsilon  \Big[ -\varepsilon \varphi^i - T_1^i(\varphi) + \lambda^i(\varphi) + O \Big(\frac{1}{\varepsilon} \Big)  \Big], \qquad \lambda^i \equiv \lambda_j^i(\varphi) \varphi^j.\nonumber
\end{equation}
Expressing $V_{i0}$ in terms of $V_i$ we get
\begin{equation}
    V_{i0} \psi^i \equiv [V_i] Z^{-1i}{}_j \psi^j, \qquad Z^{-1i}{}_j \psi^j = -\varepsilon \varphi^i - T_1^i + \lambda^i + \varphi^j \frac{\partial T_1^i}{\partial \varphi^j} + Q^i + O \Big(\frac{1}{\varepsilon} \Big),
\end{equation}
\begin{equation}
    Q^i \equiv Q_{1j}^i \varphi^j.\nonumber
\end{equation}
All $1/\varepsilon^n$, $n \ge 1$, pole terms in (5.22) must cancel out because of the finiteness of $T_a^a$. Thus finally
\begin{equation}
    \sqrt{g}\, T_a^a = [V_i] \hat{\bar{\beta}}^i, \qquad \hat{\bar{\beta}}^i = \hat{\beta}^i + \sigma^i = -\varepsilon \varphi^i + \bar{\beta}^i,
\end{equation}
\begin{equation}
    \bar{\beta}^i = \beta^i + \sigma^i, \qquad \sigma^i \equiv \lambda^i + Q^i,\nonumber
\end{equation}
where $\bar{\beta}^i$ is simply equal to the finite term in (5.22) (we have used Eq. (5.5)). The difference ($\sigma^i$) between the Weyl anomaly coefficients $\bar{\beta}^i$ and the RG $\beta^i$ functions is due to the total derivative terms in (5.11) and (5.19): $\sqrt{g} T_a^a = [V_i] \hat{\beta}^i + \partial_a (-\varepsilon \Omega_i^a \varphi_0^i + \omega_i^a \varphi_0^i)$. Since these terms drop out under the integration over 2-space we find the standard expression for the scale anomaly
\begin{equation}
    \int d^d \sigma \sqrt{g}\, T_a^a = \int d^d \sigma\  [V_i] \ \hat{\beta}^i.
\end{equation}
This relation can be easily proved directly using $\int d^d \sigma \sqrt{g}\, T_a^a = -\partial I / \partial t$ and Eqs. (5.7) and (5.8).

Since $dI/dt=0$, $T_a^a$ in (5.18) is not renormalized, i.e.
\begin{equation}
    \frac{d}{dt} T_a^a = 0.
\end{equation}
It is easy to see by applying $\partial / \partial t$ to $\frac{dW}{dt} = 0$ (Eq. (5.7)) that the equation $\frac{d}{dt} \int d^d \sigma \sqrt{g}\, T_a^a = 0$ is satisfied identically. However, (5.25) implies a nontrivial relation between the RG coefficients when the total derivative terms are taken into account. Making use of (5.23), (5.15) and (5.16) we get
\begin{equation}
    \sqrt{g}\, \frac{d}{dt} T_a^a = [V_i] \alpha^i, \qquad \alpha^i = \frac{d\hat{\bar{\beta}}^i}{dt} + \hat{\Gamma}^i{}_j \hat{\bar{\beta}}^j,
\end{equation}
\begin{equation}
    \alpha^i = \hat{\beta}^j \frac{\partial \hat{\bar{\beta}}^i}{\partial \varphi^j} - \frac{\partial \hat{\bar{\beta}}^i}{\partial \varphi^j} \hat{\bar{\beta}}^j - P^i{}_j \hat{\bar{\beta}}^j = \hat{\beta}^j \frac{\partial \sigma^i}{\partial \varphi^j} - \frac{\partial \hat{\beta}^i}{\partial \varphi^j} \sigma^j - P^i{}_j \hat{\beta}^j.
\end{equation}
The $O(\varepsilon)$ term in (5.27) vanishes identically ($\lambda^i_j$ in (5.19) and (5.20) is of zero degree of homogeneity in $\varphi^i$, i.e. $\varphi^k \frac{\partial}{\partial \varphi^k} \lambda^i_j = 0$). Thus finally we arrive at the following identity
\begin{equation}
    \alpha^i = \beta^j \frac{\partial \sigma^i}{\partial \varphi^j} - \frac{\partial \beta^i}{\partial \varphi^j} \sigma^j - P^i{}_j \beta^j = 0,
\end{equation}
where $\sigma^i$ and $P^i{}_j$ are defined in (5.23) and (5.16). Equation (5.28) must hold true for arbitrary $\varphi^i$ and it provides a relation between the $\beta$ functions and $Q_{1j}^i$. Note that (5.28) can be re-written also as
\begin{equation}
    \alpha^i = \beta^j \frac{\partial \sigma^i}{\partial \varphi^j} - \frac{\partial \bar{\beta}^i}{\partial \varphi^j} \sigma^j - P^i{}_j \bar{\beta}^j = 0.
\end{equation}

\subsection{}

 It is now straightforward to apply the above general relations to the case of the renormalizable bosonic $\sigma$ model (3.1) with the bare action
\begin{align}
    I = &\frac{1}{2\lambda} \int d^d \sigma \Big[ \sqrt{g}\, \partial^a x^\mu \partial_a x^\nu G_{0\mu\nu}(x) + \epsilon^{ab} \partial_a x^\mu \partial_b x^\nu B_{0\mu\nu}(x) + \alpha' \sqrt{g}\, \bar{R} \phi_0(x) \Big], \\
    &\qquad \lambda = 2\pi\alpha', \qquad \qquad  \bar{R} \equiv {R^{(d)} \over  d-1 } \ , \nonumber 
\end{align}
where 
$R^{(d)}$ is the scalar curvature of $g_{ab}$. In two dimensions $\epsilon^{ab} = i \varepsilon^{ab}$, $\varepsilon^{12}=1$ ($i$ is needed in order for the Minkowski-space action to be real). There are some subtleties in the treatment of the antisymmetric tensor coupling (see Ref. 54) which we shall disregard here. In the present case $\varphi^i = \{G_{\mu\nu}, B_{\mu\nu}, \phi\}$. Observing that the renormalization of the flat-space couplings $G$ and $B$ is not influenced by the curvature we have (cf. (5.4))
\begin{equation}
    G_0 = \mu^\varepsilon  \Big( G + \sum_{n=1}^\infty \frac{1}{\varepsilon^n} T_n^G(G, B)  \Big), \qquad B_0 = \mu^\varepsilon  \Big( B + \sum_{n=1}^\infty \frac{1}{\varepsilon^n} T_n^B(G, B)  \Big),
\end{equation}
\begin{equation}
    \phi_0 = Z_1 \phi + Z_2 = \mu^\varepsilon  \Big( \phi + \sum_{n=1}^\infty \frac{1}{\varepsilon^n} T_n^\phi(G, B, \phi)  \Big),
\qquad 
    T_n^\phi = \zeta_n(G, B)\phi + \kappa_n(G, B).
\end{equation}
Hence
\begin{equation}
    \hat{\beta}^G = -\varepsilon G + \beta^G, \qquad \hat{\beta}^B = -\varepsilon B + \beta^B, \qquad \beta^{G, B} = (-1 + \Delta) T_1^{G, B},
\end{equation}
\begin{equation}
    \hat{\beta}^\phi = -\varepsilon \phi + \beta^\phi, \qquad \beta^\phi = -\gamma \phi + \omega, \qquad \gamma = -\Delta \zeta_1, \qquad \omega = (-1 + \Delta) \kappa_1,
\end{equation}
\begin{equation}
    \Delta \equiv G \cdot \frac{\partial}{\partial G} + B \cdot \frac{\partial}{\partial B}.\nonumber 
\end{equation}
Here $\gamma$ is a differential operator (anomalous dimension operator for composite operators of naive dimension zero [58]). Both $\gamma$ and $\omega$ depend only on $G$ and $B$. Introducing three arbitrary functions $F_{\mu\nu}^G(y)$, $F_{\mu\nu}^B(y)$ and $F^\phi(y)$ we can represent the corresponding composite operators in the form
\begin{equation}
    V_G \cdot F^G = \frac{1}{2\lambda} \, \sqrt{g}\, \partial^a x^\mu \partial_a x^\nu F_{\mu\nu}^G(x), \nonumber
\end{equation}
\begin{equation}
    V_B \cdot F^B = \frac{1}{2\lambda} \,  \epsilon^{ab} \partial_a x^\mu \partial_b x^\nu F_{\mu\nu}^B(x), \qquad V_\phi \cdot F^\phi = \frac{\alpha'}{2\lambda} \,  \sqrt{g}\, \bar{R} F^\phi(x).
\end{equation}
In addition to the three operators $V_i$ and $E = E^\mu(x) \delta I / \delta x^\mu$ there are also two total derivative-dimension 2 operators which can mix with $V_G$ and $V_B$ under renormalization
\begin{equation}
    N = \partial_a (\sqrt{g} g^{ab} N_\mu(x) \partial_b x^\mu), \qquad K = \partial_a (\epsilon^{ab} K_\mu(x) \partial_b x^\mu).
\end{equation}
Using the bare equations of motion we can re-write (5.36) as the linear combinations of the first four bare operators
\begin{equation}
    N = V_{G0}^{\mu\nu} \cdot (2 \mathcal{D}_{0(\mu} N_{\nu)}) + V_{B0}^{\mu\nu} \cdot (\frac{1}{2} H^\lambda_{\mu\nu} N_\lambda) + V_{\phi 0} \cdot (\frac{1}{2} \mathcal{D}^\mu_0 \phi_0 N_\mu) + E,
\end{equation}
\begin{equation}
    E^\mu = -\lambda G^{\mu\nu} N_\nu, \qquad K = V_{B0}^{\mu\nu} \cdot (\partial_{[\mu} K_{\nu]}),
\qquad \ \ \ 
    H_{\lambda\mu\nu} \equiv 3 \partial_{[\lambda} B_{\mu\nu]}. \nonumber
\end{equation}
Hence the total derivative terms in (5.11) can be represented as (note that $V_\phi$ cannot mix with total derivatives)
\begin{equation}
    \partial_a \Omega_G^a = N_G + K_G, \qquad \partial_a \Omega_B^a = N_B + K_B, \qquad \partial_a \Omega_\varphi^a = 0.
\end{equation}
Then  the total derivative terms in the renormalization matrix $Z^i{}_j$ in (5.13) are parametrized by the bilocal objects $N_G^{\lambda\rho}{}_\mu$, $N_B^{\lambda\rho}{}_\mu$, $K_G^\lambda{}_{\mu}$, $K_B^\lambda{}_{\mu}$, depending on $G$ and $B$ (e.g. $Z^{G\lambda\rho}_{G\mu\nu} = \partial G_{0\mu\nu} / \partial G_{\lambda\rho} + \mathcal{D}_{0(\mu} N^{\lambda\rho}_{G\nu)}$). In general
\begin{equation}
    N_\alpha = \mu^\varepsilon \sum_{n=1}^\infty \frac{1}{\varepsilon^n} S_{n\alpha}, \qquad K_\alpha = \mu^\varepsilon \sum_{n=1}^\infty \frac{1}{\varepsilon^n} U_{n\alpha}, \qquad \alpha = (G, B).
\end{equation}
Then $Q_{nj}^i$ in (5.12) can be expressed in terms of $S_1$ and $U_1$. We get for $P^i{}_j$ in (5.16)
\begin{equation}
    (P_G^G)^{\lambda\rho}_{\mu\nu} = \mathcal{D}_{(\mu} \tilde{S}^\lambda_{\nu)}, \qquad (P_B^B)^{\lambda\rho}_{\mu\nu} = \frac{1}{2} H_{\mu\nu}^\sigma \tilde{S}^\lambda_\sigma + \partial_{[\mu} \tilde{U}^\lambda_{\nu]},
\end{equation}
\begin{equation}\nonumber
    (P_\phi^\phi)^{\lambda\rho} = \frac{1}{2} \mathcal{D}^\sigma \phi \tilde{S}^\lambda_\sigma, \qquad P_i^\phi = 0,
\end{equation}
\begin{equation}
    \tilde{S}_i \equiv \Delta S_{1i}, \qquad \tilde{U}_i = \Delta U_{1i}, \qquad \Delta \equiv G \cdot \frac{\partial}{\partial G} + B \cdot \frac{\partial}{\partial B}.
\end{equation}
The trace of the energy-momentum tensor (5.18) corresponding to (5.30) is
\begin{equation}
    \sqrt{g}\, T_a^a = V_{G0} \cdot (-\varepsilon G_0) + V_{B0} \cdot (-\varepsilon B_0) + V_{\phi 0} (-\varepsilon \phi_0) + \frac{1}{2\pi} \partial_a (\sqrt{g} \partial^a \phi_0)_0\ , 
\end{equation}
where we have used that under $\delta g_{ab} = 2 \rho g_{ab}$, $\delta (\sqrt{g} \bar{R}) = \sqrt{g}\, (\varepsilon \bar{R} \rho - 2 \nabla^2 \rho)$. 

\def \no {\nonumber}
\def \foot {\footnote}

Using the bare equations of motion we can re-write the total derivative term in (5.42) as a combination of $V_{i0}$ (and $E$). Expressing the bare operators in terms of the renormalized ones (cf. (5.19)-(5.23)) we end up with the following representation for the \textit{Weyl anomaly} of the $\sigma$ model (see Ref. 54 for details)
\begin{equation}
    2\lambda \sqrt{g}\, T_a^a = [\sqrt{g} \partial^a x^\mu \partial_a x^\nu \hat{\bar{\beta}}_{\mu\nu}^G(x)] + [\epsilon^{ab} \partial_a x^\mu \partial_b x^\nu \hat{\bar{\beta}}_{\mu\nu}^B(x)] + [\alpha' \sqrt{g}\, \bar{R} \hat{\bar{\beta}}^\phi(x)],
\end{equation}
or directly in $d=2$
\begin{equation}
    2\lambda \sqrt{g}\, T_a^a = [\sqrt{g} \partial^a x^\mu \partial_a x^\nu \bar{\beta}_{\mu\nu}^G(x)] + [\epsilon^{ab} \partial_a x^\mu \partial_b x^\nu \bar{\beta}_{\mu\nu}^B(x)] + [\alpha' \sqrt{g}\, R^{(2)} \bar{\beta}^\phi(x)], \tag{5.43'}
\end{equation}
\begin{equation}
    \bar{\beta}_{\mu\nu}^G = \beta_{\mu\nu}^G + \mathcal{D}_{(\mu} M_{\nu)}, \qquad \bar{\beta}^\phi = \beta^\phi + \frac{1}{2} M^\lambda \partial_\lambda \phi,
\end{equation}
\begin{equation}
    \bar{\beta}_{\mu\nu}^B = \beta_{\mu\nu}^B + \frac{1}{2} H_{\mu\nu}^\lambda M_\lambda + \partial_{[\mu} L_{\nu]}, \qquad M_\mu = 2\alpha' \partial_\mu \phi + W_\mu,
\end{equation}
\begin{equation}\nonumber 
    W_\mu(G, B) \equiv S_{1G\mu} \cdot G + S_{1B\mu} \cdot B, \qquad L_\mu \equiv U_{1G\mu} \cdot G + U_{1B\mu} \cdot B.
\end{equation}
The expression for the {\it  scale anomaly} (5.24) is (in $d=2$)
\begin{align}
    2\lambda \int d^2 \sigma \sqrt{g}\, T_a^a = \int d^2 \sigma \Big( &[\sqrt{g} \partial^a x^\mu \partial_a x^\nu \beta_{\mu\nu}^G(x)] + [\epsilon^{ab} \partial_a x^\mu \partial_b x^\nu \beta_{\mu\nu}^B(x)] \nonumber \\
    & + [\alpha' \sqrt{g}\, R^{(2)} \beta^\phi(x)] \Big).
\end{align}
Next let us find the explicit form of the identities which follow from the condition of nonrenormalization of $T_a^a$ (see (5.26)-(5.28)). Given $P^i_j$ (5.40) and $\sigma^i = \bar{\beta}^i - \beta^i$ (see (5.44)) it is straightforward to compute $\alpha^i$ in (5.28). Since $\int d^d \sigma \sqrt{g}\, T_a^a$ vanishes identically, $\frac{d}{dt} \sqrt{g}\, T_a^a$ must be a total derivative operator, i.e. a sum of $N$ and $K$ in (5.36). In view of (5.37) this implies
\begin{equation}
    \frac{d}{dt} \sqrt{g}\, T_a^a = [V_i] \alpha^i, \qquad \qquad \alpha_{\mu\nu}^G = \mathcal{D}_{(\mu} \Lambda_{\nu)},
\end{equation}\begin{equation}
    \alpha_{\mu\nu}^B = \frac{1}{2} H_{\mu\nu}^\lambda \Lambda_\lambda + \partial_{[\mu} \Sigma_{\nu]}, \qquad \alpha^\phi = \frac{1}{2} \mathcal{D}^\lambda \phi \Lambda_\lambda.\no
\end{equation}
Hence $\alpha^i = 0$ is equivalent to $\Lambda_\mu = 0, \Sigma_\mu = 0$. We find [54]
\begin{equation}
    \Lambda_\mu = 2\alpha' \partial_\mu \bar{\beta}^\phi +  \Big( \bar{\beta}^G \cdot \frac{\partial}{\partial G} + \bar{\beta}^B \cdot \frac{\partial}{\partial B}  \Big)
     M_\mu     -        \bar \beta^G_{\mu\nu} M^\nu - \tilde{S}_{G\mu} \cdot \bar{\beta}^G - \tilde{S}_{B\mu} \cdot \bar{\beta}^B = 0,
\end{equation}
\begin{equation}
    \Sigma_\mu =  \Big( \bar{\beta}^G \cdot \frac{\partial}{\partial G} + \bar{\beta}^B \cdot \frac{\partial}{\partial B}  \Big) L_\mu + \bar{\beta}_{\mu\nu}^B M^\nu - \tilde{U}_{G\mu} \cdot \bar{\beta}^G - \tilde{U}_{B\mu} \cdot \bar{\beta}^B = 0.
\end{equation}
Equation (5.48) (first derived in the case of $B_{\mu\nu} = 0$ in Ref. 59) can be re-written as
\begin{equation}
    2\alpha' \partial_\mu \bar{\beta}^\phi = \mathcal{O}_\mu^G \cdot \bar{\beta}^G + \mathcal{O}_\mu^B \cdot \bar{\beta}^B.
\end{equation}
Equations (5.48) and (5.49) must be true for arbitrary $G, B, \phi$. Hence the terms in (5.48) depending on $\phi$ must be separately equal to zero. This observation makes it possible to compute the dilaton $\beta$ function (5.34) (its $\gamma$ and $\omega$ parts) using only the $\beta$ functions for $G$ and $B$ and the renormalization matrices $S_{1i}, U_{1i}$ (which govern the mixing of dimension 2 composite operators with the total derivative operators (5.36)) which can all be computed in a \textit{flat} 2-space.


Let us now discuss some consequences of the above general relations (5.43)--(5.50). Given the operator expression for the Weyl anomaly (5.43) we can compute the expectation value of the trace $\langle T_a^a \rangle$. Integrating the relation $\langle T_a^a \rangle = \frac{2}{\sqrt{g}} g^{ab} \frac{\delta W}{\delta g^{ab}}$ (using that in $d=2$, $g_{ab}$ can be represented as $e^{2\rho}\hat{g}_{ab}$ where $\hat{g}_{ab}$ is a standard metric on a plane or a sphere) we can thus determine $W$.\foot{To find the dependence of $W$ on the moduli (for higher genus surfaces)  we need  to know $\langle T_{ab} \rangle$ [61]. } Clearly, $\langle T_a^a \rangle = 0$ is equivalent to $W = \text{const}$ (i.e. to the decoupling of $\rho$ in the string partition function). Computing the expectation value near $x^\mu = 0$ (here we assume that the 2-space has the topology of a plane) we get [20, 27, 34]
\begin{equation}
    \langle T_a^a \rangle = \frac{1}{4\pi} R^{(2)} \tilde{\beta}^\phi + \sum_n f_n \omega_n, \tag{5.51}
\end{equation}
where $\tilde{\beta}^\phi$ and $f_n$ are covariant functions of the fields $\varphi^i$ (taken at $x=0$) which can be represented as linear differential operators acting on $\bar{\beta}^i$ (hence they vanish if $\bar{\beta}^i = 0$). $\omega_n$ are nonlocal dimension 2 functionals of $g_{ab}$. The basic Weyl anomaly (``central charge'') coefficient is given by [34]
\begin{equation}
    \tilde{\beta}^\phi = \bar{\beta}^\phi - \frac{1}{4} \bar{\beta}_{\mu\nu}^G \zeta^{\mu\nu} + \dots, \qquad \zeta^{\mu\nu} = G^{\mu\nu} + \alpha' c_1 R^{\mu\nu} + \dots. \tag{5.52}
\end{equation}
Dots in $\tilde{\beta}^\phi$ stand for other possible terms (depending on derivatives of $\bar{\beta}^\phi$ and $\bar{\beta}^G$) which have renormalization scheme dependent coefficients. We have used that in dimensional regularization: $\langle \partial^a x^\mu \partial_a x^\nu \rangle = -\frac{\alpha'}{4} R^{(2)} G^{\mu\nu} + \dots$ [60]. The coefficients $c_1$, etc. in $\zeta^{\mu\nu}$ are renormalization scheme (RS) dependent: we can redefine $G^{\mu\nu}$ so as to ensure $\zeta^{\mu\nu} = G^{\mu\nu}$. There should exist an RS in which
\begin{equation}
    \tilde{\beta}^\phi = \bar{\beta}^\phi - \frac{1}{4} \bar{\beta}_{\mu\nu}^G G^{\mu\nu}. \tag{5.53}
\end{equation}
A remarkable consequence of (5.50) is that if $\bar{\beta}^G = \bar{\beta}^B = 0$ then $\bar{\beta}^\phi = \text{const}$. Then (5.43') reduces to
\begin{equation}
    T_a^a = \frac{1}{4\pi} R^{(2)} \bar{\beta}^\phi = \langle T_a^a \rangle, \qquad \bar{\beta}^\phi = \text{const}, \tag{5.54}
\end{equation}
and hence $\bar{\beta}^\phi = \tilde{\beta}^\phi$ plays the role of the central charge of the Virasoro algebra (see, e.g., Refs. 62, 27 and 63). In this case $W$ in (5.6) is simply given by [38]
\begin{equation}
    W = \frac{1}{16\pi} \bar{\beta}^\phi \int R^{(2)} \Delta^{-1} R^{(2)} + \text{const}. \tag{5.55}
\end{equation}
In general (for arbitrary $\varphi^i$) $\langle T_a^a \rangle$ is nonlocal and contains an infinite number of structures [34] (see also below).

Let us now add some general remarks about the above relations (5.43), (5.48) and (5.49). First, we see that there is the difference between the \textit{scale anomaly} coefficients $\beta^i$ in (5.46) (the ordinary RG $\beta$ functions) and the \textit{Weyl anomaly} coefficients $\bar{\beta}^i$ in (5.44). The ``extra'' $M_\mu$ and $L_\mu$ terms in (5.44) correspond to a diffeomorphism transformation of $G, B, \phi$ and a gauge transformation of $B$. Their role is quite important since they compensate for the corresponding ambiguities in the $\beta$ functions (the expressions for the $\beta$ functions are ambiguous since they are determined from counterterms which are given by the integrals over 2-space). 

While the $\beta$ functions change under a reparametrization of the $\sigma$ model field $x^\mu$ [58], the Weyl anomaly coefficients must be invariant (the scale anomaly relation (5.46) remains invariant since its integrand changes by a total derivative plus an equation of motion term). The energy-momentum tensor and $W$, being the ``physical'' quantities, cannot depend on a parametrization of the $\sigma$ model field space (though their form may, of course, depend on a parametrization of the coupling constant space $\{\varphi^i\}$, i.e. on a 2d renormalization scheme). While formally there exists a parametrization in which $M_\mu = L_\mu = 0$, these functions may be nonzero in the standard parametrization used for computation of the $\beta$ functions (see also Refs. 64 and 65).

Next, we note that the Weyl invariance $\bar{\beta}^i = 0$ implies the perturbative finiteness of the $\sigma$ model $\beta^i \approx 0$ (modulo diffeomorphism and gauge transformation terms), while the converse is not true ($M_\mu$ and $L_\mu$ in (2.45) have a fixed form if computed in a fixed parametrization). The condition that a $\sigma$ model is to satisfy in order to describe a consistent string vacuum is Weyl invariance (decoupling of $\rho$) rather than finiteness (2d infinities are harmless in string theory since they can be renormalized away by a re-definition of the $\sigma$ model couplings). Finiteness is equivalent to scale invariance. The condition of Weyl invariance is in general stronger than the condition of scale invariance (see also Ref. 66). 

This conclusion seems to be in contradiction with some claims in the literature that conformal invariance (on a flat world sheet) always follows from scale invariance and 2d locality (existence of an energy-momentum tensor). It appears that this common wisdom is correct only if one assumes the dilaton field to be constant and fixes the $\sigma$ model parametrization in such a way that $W_\mu = L_\mu = 0$. Note that $\phi$, which is a ``hidden'' coupling constant on a flat world sheet, is certainly important for renormalizability on a curved world sheet ($\beta^\phi \neq \text{const}$ in general). At the same time, it seems that $W_\mu$ and $L_\mu$ in (5.45) may have some particular form when computed using a standard parametrization. Perturbation theory results suggest that in the covariant parametrization corresponding to the use of the normal coordinate expansion (see Sec. 7)
\begin{equation}
    W_\mu = \partial_\mu \chi, \qquad L_\mu = 0, \qquad \chi = \chi(G, B). \tag{5.56}
\end{equation}
If this is the case $T_a^a$ can be represented as
\begin{equation}
    T_a^a = [V_i] \cdot \beta^i + \nabla^2 \eta, \qquad \eta = \eta(G, B), \tag{5.57}
\end{equation}
where the first term is the integrand of the scale anomaly and $\nabla^2$ is the 2d Laplacian (cf. in this connection the recent discussion [67] of a relation between the scale and Weyl anomalies).


\section{Sigma model Weyl invariance conditions from \\ stationarity of ``central charge'' action}
\renewcommand{\theequation}{6.\arabic{equation}}
\setcounter{equation}{0}

\def \Z {{\cal Z}}

Another miraculous property of the 2d $\sigma$ model (3.1) is that its Weyl invariance conditions $\bar{\beta}^i = 0$ are equivalent to the equations of motion for the action
\begin{equation}
    S = \int d^D y\  \sqrt{G} \ e^{-2\phi}\  \tilde{\beta}^\phi, \tag{6.1}
\end{equation}
where $\tilde{\beta}^\phi$ is the ``central charge'' coefficient (see (5.51)--(5.53)). Below we shall discuss a proof of this statement [45] trying to generalize the ``c-theorem'' of Ref. 68 to the case of the $\sigma$ model with dilaton coupling (see also Ref. 46). The observation that the equations $\bar{\beta}^i = 0$ do follow (to the leading order in $\alpha'$) from an action was first made in Ref. 27. That this action should, in general, have the form (6.1) was conjectured, e.g., in Refs. 34 and 49.

\subsection{}

 Consider the partition function of the $\sigma$ model defined on a compact 2-manifold $M^2$ with the topology of a sphere and the metric $g_{ab}$
\begin{equation}
    Z = \int [dx]\ e^{-I'}, \qquad I' = \Lambda^2 \int d^2\sigma\ \sqrt{g}\ \psi_0 + I \tag{6.2}
\end{equation}
where $I$ is the bare $\sigma$ model action (5.30). $\Lambda$ is an UV cutoff and $\psi_0$ is the bare tachyon coupling (we assume that the renormalized value of $\psi$ is set equal to zero). Since $M^2$ is compact it is natural to split $\int [dx]$ into the integrals over the constant and non-constant parts of $x^\mu$, $x^\mu = y^\mu + \xi^\mu(\sigma)$ (see below)
\begin{equation}
    Z = \int d^D y \sqrt{G_0(y)} e^{-2\phi_0(y)} \Z(y), \qquad \Z= e^{-W_0} = \int [d\xi] e^{-\tilde{I}}, \tag{6.3}
\end{equation}
\begin{equation}
    \tilde{I} = \frac{1}{2\lambda} \int d^2\sigma \sqrt{g} \Big[\partial^a \xi^m \partial_a \xi^n G_{0mn}(y+\xi) + \alpha' R^{(2)} \partial_m \phi_0(y) \xi^m + \dots\Big]. \tag{6.4}
\end{equation}
In what follows we absorb the overall $\alpha'^{-D/2}$ factor in $Z$ into $\sqrt{G}$. $[d\xi]$ contains the ``gauge condition'' $\delta^{(D)}(\int d^2\sigma \sqrt{g}\ \xi)$ [58] or $\delta^{(D)}(\xi(\sigma_0))$ [69] ($\sigma_0$ is an arbitrary point of $M^2$). We have used that for the spherical topology $\int d^2\sigma \sqrt{g} R^{(2)} = 8\pi$. $W_0 = \sum_n p_n(\varphi_0, \Lambda) \kappa_n[g]$ where $p_n$ are covariant functions of the UV cutoff $\Lambda$, bare fields $\varphi_0$ and their derivatives at the point $y$. $\kappa_n$ are dimensionless functionals of the 2-metric.

To define $Z$ to be covariant under the $D$-dimensional coordinate transformations we are to ensure that not only the action $I$, but also the measure $[dx]$ and the renormalization procedure in (6.2) are covariant. Choosing $[dx] = \prod_\sigma dx(\sigma) \sqrt{G_0(x(\sigma))}$ and assuming that the regularization is such that only the $\sqrt{G(y)}$ factor survives after splitting $x$ into $y$ and $\xi$, we indeed get the covariant expression (6.3). 

From a more general point of view, a choice of a local measure can be included into the choice of a renormalization prescription. Namely, we may start with the simplest measure $[dx] = \prod_\sigma dx(\sigma)$ but prepare the bare couplings so that the renormalized $Z$ is covariant. If $[dx]$ is trivial, the dependence of $Z$ on $G_{\mu\nu} = \text{const}$ is given by
\begin{align}
   & \int d^D y \int [d\xi] \exp\Big( -\frac{1}{2\lambda} \int d^2\sigma \xi^\mu \Delta \xi^\nu G_{0\mu\nu}(y)\Big) \sim \int d^D y\  [\det {}' G_{0\mu\nu} \Delta]^{-1/2} \nonumber \\
    &\sim \int d^D y (\det G_0)^{-{1\over 2} N_\Delta} = \int d^D y \sqrt{G_0} \exp\Big( -\frac{1}{2} N_\Delta \log G_0\Big) \tag{6.5}
\end{align}
where $N_\Delta (N_\Delta')$ is a regularized number of all (nonzero) eigenmodes of the 2d Laplacian $\Delta$, \ $N_\Delta = N_\Delta' + 1$. The ``extra'' noncovariant $N_\Delta \log G_0$ term in (6.5) must be renormalized away. Using, e.g., the heat kernel regularization for $N_\Delta$ we get
\begin{equation}
    N_\Delta = \sum_n e^{-\epsilon\lambda_n} = \text{tr } e^{-\epsilon\Delta} = \frac{B_0}{\epsilon} + B_2 + O(\epsilon), \tag{6.6}
\end{equation}
\begin{equation}
    B_0 = \frac{1}{4\pi} \int d^2\sigma \sqrt{g}, \qquad B_2 = \frac{1}{24\pi} \int d^2\sigma \sqrt{g} R^{(2)}, \qquad \epsilon = \Lambda^{-2} \to 0.\no
\end{equation}
Hence the $N_\Delta \log G_0$ term in (6.5) can be cancelled by the following (noncovariant) renormalizations of the tachyon and dilaton couplings in $I'$: $\delta\psi = a_1 \log G_0$, $\delta\phi = a_2 \log G_0, a_1 = \frac{1}{8\pi}, a_2 = \frac{1}{12}$. The values of $a_i$ depend on a regularization used for $N_\Delta$ (e.g., in dimensional regularization: $a_1 = 0, a_2 = -\frac{1}{4}$). Thus finally we get (6.3) and (6.4) where $\phi_0$ is a bare dilaton shifted by the $G_0$ term. In Eq. (6.4) $m, n$ are the vielbein indices: $\xi^m = e_\mu^m(y) \xi^\mu$, $e_\mu^m e_\nu^n \delta_{mn} = G_{0\mu\nu}$, i.e. $G_{0mn}(y+\xi) = \delta_{mn} + O(\xi)$. It is possible to make a local re-definition of $\xi^m$, re-writing (6.4) in terms of the normal coordinates $\eta^m = \eta^m(\xi, y)$.

More general treatment of the contribution of the measure to the $\sigma$ model path integral should proceed along the following lines (see also Refs. 58 and 70). Let us define the metric in the tangent space $T_x \mathcal{M}$ to the space $\mathcal{M}$ of the maps $x^\mu(\sigma): M^2 \to M^D$ by
\begin{equation}
    \gamma(\xi_1, \xi_2)(x) = \int d^2\sigma \sqrt{g} G_{\mu\nu}(x) \xi_1^\mu \xi_2^\nu.\no
\end{equation}
If $\{e_A\}$ is a basis in $T_x \mathcal{M}, \ \ \xi = \sum_{A=1}^\infty \xi^A e_A$, the invariant measure is
\begin{equation}
    [dx] \sqrt{\det \gamma_{AB}(x)}, \qquad \gamma_{AB} = \gamma(e_A, e_B)(x) \ , \no 
\end{equation}
where $[dx]$ is a naive measure. Introducing two real ``ghost'' fields $c^\mu(\sigma), \tilde{c}^\mu(\sigma)$ we can employ the following representation
\begin{equation}
    \det \gamma_{AB} = \int [d\tilde{c}\  dc] \exp\Big[ \int d^2\sigma \sqrt{g} G_{\mu\nu}(x(\sigma)) \tilde{c}^\mu(\sigma) c^\nu(\sigma)\Big].\no 
\end{equation}
Assuming that $M^2$ is compact let us separate a constant part of $x^\mu$, $x^\mu = y^\mu + \xi^\mu(\sigma)$ and introduce the normal coordinates $\{\eta^\mu\}, \xi = \xi(\eta, y)$, so that $G_{\mu\nu}(x) = G_{\mu\nu}(y) - \frac{1}{3} R_{\mu\alpha\nu\beta}(y) \eta^\alpha \eta^\beta + O(\eta^3)$. Then the $\sigma$ model partition function takes the form
\begin{equation}\no 
    Z = \int d^D y \int [d\eta] \exp(-I[\eta, y, g] - I_\gamma[\eta, y, g]),
\end{equation}
where $I = \frac{1}{4\pi\alpha'} \int d^2\sigma \sqrt{g} (\partial^a \eta^\mu \partial_a \eta^\nu G_{\mu\nu}(y) - \frac{1}{3} R_{\mu\alpha\nu\beta}(y) \eta^\alpha \eta^\beta \partial^a \eta^\mu \partial_a \eta^\nu + \dots)$ and $I_\gamma$ is the contribution of the measure
\begin{align}
    I_\gamma &= -\frac{1}{2} \log \det \gamma_{AB} = -\frac{1}{2} \log \int [d\tilde{c} \ dc] \nonumber \\
    &\qquad \times \exp\Big[ \int d^2\sigma \sqrt{g} (G_{\mu\nu}(y) \tilde{c}^\mu c^\nu - \frac{1}{3} R_{\mu\alpha\nu\beta}(y) \eta^\alpha \eta^\beta \tilde{c}^\mu c^\nu + \dots)\Big].\no 
\end{align}
The propagator of ``ghosts'' is simply $G^{\mu\nu}(y)$ times the covariant $\delta$-function $\delta^{(2)}(\sigma, \sigma')$. Assuming that the $\delta$ function is regularized using the scalar Laplacian, i.e. (cf. (6.6))
\begin{equation}\no 
    \delta_\epsilon^{(2)}(\sigma, \sigma') = \langle \sigma | e^{-\epsilon\Delta} | \sigma' \rangle, \qquad \delta_\epsilon^{(2)}(\sigma, \sigma) = \frac{1}{4\pi\epsilon} + \frac{1}{24\pi} R^{(2)}
\end{equation}
we conclude that the one-loop contribution to $Z$ is simply $\int d^D y \sqrt{G} \times (\det' \Delta)^{-D/2}$ ($\sqrt{G}$ factor comes from the integral over the constant modes of the ``ghosts'' while the $\sqrt{G}$ contributions at the integrals over nonconstant modes of $\eta$ and $\tilde{c}, c$ mutually cancel).

Since in the heat kernel regularization $(\epsilon = \Lambda^{-2})$ $\langle \eta^\alpha(\sigma) \eta^\beta(\sigma) \rangle = G^{\alpha\beta} (\frac{1}{4\pi} \log \Lambda^2 + \frac{1}{4\pi} \Delta^{-1} R^{(2)} \sqrt{g})$, $\langle c^\mu(\sigma) \tilde{c}^\nu(\sigma) \rangle = G^{\mu\nu} \times (\frac{1}{4\pi} \Lambda^2 + \frac{1}{24\pi} R^{(2)})$ we conclude that starting with the two-loop approximation the measure contributes to the logarithmically divergent ($\log \Lambda^2 \int R^{(2)} + \dots$) and Weyl anomalous ($\int R^{(2)} \Delta^{-1} R^{(2)} + \dots$) parts of the 2d effective action $W$ in (6.3) (the contribution of the measure also cancels $\Lambda^{2n} \log^m \Lambda, n=0, 1, \dots$ infinities in $W$). 

In particular, we get the additional (as compared to the result in dimensional regularization in the noncompact case in which the contribution of the measure can be ignored) $\log \Lambda^2\ R$ contribution which is necessary for the renormalization of the $\sqrt{G}$ factor in (6.3).
The regularized contribution of the measure can be rewritten as
\begin{equation}\no 
    I_\gamma = \frac{1}{2} \text{tr}(\log G e^{-\epsilon\Delta}) = \frac{1}{2} \int d^2\sigma \sqrt{g} \log G(x(\sigma)) \delta_\epsilon^{(2)}(\sigma, \sigma)
\end{equation}
or, in normal coordinates,
\begin{equation}\no 
    I_\gamma = \frac{1}{2} \int d^2\sigma \sqrt{g} \delta_\epsilon^{(2)}(\sigma, \sigma)\Big[ \log G(y) - \frac{1}{3} R_{\mu\nu}(y) \eta^\mu \eta^\nu + O(\eta^3)\Big].
\end{equation}
In splitting the path integral over $x^\mu$ into the integral over $y^\mu$ and $\xi^\mu$, it is necessary to impose a gauge condition on $\xi^\mu$ to avoid overcounting. This can be done by inserting the following ``unit'' into (6.2) (see Refs. 58, 20 and 70c)
\begin{equation}
    1 = \int d^D y \int \prod_\sigma d\xi(\sigma) \delta^{(D)}(x^\mu(\sigma) - y^\mu - \xi^\mu(\sigma)) \ \delta^{(D)}(P^\mu[y, \xi])\  Q[y, \xi]\no 
\end{equation}
\begin{equation}
    Q = \det\Big( \frac{\partial P^\mu[y-a, \xi+a]}{\partial a^\nu}\Big)_{a=0},\no 
\end{equation}
where $P^\mu=0$ is a ``gauge condition'' which fixes the symmetry $y \to y-a, \xi \to \xi+a$, $a=\text{const}$, of the 
$\delta$-function and $Q$ is the corresponding ``ghost determinant''. To make the perturbation theory manifestly covariant, we exchange $\xi^\mu$ for the normal coordinate variable $\eta^\mu(y, \xi) = \xi^\mu + \frac{1}{2} \Gamma_{\alpha\beta}^\mu(y) \xi^\alpha \xi^\beta + \dots$ and choose
\begin{equation}
    P^\mu = \int d^2\sigma \sqrt{g} \eta^\mu(\sigma), \qquad Q = \det\Big( \int d^2\sigma \sqrt{g} \lambda^\mu{}_\nu\Big),
\qquad \no
    \lambda^\mu{}_\nu = \frac{\partial \eta^\mu(y, \xi)}{\partial \xi^\nu} - \frac{\partial \eta^\mu(y, \xi)}{\partial y^\nu}.
\end{equation}
Note that $Q$ is nontrivial. $\lambda^\mu{}_\nu$ can be represented [58] as a covariant derivative $-\mathcal{D}_\nu \eta^\mu$. Computing the normal coordinate expansion of $\lambda^\mu{}_\nu$ (see Refs. 58 and 70c) we get
\begin{equation}\no 
    \lambda^\mu{}_\nu = \delta^\mu_\nu - \frac{1}{3} R^\mu{}_{\alpha\nu\beta} \eta^\alpha \eta^\beta + O(\eta^3),
\end{equation}
\begin{equation}\no
    Q = A^D \exp\Big[ -\frac{1}{3A} \int d^2\sigma \sqrt{g} R_{\mu\nu}(y) \eta^\mu \eta^\nu + \dots\Big], \qquad A = \int d^2\sigma \sqrt{g}.
\end{equation}
As discussed in Ref. 70c, the account of the contribution of $Q$ is crucial for obtaining the correct result for $Z$. For example, in the two-loop approximation one finds
\begin{equation}
    Z = a_0 \int d^D y \sqrt{G_0} e^{-2\phi_0} \Big[ 1 + \alpha'(b_1 R_0 + b_2 D^2 \phi_0) + O(\alpha'^2) \Big],\no 
\end{equation}
\begin{equation}\no 
    a_0 = (2\pi\alpha')^{-D/2} \exp\Big[ -\frac{1}{2} D\Big( \frac{1}{4\pi\epsilon} A + \frac{1}{6} \chi \log \epsilon + \text{const}\Big)\Big], \qquad \chi=2,
\end{equation}
where $b_1 = a_1 + a_2 + a_3$ is the sum of the contributions of the action, of the measure and of the determinant $Q$ (cf. Eq. (6.6))
\begin{align}
    a_1 &= \frac{\pi}{3} \int d^2\sigma \sqrt{g} \Big[(\nabla_a \nabla'^a \Delta^{-1}(\sigma, \sigma'))_{\sigma=\sigma'} \Delta^{-1}(\sigma, \sigma) - (\nabla_a \Delta^{-1}(\sigma, \sigma') \nabla^a \Delta^{-1}(\sigma, \sigma'))_{\sigma=\sigma'}\Big] \nonumber \\
    &= \frac{\pi}{3} N_\Delta' d + \bar{a}_1,\no 
\\
    a_2 &= -\frac{\pi}{3} \int d^2\sigma \sqrt{g} \delta_\epsilon^{(2)}(\sigma, \sigma) \Delta^{-1}(\sigma, \sigma) = -\frac{\pi}{3} N_\Delta d + \bar{a}_2,\no
\\
    a_3 &= -\frac{2\pi}{3A} \int d^2\sigma \sqrt{g} \Delta^{-1}(\sigma, \sigma) = -\frac{2\pi}{3} d + \bar{a}_3,\no 
\\ 
    b_2 &= -\frac{1}{4} \int d^2\sigma \sqrt{g} R^{(2)} \Delta^{-1}(\sigma, \sigma) = -\pi \chi d + \bar{b}_2, \qquad \chi=2,\no 
\\
    d &= -\frac{1}{4\pi} \log \epsilon^2, \qquad \epsilon^2 = \frac{\epsilon}{r}, \qquad r = A^{1/2},\no 
\end{align}
where $\bar{a}_k$ and $\bar{b}_k$ are the finite functionals of $g_{ab}$. Hence
\begin{equation}
    Z = a_0 \int d^D y \sqrt{G_0} e^{-2\phi_0} \Big[1 + \frac{1}{2} \alpha' \log \epsilon (R_0 + 2 D^2 \phi_0) + \alpha'(\bar{b}_1 R_0 + \bar{b}_2 D^2 \phi_0) + O(\alpha'^2)\Big].\no 
\end{equation}
Recalling the standard expressions for the one-loop $\beta$ functions (see (7.3) and (7.10)) we conclude that $Z$ is renormalizable, i.e. can be rewritten as a finite functional of the renormalized couplings $(G, \phi)$,
\begin{equation}
    G_{0\mu\nu} = G_{\mu\nu} - \alpha' R_{\mu\nu} \log \epsilon + \dots
\ , \qquad \no 
    \phi_0 = \phi - \frac{1}{6} D \log \epsilon + \frac{1}{2} \alpha' D^2 \phi \log \epsilon + \dots.
\end{equation}
Note that taking $\partial / \partial \log \epsilon$ of the above result for $Z$ we indeed get (see Eq. (4.12)) the well-known expression for the effective action
\begin{equation}\no 
    S = a \int d^D y \sqrt{G} e^{-2\phi} \Big[D - 26 - \frac{3}{2} \alpha' (R + 2 D^2 \phi) + O(\alpha'^2)\Big]
\end{equation}
(we have accounted for the $\log \epsilon$ dependence in $a_0$ and replaced $D$ by $D-26$ assuming that the contribution of the reparametrization ghosts is included).

\subsection{ }

 The renormalizability of the theory implies that $Z$ (6.3) can be re-written as
\begin{equation}
    Z = Z_R(\varphi, \mu) = \int d^D y \sqrt{G} e^{-2\phi} e^{-W}, \qquad W = W(\varphi, \mu), \tag{6.7}
\end{equation}
where $\varphi^i$ are the renormalized couplings. We shall use the following averages
\begin{equation}
     \langle\!\langle F  \rangle\!\rangle = \int [dx] e^{-I} F(x) = \int d^D y \sqrt{G} e^{-2\phi - W} \langle F \rangle, \qquad \langle F \rangle \equiv e^W \int [d\xi] e^{-\tilde{I}} F(y+\xi). \tag{6.8}
\end{equation}
The dependence of $W$ on the conformal factor of $g_{ab}$ is found by integrating the averaged Weyl anomaly relation (5.51):\footnote{Here we temporarily ``forget'' about the ``measure factor'' $\sqrt{G} e^{-2\phi}$ and assume that $W$ is equal to $-\log Z$ where $Z$ is the partition function of the $\sigma$ model defined in curved 2-space with the topology of a plane (see also the next footnote).}
\begin{equation}
    W = \int d^2\sigma d^2\sigma' (\sqrt{g} R^{(2)})_\sigma Q(\sigma, \sigma') (\sqrt{g} R^{(2)})_{\sigma'} + \text{other structures}, \tag{6.9}
\end{equation}
\begin{equation}
    Q = \Delta^{-1} \nu (\Delta), \qquad \nu  = \gamma_1 + \gamma_2 \log \frac{\Delta}{\mu^2} + \gamma_3\Big( \log \frac{\Delta}{\mu^2}\Big)^2 + \dots, \tag{6.10}
\end{equation}
\begin{equation}
    \gamma_1 = -\frac{1}{16\pi} \tilde{\beta}^\phi, \qquad \Delta = -\nabla^2 = -\frac{1}{\sqrt{g}} \Box, \qquad \Box = \partial_a (\sqrt{g} g^{ab} \partial_b) \tag{6.11}
\end{equation}
where $\gamma_n$ are covariant functions of renormalized couplings and their derivatives. If $d\nu /dt = 0 (t = \log \mu)$ we get $\nu  = \nu (\partial) = \gamma_1(\varphi(\partial)), \partial \equiv \sqrt{\Delta}$, where we have introduced the running coupling constant: $d\varphi^i/dt = \beta^i(\varphi)$. One way to justify (6.9) is to consider the expansion near a flat space
\begin{equation}
    W = \int (dk) \tilde{h}_{ab}(k) P_{abcd}(k) \tilde{h}_{cd}(-k) + O(\tilde{h}^3), \tag{6.12}
\end{equation}
$$(dk) = \frac{d^2k}{(2\pi)^2}, \qquad \qquad \tilde{h}_{ab}(k) = \int d^2\sigma e^{-ik\sigma} h_{ab}(\sigma) . $$
The tensor $P$ must be a symmetric and transverse ($k_a P_{abcd} = 0$ is necessary for invariance under the linearized coordinate transformations). It is not difficult to check that these conditions together with Lorentz invariance and dimensional considerations uniquely fix $P$ to be
\begin{equation}
    P_{abcd} = \frac{\lambda(k)}{k^2} (k_a k_b - \delta_{ab} k^2) (k_c k_d - \delta_{cd} k^2). \tag{6.13}
\end{equation}
Hence (6.12) is in agreement with (6.9) if $\lambda(k) = \nu(k), \ \nu(k) = -\frac{1}{16\pi} \tilde{\beta}^\phi(\varphi(k))$.\footnote{We assumed that the effective action of the $\sigma$ model $W$ expanded near the flat space is free from infrared infinities so that $\frac{d\lambda}{dt} = 0$ and $\lambda = \lambda(k, \mu, \varphi(\mu))$ imply $\lambda = \lambda(\varphi(k))$. It was recently claimed in Ref. 65c that IR infinities are present in $W$ thus invalidating the arguments used in Refs. 68 and 45. Infrared infinities are present in $Q$ and $\lambda$ (computed using the expansion near a flat space) which thus have more complicated structure than it is assumed above. Let us stress that to give a consistent analysis of the ``c-theorem'', it is necessary to use the expansion near a compact 2-space (e.g. $S^2$). Let us ignore these complications for a moment and return to the issue of the IR infinities at the end of this section.} 

It is interesting to note that in $d=2, \int \log \Delta \ R^{(2)} = \int R^{(2)} \Delta^{-1} R^{(2)}$ (this can be checked by using the conformal gauge or by expanding near a flat space). Hence the terms like $\int (\log \Delta)^n R^{(2)}$ can be re-written as $\int R^{(2)} (\log \Delta)^k \dots \Delta^{-1} R^{(2)}$. If not for the presence of the IR infinities, the following conjecture could be true: there are no ``other structures'' in Eq. (6.9), i.e. $W$ is exactly given by the $\int R^{(2)} Q R^{(2)}$ term (note that in the conformal gauge $R^{(2)} Q R^{(2)}$ contains terms of all powers in the conformal factor).

Now let us consider the correlators of the energy-momentum tensor $T_{ab} = \frac{2}{\sqrt{g}} \frac{\delta I}{\delta g^{ab}}$. $ \langle\!\langle T_{ab}  \rangle\!\rangle$ is finite if we account for the renormalization of the dilaton coupling. However, additional infinities may appear in
\begin{equation}
     \langle\!\langle T_{ab}(\sigma) T_{cd}(\sigma')  \rangle\!\rangle = \langle\!\langle\Big( \frac{2}{\sqrt{g}} \frac{\delta}{\delta g^{ab}}\Big)_\sigma T_{cd}(\sigma') \rangle\!\rangle -\Big( \frac{2}{\sqrt{g}} \frac{\delta}{\delta g^{ab}}\Big)_\sigma  \langle\!\langle T_{cd}(\sigma')  \rangle\!\rangle \tag{6.14}
\end{equation}
(we consider only the connected part of the correlator). The second term in (6.14) is a derivative of a finite quantity with respect to a finite quantity and hence is finite. The infinities which may appear from the first term in (6.14) are necessarily local, i.e. are proportional to $\delta^{(2)}(\sigma - \sigma')$ or $\partial_a \delta^{(2)}(\sigma - \sigma')$. Let $\langle \dots \rangle_{reg}$ denote the expectation value $\langle \dots \rangle$ in which such local terms are dropped. Then
\begin{align}
    \langle T_{ab}(\sigma) T_{cd}(\sigma') \rangle_{reg} &= -\Big( \frac{2}{\sqrt{g}} \frac{\delta}{\delta g^{ab}}\Big)_\sigma \langle T_{cd}(\sigma') \rangle = -\Big( \frac{2}{\sqrt{g}} \frac{\delta}{\delta g^{ab}}\Big)_\sigma\Big( \frac{2}{\sqrt{g}} \frac{\delta}{\delta g^{cd}}\Big)_{\sigma'} W. \tag{6.15}
\end{align}
Substituting (6.9) into (6.15) and taking the flat space limit we get
\begin{equation}
    \langle T_{ab}(\sigma) T_{cd}(\sigma') \rangle_{reg \text{ fl}} = - \bar{\nu }(\partial) \Box^{-1} (\partial_a \partial_b - \delta_{ab} \Box) (\partial_c \partial_d - \delta_{cd} \Box) \delta^{(2)}(\sigma - \sigma'), \tag{6.16}
\end{equation}
\begin{equation}
    \bar{\nu}(\partial) = -2\nu(\partial) = \frac{1}{8\pi} \tilde{\beta}^\phi(\varphi(y, \partial)).\no 
\end{equation}
In the momentum representation $(\tilde{T}(k) = \int d^2\sigma e^{-ik\sigma} T(\sigma))$
\begin{equation}
    \langle \tilde{T}_{ab}(k) \tilde{T}_{cd}(q) \rangle_{reg \text{ fl}} = (2\pi)^2 \delta^{(2)}(k+q) \tilde{P}_{abcd}(k), \tag{6.17}
\end{equation}
\begin{equation}
    \tilde{P}_{abcd} = \bar{\nu}(k) k^{-2} (k_a k_b - \delta_{ab} k^2) (k_c k_d - \delta_{cd} k^2). \tag{6.18}
\end{equation}
An analogous relation is true for $ \langle\!\langle \tilde{T}_{ab}(k) \tilde{T}_{cd}(q)  \rangle\!\rangle_{reg \text{ fl}}$ with $\tilde{\nu}$ replaced by $\bar{\nu}$,
\begin{equation}
    \bar{\nu} = \int d^D y\  \sqrt{G} \ e^{-2\phi} \ \tilde{\nu} = \frac{1}{8\pi} \int d^D y\  \sqrt{G} \ e^{-2\phi} \ \tilde{\beta}^\phi. \tag{6.19}
\end{equation}
While we have taken the flat-space limit in the correlator of energy-momentum tensors it is necessary to include the $e^{-2\phi}$ factor to account for the spherical topology of the 2-space. Introducing the complex coordinates ($z, \bar{z} = \sigma^1 \pm i\sigma^2, k_a \sigma^a = k_z z + k_{\bar{z}} \bar{z}, k^2 = 4k_z k_{\bar{z}}$) we define the objects which appear in the proof of the ``c-theorem'' [68] ($\sigma' = 0$)
\begin{equation}
    \theta = T_a^a = 4T_{z\bar{z}}, \qquad T = 4T_{zz} = T_{11} - T_{22} - 2iT_{12},\no 
\end{equation}
\begin{equation}
    U = z^4  \langle\!\langle T(\sigma) T(0)  \rangle\!\rangle_{\text{fl}}, \qquad X = z^3 \bar{z}  \langle\!\langle T(\sigma) \theta(0)  \rangle\!\rangle_{\text{fl}}, \tag{6.20}
\end{equation}
\begin{equation}
    Y = z^2 \bar{z}^2  \langle\!\langle \theta(\sigma) \theta(0)  \rangle\!\rangle_{\text{fl}}. \nonumber
\end{equation}
These correlators are UV finite because of explicit factors of $z$ and $\bar{z}$ (hence $ \langle\!\langle \dots  \rangle\!\rangle$ in (6.20) can be replaced by $ \langle\!\langle \dots  \rangle\!\rangle_{reg}$). It is easy to prove that $U, X, Y$ are real functions of $|z|$ only (note that $U$ and $X$ are invariant under $z \to e^{i\alpha} z$). This implies that they are functions of the ``running'' coupling $\varphi(1/|z|)$ (additional dependence on $|z|$ is ruled out because $U, X, Y$ are dimensionless). As was noted above, we, for a moment, ignore the presence of the IR infinities (see the end of the section). Using (6.17)--(6.19) one finds [45]
\begin{equation}
    U = 16z^4 \int (dk) e^{ik\sigma} \frac{k_z^4}{k^2} \bar{\nu}(k) =\Big( \frac{d}{d\eta} + 6\Big)\Big( \frac{d}{d\eta} + 4\Big)\Big( \frac{d}{d\eta} + 2\Big) \frac{d}{d\eta} \mathcal{O},\no 
\end{equation}
\begin{equation}
    X = -16z^3 \bar{z} \int (dk) e^{ik\sigma} \frac{k_z^3 k_{\bar{z}}}{k^2} \bar{\nu}(k) = -\Big( \frac{d}{d\eta} + 4\Big)\Big( \frac{d}{d\eta} + 2\Big) \frac{d^2}{d\eta^2} \mathcal{O}, \tag{6.21}
\end{equation}
\begin{equation}
    Y = 16z^2 \bar{z}^2 \int (dk) e^{ik\sigma} \frac{k_z^2 k_{\bar{z}}^2}{k^2} \bar{\nu}(k) =\Big( \frac{d}{d\eta} + 2\Big)^2 \frac{d^2}{d\eta^2} \mathcal{O},\no 
\end{equation}
\begin{equation}
    \mathcal{O}(|z|) = \int (dk) e^{ik\sigma} \frac{\bar{\nu}(k)}{k^2}, \qquad \eta = -\log|z|. \tag{6.22}
\end{equation}
\def \be {\begin{equation}}
 \def \ee {\end{equation}}
Introducing their combination [{68}]
\be 
C = \frac{1}{12} (U - 2X - 3Y) =\Big(\frac{d}{dt} +2\Big)^2\frac{d}{dt}\mathcal{O} \qquad \ \  \tag{6.23}
\ee
we find from (6.21)
\begin{equation}     {d C\over dt}= Y, \ \ \ \ \   {\rm i.e.} \ \ \ \ 
\beta^{i} \frac{\partial C}{\partial \varphi^{i}} = Y. \tag{6.24}
\end{equation}
The combination $C$ (6.23) was chosen exactly so as to have the property that its derivative is equal to $Y$. Since $\theta = T_{a}^{a} = [V_{i}] \cdot \bar{\beta}^{i} + E$ (where $E = E^{\mu}(x) \frac{\delta I}{\delta x^{\mu}}$ was omitted in (5.43) because $\langle E \rangle = 0$), we have
\begin{equation}
  \langle\!\langle \theta(\sigma) \theta(\sigma^{\prime})   \rangle\!\rangle_{reg} = \bar{\beta}^{i} \bar{\beta}^{j} \kappa_{ij}(\sigma, \sigma^{\prime}), \qquad \ \  \kappa_{ij} =   \langle\!\langle [V_{i}(\sigma)] [V_{j}(\sigma^{\prime})]   \rangle\!\rangle_{reg} \tag{6.25}
\end{equation}
(we have used that $  \langle\!\langle (E^{\mu}(x) \frac{\delta I}{\delta x^{\mu}})(\sigma) \theta(\sigma^{\prime})   \rangle\!\rangle_{reg} = -  \langle\!\langle \frac{\delta}{\delta x^{\mu}(\sigma)} \theta(\sigma^{\prime}) E^{\mu}(x(\sigma))   \rangle\!\rangle_{reg} = 0$). In the momentum representation (see (6.17)):
\begin{equation}
  \langle\!\langle \bar{\theta}(k) \bar{\theta}(q)   \rangle\!\rangle_{reg, fl} = (2\pi)^{2} \delta^{(2)}(k+q) k^{2} \bar{\nu}(k). \tag{6.26}
\end{equation}
According to the standard positivity argument:
\begin{equation}
  \langle\!\langle \tilde{\theta}(k) \tilde{\theta}(q)   \rangle\!\rangle_{fl} = (2\pi)^{4} \delta^{(2)}(k+q) \sum_{n} \delta^{(2)}(k - k_{n}) \Big|  \langle\!\langle 0 | \tilde{\theta}(0) | n   \rangle\!\rangle_{fl}\Big|^2 \ge 0. \tag{6.27}
\end{equation}
Because of the translational invariance in the flat space limit, $\kappa_{ij}$ in (6.25) depends on $\sigma - \sigma^{\prime}$ and hence the positivity of $\bar{\nu}$, which follows from (6.27), implies the positivity of the matrix
\begin{equation}
\tilde{\kappa}_{\alpha\beta}(k) = \int d^{2}\sigma \, e^{-ik\sigma} (\kappa_{\alpha\beta}(\sigma, \sigma^{\prime}))_{fl}, \tag{6.28}
\end{equation}
\begin{equation}
  \langle\!\langle \tilde{\theta}(k) \bar{\theta}(q)   \rangle\!\rangle_{reg, fl} = (2\pi)^{2} \delta^{(2)}(k+q) \bar{\beta}^{\alpha} \bar{\beta}^{\beta} \tilde{\kappa}_{\alpha\beta}(k), \tag{6.29}
\end{equation}
where $\alpha, \beta$ numerate only the flat space couplings $G_{\mu\nu}$ and $B_{\mu\nu}$ (the dilaton term in $I$ is proportional to $R^{(2)}$ and thus drops out in the flat space limit). Hence the vanishing of $Y$ in (6.20) implies $\bar{\beta}^{\alpha} = 0$ (i.e., the vanishing of $\theta_{fl}$). 

We conclude from (6.24) that if $\partial C / \partial \varphi^{i} = 0$, then $Y = 0$, i.e., $\bar{\beta}^{\alpha} = 0$. As discussed in Sec. 5, one consequence of the nonrenormalizability of $\theta$ is that $\bar{\beta}^{\alpha} = 0$ implies $\bar{\beta}^{\phi} = \tilde{\beta}^{\phi} = \text{const}$ (see (5.48)).

In view of (6.19), $C$ can be represented as
\begin{equation}
C = \int d^{D}y \sqrt{G} e^{-2\phi} \hat{c}(\varphi(y, \mu)), \tag{6.30}
\end{equation}
where $\hat{c}$ depends on $\varphi$ only through the derivatives with respect to $y$. Therefore, $\partial C/\partial\phi=0$ implies $\int d^{D}y \sqrt{G} e^{-2\phi} \hat{c}=0$ i.e. $C=0$. If $\overline{\beta}^{\phi}=$ const, then $\overline{\nu}=$ const and hence $W = -\frac{1}{2\pi} \overline{\nu}  \log z + \text{const}$, $C = \frac{2}{\pi} \overline{\nu}$. Now $C=0$ implies $\overline{\nu}=0$ and therefore $\overline{\beta}^{\phi}=0$ (note that $\int d^{D}y \sqrt{G} e^{-2\phi} > 0$).

This completes the proof that the Weyl invariance conditions $\overline{\beta}^{i}=0$ follow from the stationarity of the action $C$.\footnote{There is an indication that the argument we have presented is not a rigorous one.
In fact, as follows from (6.24), ${\beta}^{\, i} \approx 0$ (modulo arbitrary
diffeomorphism terms) also imply $Y = 0$ and hence $\bar{\beta}^{\alpha} = 0$.
This is a contradiction, since the diffeomorphism terms in $\bar{\beta}^{\alpha} $ are not
arbitrary (see (5.44), and (5.45)). It may be that our proof applies only in a
particular parametrization in which $\bar{\beta}^{\, i} = \beta^{\, i}$.
Another apparent problem with our proof is that we managed to derive the vanishing
of the ``central charge'' $\bar{\beta}^{\phi} = 0$ from the conformal invariance
conditions $\bar\beta^{\alpha} = 0$. The reason for the contradictions is that we assumed
the 2-space to be asymptotically flat in one part of our argument and to have the
spherical topology in another. No contradiction should appear if the argument is
consistently carried out on $S^{2}$ (then $Y$ in (6.24) contains the
$\bar{\beta}^{\phi}$ terms, etc.).    }

In view of (6.21)-(6.23) we can represent $C$ as $(k=|k|)$
\begin{equation}
C = \int_{0}^{\infty} dk \, f(k|z|) \overline{\nu}(k), \tag{6.31}
\end{equation}
where $f$ can be expressed in terms of the zeroth Bessel function [45]. As follows from (6.31), \ $\partial \overline{\nu} / \partial\varphi^{i} = 0$ implies $\partial C / \partial\varphi^{i} = 0$ and hence $\overline{\beta}^{i} = 0$. Thus the ``central charge'' $\bar \nu $ (6.19) itself can be taken as an action $S$, which has the property that
\begin{equation}
\frac{\partial S}{\partial\varphi^{i}} = 0 \tag{6.32}
\end{equation}
imply $\overline{\beta}^{i}=0$. Hence we finish with the functional
\begin{equation}
S = \int d^{D}y \sqrt{G} e^{-2\phi} \tilde{\beta}^{\phi} = \int d^{D}y \sqrt{G} e^{-2\phi} (\overline{\beta}^{\phi} - \frac{1}{4} \overline{\beta}_{\mu\nu}^{G} G^{\mu\nu}), \tag{6.33}
\end{equation}
where we have used Eq. (5.53). It is not clear yet how to prove the converse statement, i.e. that $\overline{\beta}^{i}=0$ imply the stationarity of $S$ (6.32) (cf. Ref. 68).

\subsection{} 

The action (6.33) has several remarkable properties. First, if we substitute
\begin{equation}
\bar{\beta}^{\phi} = \beta^{\phi} + \frac{1}{2} M^{\mu} \partial_{\mu} \phi,\qquad 
  \bar{\beta}_{\mu\nu}^{G} = \beta_{\mu\nu}^{G} + \mathcal{D}_{(\mu } M_{\nu)} \tag{6.34} \end{equation}
 into (6.33), then the $M_{\mu}$ terms form a total derivative and hence drop out of $S$ 
\begin{equation}
S = \int d^{D}y \sqrt{G} e^{-2\phi} (\beta^{\phi} - \frac{1}{4} \beta_{\mu\nu}^{G} G^{\mu\nu}). \tag{6.35}
\end{equation}
Next we observe that (6.35) can be re-written as
\begin{equation}
S = -\frac{1}{2} \Big( \beta^{G} \cdot \frac{\partial}{\partial G} + \beta^{\phi} \cdot \frac{\partial}{\partial \phi} \Big) \int d^{D}y \sqrt{G} e^{-2\phi}. \tag{6.36}
\end{equation}
If we formally assume that $G$ and $\phi$ are taken at the renormalization point $\mu$, then (6.36) can also be put into the following form [45]
\begin{equation}
S = \frac{d}{dt} V, \qquad \ \  V = -\frac{1}{2} \int d^{D}y \sqrt{G} e^{-2\phi}, \tag{6.37}
\end{equation}
where we have used that $\beta^{i} = d\varphi^{i}/dt$, $t =  \log \mu$. 
Note that the expression (5.53) for $\tilde{\beta}^{\phi}$ and hence the representations (6.33) and (6.35) for $S$ are true only for a particular choice of the couplings $\varphi^{i}$ (i.e. for a particular RS).

Let us now prove that $S$ (6.35) can be rewritten also as [23]
\begin{equation}
S = \frac{\partial}{\partial t} Z_{R} \big|_{t=0} \qquad \ \  \text{i.e.} \qquad \ \  S = -\beta^{i} \frac{\partial}{\partial \varphi^{i}} Z_{R} \big|_{t=0}, \tag{6.38}
\end{equation}
where $Z_{R}$ is the renormalized partition function (6.7) of the $\sigma$ model 
\begin{equation}
\Big( \frac{\partial}{\partial t} + \beta^{i} \frac{\partial}{\partial \varphi^{i}} \Big) Z_{R} = 0. \tag{6.39}
\end{equation}
Comparing (6.38) with (6.36) we see that we are to prove that there exists a choice of $\varphi^{i}$ for which
\begin{equation}
Z_{R} \big|_{t=0} = \text{const} \int d^{D}y \sqrt{G} e^{-2\phi}. \tag{6.40}
\end{equation}
In general, we can represent $W$ in (6.7) and (6.9) as 
\begin{equation}
W = F(\varphi) + H(\varphi, t), \tag{6.41}
\end{equation}
\begin{equation}
F = \sum_{n} \alpha_{n}(\varphi) K_{n}[g], \qquad \ \  H = \sum_{n} \sum_{k=1}^{\infty} \rho_{nk}(\varphi) t^{k} K_{n}[g] \tag{6.42}
\end{equation}
where $K_{n}$ are functionals of $g_{ab}$. Hence $W|_{t=0} = F(\varphi)|_{\mu=1}$. Introducing the new variable $\phi' = \phi + \frac{1}{2} F(\phi)$,\footnote{Note that this is a regular  re-definition: $F$ is given  by a power series  in $\alpha'$.} 
we get $Z_{R}|_{t=0} \sim \int d^{D}y \sqrt{G} e^{-2\phi'}$, or 
(omitting prime on $\phi$) Eq. (6.38). 

Since the actions related by field re-definitions have related perturbative solutions, the equations of motion for (6.37) should also be equivalent to $\bar{\beta}^{i} = 0$ (computed in a corresponding RS). 

Let us relax the condition $t = 0$ in (6.38) introducing formally
\begin{equation}
S = \frac{\partial}{\partial t} Z_{R}, \qquad \ \  \text{i.e.} \qquad \ \  S = -\bar{\beta}^{i} \frac{\partial}{\partial \varphi^{i}} Z_{R} \tag{6.43}
\end{equation}
($\beta^{i}$ can be replaced by $\bar{\beta}^{i}$ since $Z_{R}$ is a covariant functional).  Then $dZ_{R}/dt = 0$ and $\partial \bar{\beta}^{i} / \partial t = 0$ imply that $dS/dt = 0$, i.e. that $S$ is RG invariant ($t$-independent).  Hence (6.43) is related to (6.38) by the RG transformation: $S(\varphi(1), 1) \rightarrow S(\varphi(\mu), \mu)$, $d\varphi^{i}/dt = \beta^{i}$, $\varphi^{i}(\mu) = \varphi^{i}(1) + \sum_{k=1}^{\infty} C_{k}^{i} t^{k}$, i.e. by a field re-definition. 

As follows from (6.43)
\begin{equation}\no
\frac{\partial S}{\partial \varphi^{i}} = -\frac{\partial \bar{\beta}^{j}}{\partial \varphi^{i}} \frac{\partial Z_{R}}{\partial \varphi^{j}} - \bar{\beta}^{j} \frac{\partial^{2}}{\partial \varphi^{i} \partial \varphi^{j}} Z_{R}. 
\end{equation}
It should be possible to prove that there exists a non-degenerate operator $\chi_{ij}$ such that
\begin{equation}
\frac{\partial S}{\partial \varphi^{i}} = \chi_{ij} \bar{\beta}^{j}. \tag{6.44}
\end{equation}
This relation (suggested in Refs. 68, 46 and 34) demonstrates the equivalence between $\partial S / \partial \varphi^{i} = 0$ and $\bar{\beta}^{i} = 0$.  Note that if (6.44) is true then
\begin{equation}
\frac{\partial S}{\partial t} = -\beta^{i} \frac{\partial S}{\partial \varphi^{i}} = -\bar{\beta}^{i} \frac{\partial S}{\partial \varphi^{i}} = -\bar{\beta}^{i} \chi_{ij} \bar{\beta}^{j} \tag{6.45}
\end{equation}
or
\begin{equation}
\frac{\partial S}{\partial t} = -\frac{\partial S}{\partial \varphi^{i}} (\chi^{-1})^{ij} \frac{\partial S}{\partial \varphi^{j}} \tag{6.46}
\end{equation}
and hence if $\chi_{ij}$ is positive definite $S$ decreases along a trajectory of RG (cf. Ref. 68). 

Recently, another proof of the statement that the Weyl invariance conditions for the renormalizable $\sigma$ model are equivalent to equations of motion for some action was given in Ref. 65b.  The proof is not based on the Zamolodchikov type argument and, in particular, does not use the assumption of positivity of some ``metric'' (e.g. (6.28)) on the space of couplings.  While Refs. 68 and 45 have not proved the ``off-shell'' relation (6.44), it is, in fact, established in Ref. 65b. 

To facilitate the discussion of renormalization of composite dimension 2 operators, Refs. 65 consider the generalized $\sigma$ model
\begin{equation}
I = \frac{1}{4\pi\alpha'} \int d^{2}\sigma \sqrt{g} \Big[\psi + G_{\mu\nu} \partial_{a} x^{\mu} \partial^{a} x^{\nu} + V_{\mu}^{a} \partial_{a} x^{\mu} + \alpha' R^{(2)} \phi\Big] \tag{6.47}
\end{equation}
where the couplings $\psi, G, V, \phi$ depend not only on $x^{\mu}(\sigma)$ but also explicitly on $\sigma$.

The renormalizability of the theory (6.47) implies identities between the renormalization coefficients which relate the renormalized couplings to the bare ones. Additional constraints among the counterterms arise from symmetries of (6.47) like $\delta V_{\mu a} = \partial_{\mu} F_{a}$, $\delta \psi = -\nabla^{a} F_{a}$ and invariance under diffeomorphisms of $x^{\mu}$ (combined with the corresponding transformation of the couplings). Applying $\partial/\partial \sigma^{a}$ to the Weyl anomaly coefficients $\bar{\beta}^{i}$ it is possible to derive a relation for the functional derivatives of the $\bar{\beta}$ functions (note that $\partial_{a} \bar{\beta}^{i} = \partial_{a} \varphi^{j} (\partial \bar{\beta}^{i} / \partial \varphi^{j})$), which looks like ($\psi$ and $V_{\mu}^{a}$ are set equal to zero in the final expressions):
\begin{equation}
\frac{\partial \bar{\beta}^{i}}{\partial \varphi^{j}} = Q_{jk}^{i} \bar{\beta}^{k} + E^i_{j}, \qquad \ \  \varphi^{i} = \{G, \phi\}. \tag{6.48}
\end{equation}
Choosing the action in the form:
\begin{equation}
S_{1} = \int d^{D} y \sqrt{G} e^{-2\phi} J B^{\phi}, \tag{6.49}
\end{equation}
\begin{equation}
B^{\phi} = \bar{\beta}^{\phi} + W^{\mu\nu} \bar{\beta}_{\mu\nu}^{G} \tag{6.50}
\end{equation}
where the linear operator $W^{\mu\nu}(G)$ is determined by the renormalization coefficients, and subjecting the scalar ``measure" factor $J(G, \phi)$ to a variational condition (which essentially is necessary in order to ``compensate" for the presence of the $E_{ij}$ term in (6.48)) it is possible to prove (6.44), i.e. that $\partial S_{1} / \partial \varphi^{i} = \chi_{ij} \bar{\beta}^{j}$. $\chi_{ij}$ is related (in a rather complicated way) to the renormalization coefficients [65b]. It is easy to check that the leading term in the $\alpha'$ expansion of $\chi_{ij}$ is nondegenerate and hence $\partial S_{1} / \partial \varphi^{i} = 0$ are equivalent to $\bar{\beta}^{i} = 0$ within the perturbation theory based on loop ($\alpha'$) expansion.

The action (6.49) certainly is not the only one which satisfies (6.44) for some $\chi_{ij}$. For example, any action of the form:
\begin{equation}
S(\rho) = S_{1} + \int d^{D} y \sqrt{G} e^{-2\phi} \bar{\beta}^{i} \rho_{ij} \bar{\beta}^{j} \tag{6.51}
\end{equation}
is also a solution of (6.44) (the local operator $\rho_{ij}$ should satisfy the condition that the resulting $\chi_{ij}$ is nonsingular in the $\bar{\beta}^{i} \to 0$ limit).\foot{It is interesting to note that the ambiguity in the choice of the action   seems to be  of a more general  nature  than the usual   field re-definition  ambiguity  (Eq. (6.44) is   covariant under $\varphi\to \varphi + \delta \varphi$ in $\chi_{ij}$  transforms as a tensor).}
As was mentioned in Ref. 65b it is possible to choose $\rho_{ij}$ in such a way that (6.51) can be represented simply as:
\begin{equation}
S_{2} = \beta^{i} \frac{\partial}{\partial \varphi^{i}} \int d^{D} y \sqrt{G} e^{-2\phi} J. \tag{6.52}
\end{equation}
The crucial observation is that (6.52) is equivalent (related by a field re-definition) to our ``central charge" action (6.33) or (6.38). In fact, re-defining, e.g., the dilaton $\phi' = \phi - \frac{1}{2}  \log J$, $G' = G$ and using that $\beta^{i} (\partial / \partial \varphi^{i}) = \beta^{i'} (\partial / \partial \varphi^{i'})$ we get: $S_{2} = \beta^{i'} (\partial / \partial \varphi^{i'}) \int d^{D} y \sqrt{G'} e^{-2\phi'}$. Thus there always exists a regular (power series in $\alpha'$) field re-definition which relates the action (6.52) of Ref. 65b to the action (6.35) of Ref. 45 (this was already noticed in Ref. 65b to be true to the 4-loop order).

Let us mention once again that because of the field
re-definition the ${\beta}$ functions which appear in (6.34)
correspond to a ``nonstandard'' renormalization scheme. Still the
covariance of the relation (6.44) implies that if it is valid in one RS
(i.e.\ for (6.52)) it is also valid in all other related by coupling
re-definitions (e.g.\ for (6.35)).

It is interesting to note that though $S$ in (6.38) or (6.43) appears
to depend on a choice of the 2-metric (its conformal factor) made in the
computation of $Z$, the equivalence to (6.35) implies that this
dependence can be absorbed into a re-definition of the fields
$\phi^i$. Let us mention also that according to (6.43) $S$ is
proportional to the integral of the 2-d scale anomaly
\begin{equation}
S \sim \int d^D y \,\sqrt{G}\, e^{-2\phi}
\big\langle \int d^2 \sigma \,\sqrt{g}\, T^{a}{}_{a} \big\rangle  .\no 
\end{equation}
Using (5.51) and $\int d^2 \sigma \sqrt{g}\, R^{(2)} = 8\pi$ we can
re-write (6.46) as
\[
S \sim \int d^D y \,\sqrt{G}\, e^{-2\phi}
\big(\tilde {\beta}^{\phi} + \ldots \big).
\]
The above argument which demonstrated the equivalence of (6.40) and
(6.35) implies that the extra terms here (indicated by dots) can be
eliminated by a field re-definition.

The representation (6.33) for the action functional has the following
consequence. Suppose that the space-time is a direct product, i.e.
\[
G_{\mu\nu} =
\begin{pmatrix}
G_1(x_1) & 0 \\
0 & G_2(x_2)
\end{pmatrix},
\qquad
\phi = \phi_1(x_1) + \phi_2(x_2), \no 
\]
etc. Then the $\sigma$ model splits into two noninteracting models so
that its energy-momentum tensor and hence the ``central charge''
coefficient are given by
\[
T = T_1 + T_2,
\qquad
\hat{\beta}^{\phi}
= \hat{\beta}^{\phi}_1(G_1,\phi_1)
+ \hat{\beta}^{\phi}_2(G_2,\phi_2).
\]
Thus the action (6.32) should have the following ``additivity''
property
\begin{equation}
S = \int d^{d_1}x_1\, d^{d_2}x_2\;
\sqrt{G_1}\,\sqrt{G_2}\,
e^{-2\phi_1-2\phi_2}
\big[\tilde{\beta}^{\phi}_1(x_1)
+ \tilde{\beta}^{\phi}_2(x_2)\big]   .\no
\end{equation}
This additivity property restricts the structure of $\tilde{\beta}^{\phi}$:
it  cannot contain terms which do not split into a sum
for direct product metrics. In particular, all powers like
$R^n$, $(R_{\mu\nu}R^{\mu\nu})^n$, $(R_{\mu\nu\lambda\rho}
R^{\mu\nu\lambda\rho})^n$ are ruled out [71].
Note that  the above equation  implies that the direct product worlds ``interact''
only through the $\phi$ equation:
\[
\tilde{\beta}^{\phi}_1 = - \tilde{\beta}^{\phi}_2 = \text{const},
\]
i.e.\ through the condition of the vanishing of the total central charge.

\subsection{} 

 Let us finally discuss the issue of the IR infinities in the
$\sigma$-model and related questions. It was recently observed in Ref.~65c
that the correlator of the two energy-momentum tensors computed on a flat
background contains IR divergent terms. Their presence formally invalidates
the scaling argument used in Refs.~68, 45 and above to replace
$\frac{\partial}{\partial z}$-derivatives in (6.21) by
$\mu\frac{\partial}{\partial \mu}$ in (6.23) and (6.24). This apparently
implies [65c] that the relation (6.24) (``$c$-theorem'') should be
modified by adding an additional term on its right-hand side.

 This, however,
may not be a correct conclusion: the ``global'' version of the ``$c$-theorem''
may still be true in its original form (6.24). The point is that to derive a
``global'' statement like the ``$c$-theorem'', it is necessary to compute the
correlators on a compact 2-space background (e.g.\ on the sphere)
 [45].\footnote{The analogy with the string theory EA ($S=\int d^D y \sqrt G \ e^{-2\phi}\, L$) 
  makes it particularly clear that, while the ``local'' beta functions do not ``know''
   about the topology, the global ``action'' from which they should be derived should correspond to the spherical topology.   }
Then the IR infinities are automatically cut off at the scale of 2-space and
also the expectation values are given by the integrals over the constant part
(zero mode) of $x^{\mu}$ ($\int d^D y\sqrt{G}\ldots$, see (6.8)).

 As we shall
see below, the additional term found in Ref.~65c to be present in the ``local''
version of the ``$c$-theorem'' disappears after the integration over $y^{\mu}$.
To derive the ``$c$-theorem'', it is not sufficient to consider only ``local''
(operator) relations but it is necessary to compute finally the expectation
values on a 2-space of the spherical topology. The ``$c$-functional'' thus
should ``know'' about the IR-properties of the theory (in particular, it may
depend on a radius of $S^2$ which plays the role of a ``time'' parameter of the
RG flow, see below). However, this dependence is an ``off-shell artifact'',
disappearing in the extremal points of $C$ (corresponding to $\bar{\beta}^i=0$).

Ref.~65c computed the 3-loop terms in the coefficient $\lambda$ in (6.13) or $\nu$   in
(6.16) by using the expansion near a flat space. In general (see Eq.~(7.25),
for simplicity we set $\phi=\mathrm{const}$)
\begin{align*}
\lambda(k) = \nu (k) &= -\frac{1}{16\pi}\,\tilde{\beta}^{\phi} \\
&= -\frac{1}{16\pi}\Big[
\frac{1}{6}(D-26)
-\frac{1}{4}\alpha' R
-\frac{1}{16}\alpha'^2 R_{\lambda\mu\nu\rho}^{\,2}
+a_1\alpha'^2 R^{2}
+a_2\alpha'^2 D^{2}R
+O(\alpha'^3)
\Big] .
\end{align*}
\noindent
The coefficients $a_1$ and $a_2$ are clearly ambiguous since they change under
$G_{\mu\nu}\to G_{\mu\nu}+\alpha'(b_1 R_{\mu\nu}+b_2 G_{\mu\nu}R)$. Thus their
values should not have a physical significance, being dependent on a
renormalization prescription. As was found in Ref.~65c
\[
a_1=-\frac{1}{8}\big( \log\frac{k^{2}}{\mu^{2}}+a_0\big),\qquad
a_2=\frac{1}{16}\big( \log\frac{m^{2}}{\mu^{2}}+a_0\big),
\qquad 
a_0=\gamma_0- \log 4\pi=\text{const.}
\]
Here $\mu$ is the UV renormalization point, $m$ is the IR cutoff and
$\gamma_0$ is the Euler constant.

The dependence on $\mu$ is precisely the one prescribed by RG
(note that $R_{\mu\nu}\,\cdot \frac{\partial}{\partial G_{\mu\nu}}R
= -R_{\mu\nu}^{\,2}-\frac{1}{2}D^{2}R$). However, the presence of $ \log m^2$
instead of $ \log k^2$ in $\alpha'^2$ implies that we cannot represent $\nu(k)$ as
$v(\phi(k))$.\foot{The computation of the coefficient of the $D^2R$  term in $W $ 
is closely related to the computation of the coefficient of the R term [20]: the difference is only 
in the extra ``tadpole'' factor (or $\langle \xi\xi \rangle$) in $a_2$.  }
 The presence of the IR infinity in $W$ (6.12) can be understood
also by computing first the expectation value of the trace of the
energy-momentum tensor $\theta=T^{a}{}_{a}$. Expanding near some $\bar{x}^{\mu}$
we get (see (5.43)--(5.51))
\[
\langle\theta\rangle_{\bar{x}}
=\frac{1}{4\pi}\Big[
\tilde{\beta}^{\phi}(\bar{x})\,R^{(2)}
+\alpha'\pi\,D^{2}\tilde{\beta}^{\phi}(\bar{x})\,\langle\xi\xi\rangle\,R^{(2)}
+\ldots\Big], \qquad 
\langle\xi\xi\rangle=-\frac{1}{4\pi}\Big( \log\frac{m^{2}}{\mu^{2}}+a_0\Big).
\]

\noindent
Hence integrating the Weyl anomaly relation, we find that the coefficient of
the $ \log\frac{m^2}{\mu^2}$ term in $W$ is proportional to $D^{2}\tilde{\beta}^{\phi}+\ldots$.

As follows from (6.15),(6.16) similar $ \log\frac{m^2}{\mu^2}$ term appears in
the correlator $\langle T_{ab}\,T_{cd}\rangle_{fl, \bar{x}}$ computed on a flat
background and near some $\bar{x}^{\mu}$. If we formally define the objects
$\hat{U}$, $\hat{X}$, $\hat{Y}$ and $\hat{C}$ (cf.\ (6.20) and (6.23)) in terms
of the $\langle\ldots\rangle_{fl,\bar{x}}$ expectation values (as it is assumed in
Ref.~65c) we find that the presence of the $ \log\frac{m^2}{\mu^2}$ term in $\hat{U}$
implies that the ``local'' ($\bar{x}$-dependent) form of the ``$c$-theorem''
relation is \emph{different} from (6.24)  [65c]:
\begin{equation}
\beta^{i}\,\frac{\partial \hat{C}(\bar{x})}{\partial \phi^{i}}
= \hat{Y}(\bar{x})
-\frac{1}{8\pi^{2}}\,\alpha' D^{2}\tilde{\beta}^{\phi}(\bar{x})
+\ldots .
\tag{6.53}
\end{equation}
It is possible also to derive the operator analog of this equation [65c,65d].
Defining the operator products (cf.\ (6.20))
\begin{equation}
V_1=z^{4}T(\sigma)T(0),\qquad
V_2=z^{2}\bar{z}T(\sigma)\bar{T}(0),\qquad
V_3=z^{2}\bar{z}^{2}\theta(\sigma)\theta(0),
\tag{6.54}
\end{equation}
\begin{equation}
V_a=\sum_{n,m} z^{n}\bar{z}^{m}\,A_a^{nm}(\sigma),
\qquad
A^{nm}_a
=\sum_{p=0}^{\infty}\frac{1}{p!}\, \big[\Delta^{-1}(|{z}|^{2})\big]^{p}
\big[\Lambda^{nm} _{ap}\big](\sigma),
\tag{6.55}
\end{equation}
\[
\Delta^{-1}=-\frac{1}{4\pi}\Big( \log(\mu^{2}|{z}|^{2})+\text{const}\Big).
\]
and the operator (cf.\ (6.33))
\begin{equation}
\check{C}=\frac{1}{12}\,(\check{U}-2\check{X}-3\check{Y}),
\tag{6.56}
\qquad 
[\check{U}]=A^{00}_{1}, \qquad
[\check{X}]=A^{00}_{2}, \qquad
[\check{Y}]=A^{00}_{3},
\end{equation}
and using the RG invariance $\Big(\frac{d}{dt}V_a=0\Big)$, the conservation
of the energy-momentum tensor and the absence of the IR infinities in the OPE
relations (6.55), one finds  [65c] (cf.\ (6.24) and (6.50))
\begin{equation}
\beta^{i}\,\frac{\partial}{\partial \phi^{i}}\check{C}
= \check{Y}-\gamma\,\check{C}.
\tag{6.57}
\end{equation}

\noindent
Here $\gamma$ is the scalar anomalous dimension operator,
$\gamma=\frac{1}{2}\alpha' D^{2}+\ldots$
(see Eqs.\ (5.34), (7.4) and (7.5)). If $H$ is a scalar dimension zero
operator which depends on $\phi^{i}$, i.e.
\[
H=\int d^{D}y\, H(y(\phi))\,\delta^{(D)}(y-x(\sigma)) \equiv H\cdot 1,
\qquad [H]=H\cdot 1,
\]
then
$
\frac{d}{dt}[H]=\frac{d}{dt}H\cdot[1]
+ H\cdot\frac{d}{dt}[1]
=\big(\frac{d}{dt}+\gamma\big)H $
(we used $\frac{d}{dt}[1]=\gamma[1]$, cf.\ (5.15)).
Hence if $\frac{d}{dt}[H]=[E]$ or
$\beta^{i}\partial_{i}[H]=[E]$ then
$
\Big[\beta^{i}\partial_{i}H\Big]=[E]-[\gamma H]
$ (see also [65d]).

Computing the $\langle\ldots\rangle_{fl,\bar x}$ expectation value of (6.57) we get
back Eq.\ (6.53). Note that the presence of the additional
$\gamma\check{C}$ term in the operator relation (6.57), in principle, has
nothing to do with the presence of IR infinities in the expectation values
but is simply a consequence of defining $\check{C}$ as a local scalar operator [{65d}].

It is clear that the IR infinities are absent at the Weyl invariant point.
In fact, if $\bar{\beta}^{i}=0$ then $\tilde{\beta}^{\phi}=\text{const}$
and hence
\[
\langle \theta \rangle_x
=\frac{1}{4\pi}\,\tilde{\beta}^{\phi} R^{(2)},
\qquad
W=-\frac{1}{16\pi}\,\tilde{\beta}^{\phi}
\int R^{(2)}\Delta^{-1}R^{(2)}
\]
(see Eqs.\ (5.54) and (5.55)). Hence the effects related to their presence
should be ``off-shell artifacts''. 

In particular, we may expect that the
presence of IR infinities on a flat background should not be important for a
proper ``global'' formulation of the ``$c$-theorem''. Let us consider again
the $\langle\ldots\rangle$ expectation values (6.8) and define $U,X,Y$
(6.20) and $C$ (6.23) in terms of the expectation values
$\langle\!\langle\ldots\rangle\!\rangle_{\text{sph}}$
computed on the $S^{2}$ background. Then the radius of $S^{2}$ plays the role
of an IR cutoff and the ``local'' function $\check{C}(y)$ (defined in terms
of $\langle\ldots\rangle_{\text{sph}}$, see (6.8)) again satisfies Eq.\ (6.50).

The crucial point is that to get the equation for the global functional $C$,
we are still to integrate (6.50) over $y$ with the measure
$\sqrt{G}\,e^{-2\phi}$.\footnote{We assumed that $\phi=\text{const}$.
If we carefully account for the dependence on $\phi$ then
$\gamma=\tfrac{1}{2}\alpha'(\D_\mu -2\partial_\mu\phi) D^\mu +\ldots$
and hence $\int d^{D}y\,\sqrt{G}\,e^{-2\phi}\,\gamma\tilde{\beta}^{\phi}$
again vanishes (in the leading approximation for $\gamma$).}
Since $D^{2}\tilde{\beta}^{\phi}$ is the total covariant derivative it does
not contribute and we get precisely the original formulation of the
``$c$-theorem'' (6.24), avoiding the objection raised in Ref.~65c (at least
within the 3-loop approximation considered there). It is likely that the
$\gamma\check{C}$ term in (6.54) always gives zero, being integrated over $y$.

In the above argument, we have implicitly assumed that taking
$\beta^{i}\partial_{i}$ ``commutes'' with computing the expectation value
$\langle\!\langle\ldots\rangle\!\rangle_{\text{sph}}$.
Apparently, this is not so since $\beta^{i}\partial_{i}$ should act on
$\sqrt{G}\,e^{-2\phi}$  [65d].

Suppose, however, that we start with
\[
\partial_{\eta }[\check{C}]
=\beta^{i}\partial_{i}[\check{C}]
=[\check{Y}], \qquad \eta= \log|z|.
\]
Since $\langle\!\langle\ldots\rangle\!\rangle_{\text{sph}}$ obviously
``commutes'' with $\partial/\partial z$ we get
\[
\partial_{\eta}C=Y, \qquad
C=\langle\!\langle\check{C}\rangle\!\rangle_{\text{sph}},
\qquad
Y=\langle\!\langle\check{Y}\rangle\!\rangle_{\text{sph}}.
\]
The functions $C$ and $Y$ should be finite, i.e.\ should satisfy
$\frac{d}{dt}C=0$, $\frac{d}{dt}Y=0$.
The only complication is that $\partial C/\partial\eta$ may not be equal to
$\partial C/\partial t$ because of  possible dependence of $C$ on the radius
$a$ of $S^{2}$ ($C$ may depend on $u|z|$ {\it and} on $|z|/a$). What we expect is
that $\partial C/\partial a$, in fact, vanishes since the corresponding
``IR anomalous'' term present in the ``local'' expression (6.50) is a total
derivative. To justify this reasoning, it is clearly important to get a deeper
understanding of the structure of
$\langle\!\langle\check{C}\rangle\!\rangle_{\text{sph}}$.

Let us now add some general remarks about the structure of the 2-d effective
action $W$. When $W$ is computed on a flat background, IR infinities appear
from the ``tadpole'' (``simple'' loop) parts of the diagrams. They are present
already in one-loop approximation on a nonconstant $\bar{x}$ background,
\[
W \sim \int d^{2}\sigma\,\bar{\partial}\bar{x}^{\mu}\bar{\partial}\bar{x}^{\nu}
\,R_{\mu\nu}(\bar{x}) \log\, ({m^{2}}/{\Lambda^{2}})+\ldots
\qquad (\Lambda \text{ is a UV cutoff}).
\]
In view of the above discussion, the $\bar{x}$-independent part of the
unrenormalized $W$ can be represented as (we ignore power UV infinities)
\begin{equation}
W_{0}
=\int R^{(2)}\Delta^{-1}
\Big[\lambda_{0}+\lambda_{1} \log(\Delta/\Lambda^{2})
+\lambda_{2} \log (m^{2}/\Lambda^{2})
+\ldots\Big]R^{(2)}+\ldots ,
\tag{6.58}
\end{equation}
where $\lambda_{0}=-\frac{1}{16\pi}\tilde{\beta}^{\phi}$ and $\lambda_{2}$
starts with the 3-loop contribution ($\sim D^{2}R$). Suppose now that we do
not use the expansion near a flat space. Then the IR infinities should be cut
off by the 2-curvature
\begin{equation}
W_{0}
=\int R^{(2)}\Delta^{-1}
\Big[\lambda_{0}+\lambda_{1} \log(\Delta/\Lambda^{2})
+\lambda_{2} \log (R^{(2)}/\Lambda^{2})
+\ldots\Big]R^{(2)}+\ldots \ , 
\tag{6.59}
\end{equation}
i.e.\ $W$ may contain terms {\it nonanalytic} in the 2-curvature. 

The dependence on
the UV cutoff is of course determined by the renormalizability (RG-invariance)
condition $\Lambda\,\frac{dW_0}{d\Lambda}=0$. Note that the covariant nature of
the cutoff ($g_{ab}\Delta\sigma^{a}\Delta \sigma^{b}\ge \Lambda^{-2}$) implies
that we can absorb the $\Lambda$ dependence into $g_{ab}$ by making the
rescaling $g_{ab}\to\Lambda^{-2}g_{ab}$ in $W$.

Let
$A=\int d^{2}\sigma\sqrt{g}$ and $\bar{g}_{ab}=A^{-1}g_{ab}$,
$\int d^{2}\sigma\sqrt{\bar{g}}=1$. If $\varphi(\Lambda)$ are the bare fields
we should thus have \ldots
\begin{equation}
W_0[\varphi_0(\Lambda),\Lambda,g_{ab}]
= W'[\varphi_0(\Lambda),\Lambda^{2}g_{ab}]
= W''[\varphi_0(\Lambda),\Lambda^{2} A \bar g_{ab}]
= W[\varphi_0(A^{-1/2}),\bar g_{ab}] .\no 
\end{equation}
Hence the 2-volume $A$ plays a role of a natural renormalization scale
($\mu = A^{-1/2}$). Let us consider, in particular, the case when $g_{ab}$
is the standard metric on $S^{2}$ with the radius $a$. Ignoring power
infinities we should get\foot{As was already discussed above, in the case of a compact 2-space
$Z=\int d^{D}y \sqrt{G_0}\,e^{-W_0}$ and hence the renormalizability of $Z$
implies $d\bar W_0/d\Lambda=0$, $\bar W_0 = W_0 - \frac{1}{2}\log G_0$. For notational
simplicity, we use $W_0$ instead of $\bar W_0$ in what follows.}
\begin{equation}
W_0 = \rho_0 + \rho_1 \log(a\Lambda) + \rho_2 [\log(a\Lambda)]^{2} + \ldots\no 
\end{equation}
Here $\rho_i$ depends on the bare couplings. Note that in contrast to (6.56),
all logarithmic terms are ``on an equal footing'' ($\log a$ terms
appear both from $\log\Lambda$ and in $R^{(2)}$ terms). The renormalizability
implies that $W_0$ can be rewritten as
\begin{equation}
W_0[\varphi_0(\Lambda),a\Lambda]
= W[\varphi(a^{-1})]
= \rho_0[\varphi(a^{-1})],
\tag{6.60}
\end{equation}
\[
\varphi(a^{-1}) = \varphi_0(\Lambda) - \beta \log(a\Lambda) + \ldots .
\]
It is the dimensionless combination of the UV and IR cutoffs that plays the
role of a physical cutoff. The radius of $S^{2}$ can thus be interpreted as a
``time'' parameter of the RG flow. Note that the ``global'' functionals like
$C$ (6.30) and $S$ (6.37) all have $\varphi(a^{-1})$ as their argument and hence
are functions of $a$ (their values at the stationary points are of course
$a$-independent).

The idea to consider the ``off-shell'' dependence of the $\sigma$ model
couplings on a scale of a 2-space is very natural. In fact, the $\sigma$ model
fields $x^{\mu}$ and $g_{ab}$ both have constant ``zero modes''
($\gamma$ and $A=\int d^{2}\sigma \sqrt{g}=\pi a^{2}$) if the 2-space is
compact. This suggests that $\varphi$ may depend not only on $y^{\mu}$ but also
on $a$. While the resulting $\sigma$ model action becomes effectively nonlocal
with respect to the 2-metric, this should not be a problem since we do not
consider the 2-metric as a quantum (integration) variable.\foot{On higher genus surfaces, we should rescale the moduli parameters (e.g.\
lengths of nontrivial geodesics) by $a$ and take $\varphi$ to depend only on $a$
(but not on the moduli). Then ``modular infinities'' will correspond to a part
of $a \to 0$ singularities and produce ``modular'' contributions to the
$\beta$ function ($d\varphi/d\log a$).} 

This
suggestion is also natural from a string field theory point of view: the string
field should depend not only on $x^{\mu}$ but also on an additional bosonic
variable(s) (e.g.\ ``1-d metric''). Hence its components should be functions
$\varphi(y,a)$ of the zero modes $a$ and of its arguments. The classical vacuum
configurations are independent of $a$ (i.e.\ are scale invariant) but the
``instantons'' which interpolate between different classical string vacua
(i.e.\ solutions of $\dot{\varphi}=\beta$) should correspond to $a$-dependent
fields. It remains to understand if these remarks have some interesting
implications.\foot{Note, in particular, that Eq.~(6.37) implies that the effective action can be
represented as the derivative of the ``running'' space-time volume over the
logarithm of the world-sheet volume ($\sim \log a$).
}

\

In conclusion, let us make the following remark. It appears that the
``$c$-theorem'' cannot be true in the general 2-$d$ $\sigma$ models (and hence
in string theory) in its original formulation given in Ref.~68: there exists
such $C(\varphi)$ that

 (1) $\partial C / \partial \varphi^{i} = 0$ is equivalent to
$\beta^{i}=0$;

 (2) $C$ decreases along the RG and is equal to
the central charge at fixed points.

\noindent 
 While the first statement (and even
$\partial C / \partial \varphi^{i} = \kappa_{ij}\bar \beta^{j}$) is certainly true in
the perturbation theory (see Ref.~65b and references therein) with $C$ being a
functional, essentially equivalent to the string theory effective action $S$,
the second statement cannot be true. 

The reason is that since the dilaton
coupling must be introduced into the theory (to have renormalizability on a
curved 2-$d$ background), it participates in the RG flow, the fixed points of
which are those $\beta^{a}=0$ {\it and} $\beta^{\phi}=0$. It is clear that if
$\beta^{a}=0$ and $\beta^{\phi}=0$, the ``central charge'' vanishes. $C$ is
thus the same (equal to zero) in all fixed points, i.e.\ does not decrease.

This is, in fact, what should be for the string theory tree vacua, which are
solutions of
\[
\frac{\delta S}{\delta \varphi^{\alpha}}=0 \qquad  {\it and } \qquad 
\qquad
\frac{\delta S}{\delta \phi}=0 .
\]
Thus the condition of decreasing along the RG trajectories is inconsistent
with the requirement of being equal to the central charge at the critical
points.

\def \CC {\mathbb{C}}
We suggest that in the general ``string'' models the ``$c$-theorem'' is
replaced by the following statement: 

(1)  there exists a functional $\mathbb{C}$ such that \[
{{\partial \CC}\over {\partial \varphi^{i}} }= \kappa_{ij}\beta^{j},
\]
where $\varphi^{i}=(G,\phi)$, $\beta^{i}$ are the Weyl anomaly coefficients and
$\kappa_{ij}$ is nondegenerate;

(2)
there exists a choice of $\varphi^{i}$ such that
\[
\CC=S_{0}
=\int d^{D}y\,\sqrt{G}\,e^{-2\phi}\,\tilde{\beta}^{\phi},
\qquad
\tilde{\beta}^{\phi}
=\bar \beta^{\phi}-\frac{1}{4}G^{\mu\nu}\bar \beta^{G}_{\mu\nu},
\]
where $\tilde{\beta}^{\varphi}$ is the basic Weyl anomaly coefficient
(``central charge''). Then we get $S_{0}=0$ at the critical points. 

The
representation (4.12) implies that $S_{0}$ is equivalent to the string theory
effective action. As was noted in Ref.~65d,
\[
\beta^{i}\frac{\partial S_{0}}{\partial \varphi^{i}}
= A^{2}-B^{2} \le 0,
\]
where $A$ and $B$ are some functionals of $\varphi^{i}$. Hence $S_{0}$ does not
decrease (or increase) along the trajectories of RG. In general, we consider
it to be unlikely that there exists such choice of $\CC$ and $\varphi$ that
$\kappa_{ij}$ is negative definite.

\def \vphi  {\varphi}

\section{Perturbation theory results for Weyl anomaly\\ coefficients}
\renewcommand{\theequation}{7.\arabic{equation}}
\setcounter{equation}{0}

In this section we summarize some results of perturbative computations of
renormalization of couplings and composite operators in the bosonic
$\sigma$ model (3.1). Computing the corresponding expressions for the Weyl
anomaly coefficients we can thus check the statements made in Secs.~6 and 4,
namely, that $\bar{\beta}^{i}=0$ are equivalent to 
the equations of motion for the action
$\int d^{D}y\,\sqrt{G}\,e^{-2\phi}\,\tilde{\beta}^{\phi}$ and that this
action provides a particular representation for the closed string theory tree
effective action.

Before quoting the results of explicit computations of the RG coefficients it
is useful to discuss the renormalization scheme dependence of the
$\sigma$-model objects. The RS ambiguity corresponds to the possibility of
making finite re-definitions (renormalizations) of the $\sigma$-model couplings
\begin{equation}
\varphi^{\prime i} = \varphi^{i}
+ \sum_{k=1}^{\infty} \alpha'^k f^{\,i}_{k}(\varphi),
\tag{7.1}
\end{equation}
(since we work within the perturbation theory re-definitions must be regular
in $\alpha'$). The $f^i_k$ are local covariant functions of $\varphi^{i}$ and their
derivatives. Considering (7.1) as a coordinate transformation, we conclude
that $\beta^{i}=d\varphi^{i}/dt$ transforms as a contravariant vector
\begin{equation}
\beta^{\prime i}(\vphi')
= \frac{\partial \varphi^{\prime i}}{\partial \varphi^{j}}\,\beta^{j}(\vphi),
\qquad
\delta\beta^{i}
= \beta^{\prime i}(\vphi)-\beta^{i}(\vphi)
= \delta\varphi^{j}\,\partial_{j}\beta^{i}
- \beta^{j}\,\partial_{j}\delta\varphi^{i}.
\tag{7.2}
\end{equation}
Equation (7.2) implies that higher loop terms in $\beta^{i}$ contain some
structures which have ambiguous (RS dependent) coefficients (see Refs.~34 and
52 for details). It is useful to recall that in general the second leading
coefficient in the $\beta$ function in a theory with several couplings is not
RS-invariant.

\def \D {{\cal D}}
Let us first consider the case of $B_{\mu\nu}=0$. The 2-loop
$\beta$-function and the 1-loop ``dimension zero'' anomalous dimension
operator (see (5.34)) were found in Ref.~58
\begin{equation}
\beta^{G}_{\mu\nu}
= \alpha' R_{\mu\nu}
+ \tfrac{1}{2}\alpha'^2
R_{\mu\lambda\rho\sigma} R_{\nu}{}^{\lambda\rho\sigma}
+ O(\alpha'^3),
\tag{7.3}
\end{equation}
\begin{equation}
\gamma = \tfrac{1}{2}\alpha' \D^{2} + O(\alpha'^2).
\tag{7.4}
\end{equation}
In general [53]
\begin{equation}
\gamma = \tfrac{1}{2}\alpha' \D^{2}
+ \sum_{n=3}^{\infty} \Omega_n^{\mu_1...\mu_n }(\phi)\, \D_{\mu_1}...\D_{\mu_n },
\tag{7.5}
\end{equation}
\begin{equation}
\Omega_2^{\mu\nu}
= c_1 \alpha'^2 R^{\mu\nu}
+ c_2 \alpha'^3 R^{\mu}{}_{\lambda\rho\sigma}
  R^{\nu\lambda\rho\sigma}
+ \ldots,
\tag{7.6}
\qquad 
\Omega_3^{\mu\nu\rho}
= \alpha'^3 b_1\, \D_{\alpha } R^{\ \ \mu\nu }_{\beta \ \ \  \gamma}  R^{\alpha \beta \gamma \rho} + ... 
\end{equation}
The coefficients $c_i$ are obviously RS dependent (they change under a
re-definition of $G_{\mu\nu}$). $c_1=0$ in the minimal subtraction scheme [{53,72}].
In Ref.~53 the 2-loop expression for the renormalization function
$S_{1\,G}$ in (5.39) was computed ($B_{\mu\nu}=0$)
\begin{equation}
S_{1\,G\mu }\!\cdot\! F^{G}
= \alpha'\!\Big(\D^\lambda F^{G}_{\lambda \mu}
- \tfrac{1}{2} \D_\mu \,F^{G\lambda}_\lambda\Big)
+ \tfrac{1}{2}\alpha'^2
R_{\mu}^{\  \lambda\rho\sigma} \D_{\sigma } F^{G}_{\lambda \rho}
+ O(\alpha'^3).
\tag{7.7}
\end{equation}
The 2-loop results (7.3) and (7.7) are sufficient in order to establish the
3-loop expression for the dilaton $\beta$-function using the identity (5.48) [{59}]. 
Equation (7.7) implies that $W_{\mu}$ in (5.45) vanishes to the given
order. The $\phi$-dependent part of (5.48) is thus satisfied (to the order
$O(\alpha'^2)$) if
\begin{equation}
\gamma = \tfrac{1}{2}\alpha' \D^{2} + O(\alpha'^3),
\qquad
W_{\mu} = O(\alpha'^3),
\tag{7.8}
\end{equation}
while the $\phi$-independent part reduces to
$2\alpha'\partial_{\mu}\omega
= \tilde{S}_{G\mu}\cdot \beta^{G}$.
Recalling that
$\tilde{S}_{G\mu}
= G\cdot \,{\partial\over \partial G }S_{1\,G\mu }
= -\alpha' {\partial\over \partial \alpha'} S_{1\,G\mu }$,
one finds
\begin{equation}
\omega
= \omega_{0}
+ \tfrac{1}{16}\alpha'^2 R_{\lambda \rho \mu\nu}^{2}
+ O(\alpha'^3),
\qquad
\omega_{0}=\text{const}.
\tag{7.9}
\end{equation}
The same results (7.8) and (7.9) were previously found by the direct
computation of the $h^{2}$ term ($h_{ab}=g_{ab}-\delta_{ab}$) in curved space
effective action [{34,53}]  (they were also derived in Ref.~65 using the
manifestly covariant curved space perturbation theory).
 Thus the 3-loop
dilaton $\beta$ function is given by
\begin{equation}
\beta^{\phi}
= \tfrac{1}{6}D
- \tfrac{1}{2}\alpha' \D^{2}\phi
+ \tfrac{1}{16}\alpha'^2 R_{\lambda \rho \mu\nu}^{2}
+ O(\alpha'^3).
\tag{7.10}
\end{equation}
We have fixed the constant in (7.9) equal to its standard one-loop value (see,
e.g., Ref.~73). 

Note that there are no $R$ and $R_{\mu\nu}^{2}$ terms in
(7.10). The absence of the $R$ term is an RS-independent fact which is easy
to prove by the explicit 2-loop calculation (carefully accounting for the
contribution of the one-loop counterterm which appears to cancel against that
of the genuine 2-loop graph). The coefficient of the $R_{\mu\nu}^{2}$ term in
(7.10) is RS-dependent (if $\delta\phi=aR$,
$\delta\beta^{\phi}=-a\beta^{G}\cdot {\partial\over \partial G} R 
= a\alpha' R_{\mu\nu}^{2}+\ldots$, see (7.2)). 
In general, higher loop terms
in the $\beta$ functions (computed in the minimal subtraction scheme) cannot
contain the Ricci tensor (the latter can only appear from the tadpole graphs
whose simple poles must be cancelled out by the counterterms). 

\def \ve {\varepsilon}
\def \r {\rho} \def \m {\mu} \def \n {\nu} \def \l {\lambda}
\def \a {\alpha} \def \b {\beta} \def \g {\gamma} \def \d {\delta} \def \s {\sigma}\def \k {\kappa} 
\def \te {\textstyle}

The 3-loop term
in $\beta^{G}$ (7.3) was recently computed in Ref.~74
\begin{align}
\beta^{G(3)}_{\mu\nu}
= \alpha'^3\!\Big[ &
\tfrac{1}{8} \D_{\r}R_{\m\a\b\g}
           \D^{\r}R_\n^{\ \a\b\g}
- \tfrac{1}{16}
  \D_{\mu}R_{\lambda\rho\sigma\tau}
  \D_\n R^{\lambda\rho\sigma\tau}
\no \\ &+ \tfrac{1}{2}
  R_{\a\rho\sigma\b}
  R_{\mu}{}^{\ \s\r\g}
  R_{\n\ \ \ \g}^{\ \ \alpha\beta}
- \tfrac{3}{8}
  R_{\mu\alpha\beta\n}
  R^{\a\s\r\kappa}
  R^\b_{\ \s\r\k}
\Big].
\tag{7.11}
\end{align}
It is remarkable that applying the identity (5.48) it is possible to determine
the 4-loop dilaton $\beta$ function and $W_\m$ using only the 2-loop
renormalization function (7.7) and the 3-loop $\beta^{G}$ (7.11)  [{75}]
\begin{align}
\beta^{\phi(3)} =\te  - \gamma^{(3)}\phi + \omega^{(3)}
= \alpha'^3 \Big[
&\te -\frac{3}{16} R^\m_{\ \a\rho\sigma} R^{\nu\a\rho\sigma} \D_\m \D_n \phi\no \\
&\te + \frac{1}{32} R_{\mu\nu\rho\sigma} R^{\a\b\rho\sigma} R_{\a\b}^{\ \ \m\n}
+ \frac{1}{24} R_{\mu\nu\a\b} R^{\g\b\r\n} R_{\g \ \ \ \r}^{\ \a\m}\tag{7.12}\\
W^{(3)}_{\mu} = & \partial_{\mu}\chi^{(3)},\ \ 
\qquad
\chi^{(3)}\te  = \frac{1}{32}\alpha'^3
R_{\mu\nu\rho\sigma} R^{\mu\nu\rho\sigma}.
\tag{7.13}
\end{align}
Hence $c_2$ in (7.6) is equal to $-3/16$ (as we have already noted,
$c_2$ is RS-dependent) and $W_{\mu}$ is indeed nonvanishing (if
$B_{\mu\nu}\neq 0$, $W_{\mu}$ is nonvanishing already in the 2-loop
approximation [{52}], see below).

Let us now consider the general case of $B_{\mu\nu}\neq 0$. The
corresponding 1-loop $\beta^{G}$ and $\beta^{B}$ functions were found
in Ref.~76
\begin{equation}
\beta^{G}_{\mu\nu}\te 
= \alpha' \Big(
R_{\mu\nu} - \frac{1}{4} H_{\mu\rho\sigma} H_{\nu}{}^{\rho\sigma}
\Big) + O(\alpha'^2),
\qquad
\beta^{B}_{\mu\nu}
= -\frac{\alpha'}{2} \D^{\rho} H_{\rho\mu\nu}
+ O(\alpha'^2).
\tag{7.14}
\end{equation}
The 1-loop terms in the renormalization functions $S_{1}$, $U_{1}$
in (5.39) were computed in [54]
\begin{align}
S_{1G\m}\!\cdot\! F^{G}
&= \alpha' \Big(
\tfrac{1}{2} \D^\l  F^{G}_{\l\m}
- \tfrac{1}{2} \D_\m F^{G\l}_\l \Big) + O(\alpha'^2),
\nonumber\\
S_{1B\m}\!\cdot\! F^{B}
&= -\tfrac{1}{2}\alpha' H_{\mu}^{\ \rho\sigma} 
F^{B}_{\r\s} + O(\alpha'^2),
\tag{7.15}
\\
U_{1G\m}\!\cdot\! F^{G} &= O(\alpha'^2),
\qquad
U_{1B\m}\!\cdot\! F^{B} = \alpha' \D^{\l}  F^{B}_{\l\nu}
+ O(\alpha'^2).
\nonumber
\end{align}
In constructing $W_{\mu}$ and $L_{\mu}$ from (7.15) one should add the
counterterms which cancel the terms which are not invariant under
$\delta B_{\mu\nu}=\partial_{[\mu}\lambda_{\nu]}$
(this is the price we have to pay for the use of the dimensional
regularization  [{54}]). The gauge-invariant vectors can be dependent only on
$G_{\mu\nu}$, $R_{\mu\nu\rho\sigma}$, $\D_\m$  and $H_{\mu\nu\rho}$. In general,
\begin{equation}
W_{\mu}
= \alpha'^2 \Big(
a_1 \D_{\mu} R
+ a_2 \D_{\mu} H^2{\a\b\g}
+ a_3 H_{\mu\a\b} \D_\s H ^{\sigma\a\b}
\Big) + O(\alpha'^3),
\tag{7.16}
\end{equation}
\[
L_{\mu} = \alpha'^2 b_1  H_{\m\a\b} \D_{\s} H^{\s\a\b} + O(\alpha'^3).
\]
Here the $\a'$ appear according to the rule:
$W_{\mu}, L_\m \sim \a' f(\a' G, \a'^{-1} B)$
and we have dropped the trivial gradient terms in $L_{\mu}$. Hence $W_\m$
and $L_{\mu}$ necessarily vanish in the 1-loop approximation. Since,
as discussed above, $W_\m(B=0)=O(\alpha'^3)$, \ \ 
$a_1=0$  [53].
Given (7.14), (7.15) and $W_\m,L_{\mu}=O(\alpha'^2)$
it is easy to check that the identities (5.48), (5.49) are satisfied if the
2-loop dilaton $\beta$-function is given by [{54}]
\begin{equation}
\beta^{\phi}\te 
= \frac{1}{6}D
- \frac{1}{2}\alpha' \D^2 \phi
- \frac{1}{24}\alpha' H_{\mu\nu\rho}^2
+ O(\alpha'^2).
\tag{7.17}
\end{equation}
The expressions for the 2-loop terms in $\beta^{G}$ and $\beta^{B}$
(correcting the result of Ref.~77) were found in Refs.~78, 52 (see also
Ref.~79). For example,
\begin{align}
\beta^{G(2)}_{\mu\nu}
= \tfrac{1}{2}\alpha'^2 \Big[
& R_{\mu\alpha\beta\gamma} R_{\nu}{}^{\alpha\beta\gamma}
+ \tfrac{1}{2} R_{\mu\alpha\beta\nu} (H^2)^{\alpha\beta}
+ \tfrac{1}{8} H_{\mu\alpha\l} H_{\nu\b }^{\ \ \l}  (H^2)^{\alpha\beta}
- \tfrac{1}{2} R_{(\mu}^{\ \alpha\beta\g} H_{\n\a\l} 
  H_{\beta\g}^{\  \ \l}   
  \nonumber\\
& - \tfrac{1}{4} \D_{\l}H_{\alpha\beta\gamma}
  \D_{\nu}H^{\alpha\beta\gamma}
+ \tfrac{1}{12} \D_{\alpha}H_{\mu\beta\gamma}
  D^{\alpha}H_{\nu}{}^{\beta\gamma}
+ \tfrac{1}{8} (H^4)_{\mu\nu}
\Big],
\tag{7.18}
\end{align}
\[
(H^2)_{\mu\nu} = H_{\mu\alpha\beta} H_{\nu}{}^{\alpha\beta},
\qquad
(H^4)_{\mu\nu} =
H^{\alpha\r\beta} H_{\r\s}{}^{\l}
H^\s_{\ \b(\m} H_{\n) \l\a}.
\]
This expression is given for a particular choice of RS corresponding to the
following prescription for the $\ve^{ab}\ve^{cd}= (d-1)^{-1} ( \d^{ac} \d^{bd} - \d^{ad} \d^{bc})$, \ $d=2+\ve$, i.e.
$\ve^{ab} \ve_{ac}= \d^b_c$.
This is the unique prescription  in which  all the terms  in $\b^G$ and $\b^B$   contain at least one power 
of  the generalized curvature with torsion
$\hat R$  for the connection $\Gamma^\l_{\m\n} = \Gamma^\l_{\m\n} - {1\over 2} H^\l_{\ \m\n}$ (see Refs.~80, 76)
\begin{equation}
\beta^{G}_{\mu\nu} = \beta_{(\mu\nu)},
\qquad
\beta^{B}_{\mu\nu} = \beta_{[\mu\nu]},
\tag{7.19}
\end{equation}
\[
\beta_{\mu\nu}
= \alpha' \hat R_{\mu\nu}
+ \tfrac{1}{2}\alpha'^2
\Big(
\hat R_{\mu\alpha\beta\gamma}
\hat R_{\nu}{}^{\alpha\beta\gamma}
- \tfrac{1}{2} \hat R_{\mu\alpha\b\g}
\hat R_{\ \ \ \nu}^{\b\g\alpha}
+ \tfrac{1}{2} \hat R_{\a\mu\n\beta}
(H^2)^{\alpha\beta}
\Big)
+ O(\alpha'^3).
\]
Hence it is in this RS that $\beta^{G}$ and $\beta^{B}$ vanish for
parallelizable spaces $\hat R=0$  [76,81,52].

 An indirect argument shows
that$^{52}$
\begin{equation}
W_\m = \partial_\m  \chi^{(2)} + O(\alpha'^3),
\qquad
\chi^{(2)} =- \tfrac{\alpha'^2}{24} H^2_{\l\mu\nu},
\qquad
L_{\mu} = O(\alpha'^3),
\tag{7.20}
\end{equation}
\begin{align}
\beta^{\phi(2)} =
- &
\tfrac{5}{8}\alpha'^2 (H^2)^{\mu\nu} \D_{\mu}\D_{\nu}\phi
+ \tfrac{1}{16}\alpha'^2
\Big(
R_{\mu\nu\rho\sigma}^2
- \tfrac{11}{2} R^{\a\b\r\s} H_{\alpha\beta\l}
H_{\r\s}^{\ \ \l}
+ \tfrac{5}{24} H^4
\no \\  &+ \tfrac{11}{8}(H^2_{\m\n})^2
+ \tfrac{4}{3} \D_{\mu}H_{\a\rho\sigma}
D^{\mu}H^{\a\rho\sigma}\Big) .
\tag{7.21}
\end{align}
Equation (7.21) corresponds to the same RS as Eqs.~(7.18) and (7.19).
Notice that in this RS $\Omega_2^{\mu\nu}$ in (7.5) contains the $(H^2)^{\mu\nu}$
term. It should be possible to derive (7.21) employing again the identities
(5.48) and (5.49) and Eqs.~(7.15) and (7.19). Combining (7.17) with (7.21)
we find that for the group manifolds
\begin{equation}
\hat R_{\mu\nu\rho\sigma}=0,
\qquad
H^4 = \tfrac{1}{2}(H^2_{\mu\nu})^2 = 8 R^2_{\m\n}  = 8 D^{-1} R^2,
\tag{7.22}
\end{equation}
\[
\beta^{\phi}\big|_{\phi=\text{const}}
= \tfrac{1}{6}\Big[
D - \alpha' R + \alpha'^2 D^{-1} R^2
+ O(\alpha'^3)
\Big].
\]
Equation (7.20) is in agreement with the expansion of the exact result for
the central charge of the $\sigma$ model on the group manifold   [{82}]
\begin{equation}
c=\frac{D}{1+\tfrac{1}{2}c_V K^{-1}}
= D-\alpha' R + \alpha'^2 D^{-1} R^2 + O(\alpha'^3),
\qquad
\alpha' K = 1, \quad c_V D = 4R .
\tag{7.23}
\end{equation}
Note that for the group manifolds $W_\m$ in (7.20) is equal to zero
($R_{\mu\nu}=\tfrac14 H^2_{\mu\nu}$, $R=\tfrac14 H^2$, $\hat R=0$) and hence
for $\phi=\text{const}$,
$\bar \b^G=\beta^G_{\mu\nu}=0, \ \  \bar \b^B=\beta^B_{\mu\nu}=0$,
$\bar \beta^{\phi}=\beta^{\phi}=\tilde \beta^{\phi}=\text{const}$.

In addition to (5.56), the perturbation theory results (7.12), (7.21) suggest
that (for a proper choice of $G_{\mu\nu}$) the dependence of $\beta^{\phi}$
on $\phi$ may be given simply by
\begin{equation}
\beta^{\phi} = -\tfrac{1}{2}\alpha' \D^2\phi + \omega(G,B),
\qquad
\text{i.e. }\qquad \gamma=\tfrac{1}{2}\alpha' \D^2,
\tag{7.24}
\end{equation}
to all orders in the loop expansion.

The above results provide the possibility to check that there exists such an
RS in which the Weyl invariance conditions $\bar{\beta}^{i}=0$ are
equivalent (order by order in $\alpha'$) to the equations of motion for the
action (6.33). Consider first the $\alpha'^2$ approximation in the absence of
$B_{\mu\nu}$. We find from (7.3)–(7.9)   [34,53]
\begin{equation}
\tilde{\beta}^{\phi}
= \bar \beta^{\phi}-\tfrac14\bar  \beta^G_{\mu\nu} G^{\mu\nu}
= \omega_0
-\tfrac14 \alpha'
\Big(
R + 4 \D^2\phi - 4 \partial_\mu\phi\,\partial^\mu\phi
+ \tfrac14 R_{\mu\nu\rho\sigma}^2
\Big)
+ O(\alpha'^3),
\qquad
\omega_0=\tfrac16 D ,
\tag{7.25}
\end{equation}
(in string theory $\omega_0=\tfrac16(D-26)$).
A straightforward computation shows  [34] that the action (6.1) with
$\tilde{\beta}^{\phi}$ given by (7.25) satisfies the relation (6.44) where
$\kappa_{ij}$ is nondegenerate. The same statement remains true when we
account for the $\alpha'^3$ terms (7.11), (7.12). After the field
redefinition which eliminates the
$R^\m_{\ \a\b\g } R^{\nu\a\b\g} \D_\m\D_\n\phi$ term present in
$\beta^{\phi}$ (7.12) we find that $S$ (6.34) takes the following form   [{83,84}]
\begin{align}
S = & \int d^D y\,\sqrt{G}\,e^{-2\phi}
\Big\{
 \tfrac16(D-26)
- \tfrac14 \alpha'
\Big[
R + 2\D^2\phi
+ \tfrac14 R_{\mu\nu\rho\sigma}^2
\nonumber\\
& + \alpha'^2 \Big(
\tfrac{1}{16}
R_{\mu\nu\rho\sigma} R^{\rho\sigma\alpha\beta}
R_{\alpha\beta}{}^{\mu\nu}
- \tfrac{1}{12}
R^{\mu\nu}_{\ \ \a\b}    R_{\n\l}^{\ \ \b\g}    R^{\l\ \ \a}_{\ \m \g}
\Big)
+ O(\alpha'^3)\Big]
\Big\}.
\tag{7.26}
\end{align}
As was found in Ref.~84 the relation (6.44) can be satisfied only if
$\kappa_{ij}$ contains differential operators acting on $\bar\beta^{i}$.

Computing ${\beta}^{\phi}-\tfrac14 \beta^G_{\mu\nu}G^{\mu\nu}$ for the
case of $H_{\mu\nu\rho}\neq 0$ we find that after a field redefinition it is
possible to transform $S$ (6.34) into
\begin{align}
S = \int d^D y\,&\sqrt{G}\,e^{-2\phi}
\Big\{ 
 \tfrac16(D-26)
- \tfrac14 \alpha'
\Big[
R + 2\D^2 \phi  - \tfrac{1}{12} H_{\mu\nu\rho}^2
\nonumber\\
&+ \tfrac14 \alpha'^2 \Big(
R_{\mu\nu\rho\sigma}^2
- \tfrac12 R_{\mu\nu\rho\sigma}
H^{\mu\nu}{}_{\lambda} H^{\rho\sigma\lambda} \nonumber\\  & 
+ \tfrac{1}{24} H_{\mu\nu\l}
H^{\nu}_{\ \r\a}
H^{\r\s\l}
H_\s ^{\ \m\a} 
- \tfrac18
H_{\mu\a\b} H_{\n}^{\ \a\b}  
H^{\m\r\s}
H_{\r\s}^{\ \ \n} 
\Big)
+ O(\alpha'^3)\Big]
\Big\}.
\tag{7.27}
\end{align}
One can prove that the corresponding equations of motion
$\delta S/\delta\varphi^{i}=0$ are perturbatively equivalent to
$\bar{\beta}^{i}=0$, where $\bar{\beta}^{i}$ are given by
(7.18)–(7.21)  [52,85].
 It was also checked that it is possible to satisfy
the off-shell relation (6.44)  [85,86].

It is remarkable that the action (7.26), (7.27) possesses also another
property: it reproduces the massless sector of the bosonic string $S$ matrix (to
the given order in $\alpha'$). Hence it coincides (modulo field
redefinitions) with the bosonic string theory effective action (see
Refs.~18, 87, 83 and 52). As we have already discussed in Secs.~4 and~6 an
explanation of this coincidence is based on the possibility to re-write the
``central charge'' action (6.1) in the form (6.38) and (6.43) and on the
existence of the $\sigma$ model representation for the effective action
(4.12), (4.15). The perturbative results for the $\beta$ functions and for the
EA support, in turn, the validity of the $\sigma$ model extension of the first
quantized string theory described in Sec.~4.

The $\sigma$ model approach together with the conjecture (7.24) implies that
there exists a choice of fields for which the EA is given by
\begin{equation}
S = \int d^D x\,\sqrt{G}\,e^{-2\phi}
\Big\{
\tfrac16(D-26)
- \tfrac14 \alpha'
\Big[
R + 4(\partial\phi)^2
- \tfrac{1}{12} H_{\mu\nu\rho}^2\Big]
+ f(G,B)
\Big\}.
\tag{7.28}
\end{equation}
If $f$ depends only on $G$ and $B$ then  all the dependence of the EA on the
dilaton is known explicitly. An obvious consequence of (7.28) is that
\[
\frac{\delta S}{\delta \phi}
= -2 \sqrt{G}\,e^{-2\phi}\,\tilde{\beta}^{\phi}
\]
 and hence the effective Lagrangian vanishes in the stationary point.

In conclusion, let us emphasize once more that the $\sigma$-model approach is
more general than the $S$-matrix one: the effective action reconstructed from
the on-shell amplitudes is defined only for $D=26$ and hence does not admit
some conformal $\sigma$ models (e.g.\ group spaces) as its stationary points.

\newpage

\section{Concluding Remarks}

Let us summarize some ideas of the $\sigma$-model approach to
``first quantized'' string theory. If a 2$d$ theory describing string
propagation on a background is Weyl invariant (so that the conformal factor of
a 2-metric decouples from the Polyakov path integral) then we know how to
compute the amplitudes as expectation values of the corresponding vertex operators. The path integral approach reproduces the
standard dual model amplitudes in the case of the trivial ``tree'' vacuum
$G_{\mu\nu}=\delta_{\mu\nu}$, $D=26$. 

To avoid the infinite factors in the
amplitudes one may divide the path integral measure by the volumes of the
classical symmetry groups, i.e.\ the diffeomorphism group {\it and} the Weyl group.
Having computed the massless on-shell amplitudes on the flat background we
can reconstruct an effective field theory action, which reproduces them. Then
we find that the following remarkable fact is true. The extremals of the
effective action correspond to Weyl invariant $\sigma$ models. Hence it is
encoded in some miraculous way in the string amplitudes on a flat background
that it is the Weyl invariance principle that selects nearby tree vacua.

To know a theory is to know its  action or equations of motion and not only a
set of its vacua. The existence of the effective action suggests that it
should be possible to make a smooth ``interpolation'' between the path
integrals corresponding to different Weyl invariant backgrounds (what is
commonly understood by the Polyakov path integral is defined only for a Weyl
invariant background). As we have suggested above, such an interpolation is
provided by the ``string'' $\sigma$ model path integral which is computed with
a Weyl gauge fixed, even though the theory is not Weyl invariant ``off-shell''.
Such an ``off-shell'' extension seems reasonable since
\begin{itemize}
\item[(i)] it reproduces the standard results in the case of a Weyl invariant background, and
\item[(ii)] it provides a path integral representation for the effective action which is perfectly consistent with the effective action -- $\sigma$-model equivalence.
\end{itemize}
We emphasize that our proposal is to consider the ``off-shell'' path integral
without integration over the conformal factor of a 2-metric. This, in
principle, may not be a unique possibility. The original Polyakov’s sum over
surfaces approach [38] was based on integrating over the conformal factor and hence
was defined for arbitrary backgrounds, e.g.\ for $G_{\mu\nu}=\delta_{\mu\nu}$,
$D\le 26$ (recently an important progress was made in this approach  [{88}]).
It may seem that going out of a Weyl invariant point we are to account for the
dynamics of the conformal factor. Our point of view is that the approach based
on integrating over the conformal factor, while being natural from a 2-d
quantum gravity and ``noncritical'' string theory point of view, is not
relevant for the ``critical'' string theory in which the ``on-shell'' Weyl
invariance is one of the basic dynamical principles.

If we integrate over the conformal factor, there is no reason to anticipate
that the Weyl invariance will happen to be a dynamical principle. We also get
the infinite Weyl group volume factors in the amplitudes corresponding to Weyl
invariant backgrounds.\foot{This situation may be compared with an attempt to compute the off-shell
amplitudes in a gauge theory without fixing a gauge. It is difficult to have a
smooth on-shell limit in this case.}

 Next, if we integrate over the full 2-metric,
the corresponding 2-d theory is practically intractable for nontrivial
backgrounds (the recent progress  [{88}] was made only for the flat
background). A compelling argument against the integration over the conformal
factor is provided by the equivalence between the effective equations of motion
(reconstructed from the string amplitudes on a flat background) and the
conditions of Weyl invariance of the ordinary $\sigma$ model (defined on a
2-space with a {\it fixed}  2-metric).

While reasonable in the tree approximation the prescription of fixing a Weyl
gauge in computing the ``off-shell'' path integral appears to be ambiguous at
higher loop orders. In fact, we need an additional prescription of how to fix
the metrics on higher genus surfaces (i.e.\ how to split them into a conformal
factor and moduli) in a mutually consistent way. It seems  that some
reasonable prescription should exist. 

A satisfactory solution to this problem
would be to represent the sum of path integrals over surfaces of fixed genera
as a single path integral over infinite genus surfaces. Since the boundary of
the moduli space of genus $n+1$ surfaces is the moduli space of genus $n$
surfaces this would relate the choices of moduli on surfaces of different
genera. If the integrals over the infinite number of moduli can be represented
as a functional integral over a new (``topological'') 2-d quantum field\footnote{This field may essentially
coincide with the 2-metric. Then the approach of Ref.~88 may in fact be
relevant for the problem of determining nonperturbative string vacua.}
 then
we would get a nontrivial generalization of the $\sigma$ model. A generalized
conformal invariance condition for this $\sigma$ model may be equivalent to
loop corrected string equations of motion and hence may provide information
about nonperturbative string vacua.

\subsection*{Acknowledgments}
The author would like to thank R.E. Kallosh, A.Yu. Morozov, A.A. Rosly, A.S. Schwarz, A.M. Semikhatov, I.V. Tyutin, M.A. Vasiliev and B.L. Voronov for their encouragement and useful conversations.

\end{document}